%% file: main.tex
\documentclass[paper,notoc]{JHEP3}
\usepackage{amsmath,amssymb}
\usepackage{epsfig,cite}
\usepackage{graphicx}
\usepackage{axodraw4j}
\usepackage{comment}
\usepackage{xkvltxp}

\def\beq{\begin{equation}}   
\def\eeq{\end{equation}}
\def\bea{\begin{eqnarray}}  
\def\eea{\end{eqnarray}} 
\def\f21{{}_2F_{1}}
\def\eps{\epsilon}
\def\dhalf{\frac{d}{2}}

\def\beq{\begin{equation}}
\def\eeq{\end{equation}}
\def\bsp#1\esp{\begin{split}#1\end{split}}

\title{
Real-Virtual contributions to the inclusive Higgs cross-section at N$^3$LO
}

\author{Charalampos Anastasiou\\
  ETH Zurich, 8093 Zurich, Switzerland\\
  E-mail: \email{babis@phys.ethz.ch}}

\author{Claude Duhr\\
Institute for Particle Physics Phenomenology, University of Durham,\\
Durham, DH1 3LE, United Kingdom\\
  E-mail: \email{claude.duhr@durham.ac.uk}}

\author{Falko Dulat\\
  ETH Zurich, 8093 Zurich, Switzerland\\
  E-mail: \email{dulatf@phys.ethz.ch}}

\author{Franz Herzog\\
  CERN Theory Division, CH-1211,
Geneva 23, Switzerland.\\
  E-mail: \email{franz.herzog@cern.ch}}

\author{Bernhard Mistlberger \\
  ETH Zurich, 8093 Zurich, Switzerland\\
  E-mail: \email{bmistlbe@phys.ethz.ch}}

\preprint{IPPP/13/92, DCTP/13/184, CERN-PH-TH/2013-258, MCnet-13-18}	
\abstract{ 
We compute the contributions to the N$^3$LO inclusive Higgs boson
cross-section from the square of one-loop amplitudes with a Higgs
boson and three QCD partons as external states. Our result is a Taylor
expansion in the dimensional regulator $\epsilon$, where the
coefficients of the expansion are analytic functions of the ratio of
the Higgs boson mass and the partonic center of mass energy and they are 
valid for arbitrary values of this ratio.
We also perform a threshold expansion around the limit of
soft-parton radiation in the final state. The expressions for the coefficients of the
threshold expansion are valid for arbitrary values of the dimension.
As a by-product of the threshold expansion calculation, we have
developed a soft expansion method at the integrand level by identifying
the relevant soft and collinear regions for the loop-momentum.
}


\begin{document}

\catcode`\@=11
\font\manfnt=manfnt
\def\Watchout{\@ifnextchar [{\W@tchout}{\W@tchout[1]}}
\def\W@tchout[#1]{{\manfnt\@tempcnta#1\relax%
  \@whilenum\@tempcnta>\z@\do{%
    \char"7F\hskip 0.3em\advance\@tempcnta\m@ne}}}
\let\foo\W@tchout
\def\dubious{\@ifnextchar[{\@dubious}{\@dubious[1]}}
\let\enddubious\endlist
\def\@dubious[#1]{%
  \setbox\@tempboxa\hbox{\@W@tchout#1}
  \@tempdima\wd\@tempboxa
  \list{}{\leftmargin\@tempdima}\item[\hbox to 0pt{\hss\@W@tchout#1}]}
\def\@W@tchout#1{\W@tchout[#1]}
\catcode`\@=12

\input{masterpics.tex}

\section{Introduction} 
\label{jntro.tex}

The theoretical prediction of the Standard Model inclusive Higgs boson
cross-section is an important reference for the experimental
verification of the model at the energies probed by the Large Hadron
Collider.  The uncertainty which is associated with the truncation of
the perturbative expansion at NNLO is currently estimated at the order of $\pm
10\%$ (see, for example, ref.~\cite{Anastasiou:2012hx}). With the accumulation of more 
data by ATLAS and CMS, this perturbative uncertainty  will be ultimately one of the largest 
systematic uncertainties entering the extraction of the coupling
strengths of Higgs boson interactions. A calculation of the N$^3$LO
cross-section may reduce it to about $\pm 5\%$~\cite{Buehler:2013fha}.  

A computation of the N$^3$LO cross-section is therefore very much
desired. Many important steps have been taken towards this objective. 
The three-loop $gg \to h$ amplitude has been computed in refs.~\cite{Baikov:2009bg,Gehrmann:2010ue,Gehrmann:2010tu}.  
Contributions which are associated with collinear/ultraviolet counter-terms and 
the partonic cross-sections at lower orders have been computed in refs.~\cite{Buehler:2013fha,Pak:2011hs,Anastasiou:2012kq,Hoschele:2012xc}. 
N$^3$LO corrections due to processes with triple real radiation were computed in ref.~\cite{Anastasiou:2013srw} 
as a threshold expansion around the limit of soft-parton emissions. An important ingredient for the threshold expansion 
of the Higgs boson cross-section with both real and virtual radiation, the two-loop soft current, 
was presented in refs.~\cite{Duhr:2013msa,Li:2013lsa}.  

In this article, we focus on a different contribution to the N$^3$LO
cross-section, the integration over phase-space of the squared one-loop amplitudes 
for partonic processes for Higgs production in association with a 
quark or  gluon in the final-state.  We  denote these squared
real-virtual contributions as ${\rm (RV)^2 }$. 
We have been able to perform many independent calculations 
for the (RV)$^2$ cross-sections  by employing different methods.
  
First, we have two different methods that are suited for a threshold expansion. 
In method (Ia), we reduce the one-loop amplitudes to bubble
and box master integrals and find appropriate representations of the
box master integrals which allow for a trivial integration over
phase-space order by order in the threshold expansion.  
Method (Ib) is also a threshold expansion method and relies on expanding the amplitudes at the 
integrand level in the regions where particles in the loops are soft or collinear to an
external particle.  We have identified all such regions and we have proven 
that the expansion by regions yields the same result as our
first method for an arbitrary order in the threshold expansion. 
The coefficients of the threshold expansion are exact in the dimensional regulator $\epsilon=2-\dhalf$. 

Methods (II) and (III) yield expressions of the partonic
cross-sections which are valid for arbitrary values of the partonic center of
mass energy.  Method (II) treats the combined loop and phase-space integrations as a three-loop forward
scattering amplitude, which is reduced to master integrals with the reverse-unitarity 
method~\cite{Anastasiou:2002yz,Anastasiou:2002qz,Anastasiou:2003yy,Anastasiou:2003ds}. 
The master integrals themselves are evaluated by solving the differential equations they satisfy 
in terms of  generalized hypergeometric functions. The solution of the differential equations has been 
achieved by observing recursion patterns for the 
hypergeometric integrals, similar to the ones for in multiple polylogarithms~\cite{Goncharov:1998kja}. Due to the hypergeometric representation of our 
master integrals a threshold expansion is easily performed, thus providing an independent verification of the results of Methods (Ia) and (Ib).
To obtain the master integrals with full kinematical dependence we solve the differential equations as an $\epsilon$ expansion in terms of 
harmonic polylogarithms. Method (III) identifies counterterms for combined loop and phase-space integrals 
and proceeds with an expansion in $\epsilon$ followed by a direct integration over phase-space variables. This is made possible by embedding the classical polylogarithms that arise in the $\eps$ expansion into a larger space of multiple polylogarithms where the integration over phase space is trivial. This embedding is achieved by deriving the required functional equations using symbols~\cite{symbolsC,symbolsB,symbols,Goncharov2010jf,Duhr2011zq} and the Hopf algebra structure of multiple polylogarithms~\cite{GoncharovGalois,Brown2011ik,Duhr2012fh}.

This article is organized as follows. In Section~\ref{sec:setup} we
introduce the partonic cross-sections we are computing and define
our notation. In Section~\ref{sec:results} we present our results for 
the partonic cross-sections. In Section~\ref{sec:methods} we detail the methods employed for the purposes of our calculation. 
We present our conclusions in Section~\ref{sec:conclusions}.

\section{The real-virtual squared cross section} 
\label{sec:setup} 

We consider the partonic-processes for associated Higgs production,
\beq\bsp
g(p_1)+g(p_2) &\,\to  g(p_3) + H(p_4) \\ 
q(p_1)+ g(p_2) &\,\to  q(p_3) + H(p_4) \\ 
q(p_1)+ \bar q (p_2) &\,\to  g(p_3) + H(p_4)   
\esp\eeq
where $g,q,\bar q$ are symbols for gluon, quark and anti-quark partons correspondingly
and $H$ for the Higgs boson. The brackets refer to the momenta of the
particles. We define the kinematic invariants,
\begin{equation}
\bsp
s &\,\equiv 2 p_1 \cdot p_2 + i0, \quad  t \equiv 2 p_2 \cdot p_3-i0,  \quad  u
\equiv 2 p_1 \cdot p_3-i0,\\
p_4^2 &\,= (p_1+p_2-p_3)^2= s-t-u=M_h^2+i0\,,
\esp
\end{equation}
where we indicated explicitly the small imaginary parts carried by them.
The partonic cross-sections for these processes are given by
\beq\label{eq:xsec0}
\sigma_{X} = \frac{\mathcal{N}_X}{2s}\,\int d\Phi_2\,\sum\left| {\cal A}_{X}\right|^2\,,
\eeq
where $X\in\{gg\to Hg, gq\to Hq, q\bar{q}\to Hg\}$ labels the different subprocesses\footnote{In the following we will suppress the dependence on the final state as it can always be inferred from the initial state for the processes we consider.} and the sum symbol denotes a summation over
colors and polarizations of the initial and final state
particles. We work in conventional regularization in $d=4-2\eps$ for both the phase space and the matrix element. The $d$-dimensional phase-space measure is given by
\beq
d\Phi_2= (2\pi)^d\,\delta^{(d)}\big(p_1+p_2-p_3-p_4) \,\frac{d^dp_3}{(2\pi)^{d-1}}\,\delta_+(p_3^2)\,\frac{d^dp_4}{(2\pi)^{d-1}}\,\delta_+(p_4^2-M_h^2)\,.
\eeq
$\mathcal{N}_X$ denote the averaging over initial state spins and colors in $d$ dimensions,
\beq
\mathcal{N}_{gg} = \frac{1}{4 V^2 (1-\epsilon)^2 }\,,\qquad
\mathcal{N}_{gq} = \frac{1}{4 VN (1-\epsilon)^2 }\,,\qquad
\mathcal{N}_{q\bar{q}} =  \frac{1}{4 N^2}\,,
\eeq
with $N$ and $V\equiv N^2-1$ the number of quark and gluon colors respectively.

In the following it will be convenient to parametrize the invariants as
\beq
\label{eq:mandelstam}
s = \frac{M_h^2}{z}, \quad t = s\, \delta \, \lambda, \quad u = s\,\delta \, (1-\lambda)\,,
\end{equation}
with $\delta = 1-z$. Note that a physical scattering process corresponds to $s>0$ and $0<z,\lambda<1$, and the limit $\delta \to1$ corresponds to the threshold region where the additional final state parton is soft.
Using the parametrization~\eqref{eq:mandelstam} the phase-space measure becomes
\beq\label{eq:PS_measure}
d\Phi_2 = \frac{(4\pi)^\eps\,s^{-\eps}\,\delta^{1-2\eps}}{8\pi\,\Gamma(1-\eps)}\,d\lambda\,[\lambda\,(1-\lambda)]^{-\eps}\,\Theta(\lambda)\,\Theta(1-\lambda)\,,
\eeq
where $\Theta(x)$ is the Heaviside step function. Equation~\eqref{eq:xsec0} then reads
\beq\label{eq:xsec}
\sigma_{X} = s^{-1-\eps}\,\frac{\mathcal{N}_X\,(4\pi)^{\eps}}{16\pi\,\Gamma(1-\eps)}\,\delta^{1-2\eps}\,\int_0^1 d\lambda\,[\lambda\,(1-\lambda)]^{-\eps}\,\sum\left| {\cal A}_{X}\right|^2\,.
\eeq

In the rest of this paper we compute in an effective theory of the Standard Model
where the top-quark is decoupled. The effective unrenormalized
Lagrangian reads: 
\begin{equation}
{\cal L} = {\cal L}^{\rm eff}_{\rm QCD} - \frac{1}{4}\,C\, H\, G_{\mu \nu}^a G^{a,\mu \nu},   
\end{equation}
where the first term corresponds to an effective QCD Lagrangian with
$N_f=5$ flavors.
The Wilson coefficient $C$ can be cast as a function of the QCD coupling, the bare heavy-quark masses and the Higgs
field vacuum expectation value~\cite{Baikov:2006ch,Furlan:2011uq,Anastasiou:2010bt,Chetyrkin:1997iv,Kramer:1996iq}.  

Performing a loop-expansion of the amplitudes ${\cal
  A}_X=\sum_{j=0}^{\infty} {\cal A}_X^{(j)}$ in the effective theory
with $j$ being the number of loops, we have: 
\begin{equation}
\left| {\cal A}_X \right|^2 = \left| {\cal A}_X^{(0)} \right|^2 + 
2 \Re \left( {\cal A}_X^{(0)} {{\cal A}_X^{(1)}}^*\right) 
+\left[ 
\left| {\cal A}_X^{(1)} \right|^2 + 
2 \Re \left( {\cal A}_X^{(0)} {{\cal A}_X^{(2)}}^*\right)
\right]
+\ldots
\end{equation}
The first two terms of the above expansion enter the already known 
inclusive Higgs boson cross-section through NNLO~\cite{Graudenz:1992pv,Dawson:1990zj,Djouadi:1991tka,Spira:1995rr,Harlander:2002wh,Anastasiou:2002yz,Ravindran:2003um}. The  third term 
in square brackets contributes to the N$^3$LO coefficient. 
In this article, we compute the part of the
partonic cross-section due to the square  of the one-loop
amplitudes, namely, 
\beq\label{eq:xsec1}
\sigma_{X}^{1 \otimes 1} = s^{-1-\eps}\,\frac{\mathcal{N}_X\,(4\pi)^{\eps}}{16\pi\,\Gamma(1-\eps)}\,\delta^{1-2\eps}\,\int_0^1 d\lambda\,[\lambda\,(1-\lambda)]^{-\eps}\,\sum\left| {\cal A}_{X}^{(1)}\right|^2\,.
\eeq

\section{Results}
\label{sec:results}

The computation of the cross-sections~\eqref{eq:xsec1} for the different subprocesses is the main subject of this paper. 
We evaluated the phase-space and loop integrals in different ways that are detailed in Section~\ref{sec:methods}. 
In this section, we summarize first our main findings. 

We find for the partonic cross-sections:
\beq\label{eq:xsections_master}
\sigma_{X}^{1 \otimes 1} = \frac{\pi\, \omega_\Gamma\, \left| C \right|^2}{256\,s}\,
\left( \frac{4 \pi}{s} \right)^{3 \epsilon} \left( \frac{\alpha_s}{ \pi}\right)^3   
\mathcal{C}_X\,\Sigma_X(z;\eps)\,,
\eeq
where $g_s^2 \equiv 4 \pi \alpha_s$, and we have 
\beq\bsp
\mathcal{C}_{gg} = \frac{N}{V\,(1-\eps)^2}\,,\qquad
\mathcal{C}_{qg} = \frac{1}{N\,(1-\eps)}\,,\qquad
\mathcal{C}_{q\bar{q}} = \frac{V}{N^2}\,.
\esp\eeq
In eq.~\eqref{eq:xsections_master} we have introduced the quantity
\begin{equation}
\omega_\Gamma \equiv \frac{c_{\Gamma}^3}{\Gamma(1+\eps)\,\Gamma(1-\eps)} = \frac{\Gamma^2(1+\epsilon)\, \Gamma^5(1-\epsilon)}{\Gamma^3(1-2 \epsilon)}\,,\qquad c_\Gamma=\frac{\Gamma(1+\eps)\,\Gamma^2(1-\eps)}{\Gamma(1-2\eps)}\,.
\end{equation}

The function $\Sigma_{gg}(z;\epsilon)$ has a pole as $z \to 1$. This pole constitutes the soft singularity of the gluon initiated cross-section and it will only be 
remedied by integrating the partonic cross-section with parton distribution functions. We  separate this singular part manifestly from the remainder and write:
\begin{equation}
\Sigma_{gg}(z;\eps) = \Sigma_{gg}^{{\rm sing}}(z;\epsilon)  +\Sigma_{gg}^{{\rm reg}}(z;\epsilon) 
\end{equation}
where the $z \to 1$ singular part is given by 
\beq\bsp
\Sigma_{gg}^{{\rm sing}}(z;\epsilon) &\,=
-(1-z) ^{-1-2\epsilon} \frac{8 N^2 \left(\epsilon ^3+2 \epsilon ^2-3 \epsilon +1\right)^2}{(1-2 \epsilon )^2 (1-\epsilon ) \epsilon ^5} \\
&-(1-z) ^{-1-4\epsilon} \frac{4 N^2   \left(\epsilon ^3+2 \epsilon ^2-3 \epsilon +1\right) \Gamma (\epsilon +1) \Gamma (2 \epsilon +1) \Gamma (1-2 \epsilon )^4}{ (1-2 \epsilon )   \epsilon ^5 \Gamma (1-4 \epsilon )^2 \Gamma (1-\epsilon ) \Gamma (4 \epsilon +1)} \\
&-(1-z) ^{-1-6\epsilon} \frac{2 N^2 (1-\epsilon ) \Gamma (1-3 \epsilon )^2 \Gamma (\epsilon +1)^2 \Gamma (1-2 \epsilon )}{3  \epsilon ^5 \Gamma (1-6 \epsilon)} .
\esp\eeq
The cross-sections of the $q \bar q$ and $q g$ channels are regular in the limit $z \to 1$: 
\beq\bsp
\Sigma_{qg}(z;\eps) &\,= \Sigma_{gq}^{{\rm reg}}(z;\epsilon)\,,\\
\Sigma_{q\bar q}(z;\eps) &\,= \Sigma_{q\bar{q}}^{{\rm reg}}(z;\epsilon)\,.
\esp\eeq

We have calculated the regular functions $\Sigma_X^{{\rm reg}}(x;\epsilon)$ as an expansion in the dimensional regulator $\epsilon$ through ${\cal O}(\epsilon^0)$. 
The expressions are composed of multiple polylogarithms (MPLs) ~\cite{Goncharov:1998kja} with weights up to five. 
Specifically, the MPLs are defined as iterated integrals via
\beq
G(a_1,\dots,a_n;z) = \int_0^z\frac{dt}{t-a_1}G(a_2,\dots,a_n;t),
\end{equation}
where $G(z)=1$ and $z,a_i \in \mathbb{C}$ and the weight is the number of indices $a_i$.
All our MPLs have argument $z$ and all $a_i$ are elements of the set $\{-1,0,1\}$. The last property allows to relate our MPLs to a more 
restrictive class of functions, the so-called harmonic polylogarithms (HPLs)~\cite{Remiddi:1999ew}. The relation is given explicitly by 
\begin{equation}
H(a_1,\dots,a_n;z) = (-1)^{\sigma(a_1,\dots,a_n)}G(a_1,\dots,a_n;z),
\end{equation}
where $\sigma(a_1,\dots,a_n)$ denotes the number of elements in the set $\{a_1,\dots,a_n\}$ that are equal to $1$. 
Software to evaluate HPLs numerically in a fast and accurate way (at least up to weight four) is publicly available~\cite{Gehrmann:2001pz,Vollinga:2004sn,Maitre:2007kp,Maitre:2005uu,Buehler:2011ev}.
Due to the magnitude of the expressions obtained for the functions $\Sigma_X^{{\rm reg}}(x;\epsilon)$ we refrain from stating them here explicitly and make them publicly 
available together with the arXiv submission of this article and on the web-page \cite{xsdownload}. In the file \textsf{results.txt} we present the formulae for the $\Sigma_X^{{\rm reg}}$ 
formatted in Maple input form.


Given that the triple-real contributions to the inclusive Higgs boson cross-section at N$^3$LO~\cite{Anastasiou:2013srw} have only been computed as 
an expansion around $ z = 1-\delta  \to 1$, we also provide the same expansion for the ${\rm (RV)}^2$ cross-section functions $\Sigma_X(z,\eps)$ of this article.  
We have discovered a characteristic structure for $\Sigma_X(z,\eps)$ in the $\delta \to 0$ limit; 
all logarithmic contributions of the form $\log \delta$ exponentiate into factors of $\delta^{-a \eps} $, where $a$ is an integer in the interval $[2,6]$.  
Namely, 
\beq\label{eq:Sig_eta}
\Sigma_X(z,\eps) = \sum_{a=2}^6\delta^{-a\eps}\,\eta_X^{(a)}(\delta;\eps)\,,
\eeq
where the functions $\eta_X^{(a)}(\delta;\eps)$ are meromorphic functions of $\delta$.  
We further decompose
\begin{equation}\label{eq:eta_not_4}
\eta_X^{(a)}(\delta;\eps) = \phi_\Gamma^{(a;1)}(\eps)\, \hat \eta_X^{(a;1)}(\delta;\eps)\,, 
\end{equation}
for $a \neq 4$, while for $a=4$ we write, 
\begin{equation}\label{eq:eta_4}
\eta_X^{(4)}(\delta;\eps) = \sum_{j=1}^3 \phi_\Gamma^{(4;j)}(\eps)\, \hat \eta_X^{(4;j)}(\delta;\eps)\,, 
\end{equation}
with 
\beq\bsp
\phi_\Gamma^{(2;1)}(\eps) &\,= 1\,,\\
\phi_\Gamma^{(3;1)}(\eps) &\,=\frac{\cos (\pi  \epsilon ) \Gamma (1-2 \epsilon )^2}{\Gamma (1-3  \epsilon ) \Gamma (1-\epsilon )}\,, \\
\phi_\Gamma^{(4;1)}(\eps) &\,={\frac {  \Gamma^3 \left( 1-2\epsilon \right) }{\Gamma \left( 1-4\epsilon \right)\,   \Gamma^2 \left( 1-\epsilon \right) }}\,,  \\
\phi_\Gamma^{(4;2)}(\eps) &\,=\frac{\cos (2 \pi  \epsilon ) \Gamma (1-2 \epsilon )^3 \Gamma   (\epsilon +1)}{\Gamma (1-4 \epsilon ) \Gamma (1-\epsilon )}   \,, \\
\phi_\Gamma^{(4;3)}(\eps) &\,= {\frac {\Gamma \left( 1-2\epsilon \right)\, \Gamma \left( 1-3\epsilon \right) }{\Gamma \left( 1-4\epsilon \right)\, \Gamma \left( 1-\epsilon \right) }}\,, \\
\phi_\Gamma^{(5;1)}(\eps) &\,=\frac{\cos (\pi  \epsilon ) \Gamma (1-3 \epsilon ) \Gamma (1-2   \epsilon )^2 \Gamma (\epsilon +1)}{\Gamma (1-5 \epsilon ) \Gamma   (1-\epsilon )}\,,  \\
\phi_\Gamma^{(6;1)}(\eps) &\,={\frac {\Gamma \left( 1-2\epsilon \right)\,   \Gamma^2 \left( 1+\epsilon \right)\, \Gamma^2 \left( 1-3\epsilon \right)}{\Gamma \left( 1-6\epsilon \right) }}\,.
\esp\eeq
The structure of Eqs.~\eqref{eq:Sig_eta}--\eqref{eq:eta_4} originates from loop integrations over distinct kinematic configurations, 
where the loop momentum can be either soft ($s$), or collinear to the first incoming parton ($c_1$), or collinear to the second incoming parton ($c_2$), or, otherwise, hard $(h)$. 
Every term in the squared amplitude may be thought of as associated to a product of two regions $r_1$ and $r_2$, $r_i\in\{s,c_1,c_2,h\}$, one from the amplitude itself and 
the other from its complex conjugate. In the following we denote the contribution from such a product of regions by $(r_1,r_2)$.
The coefficients $\phi_\Gamma^{(i;j)}(\eps)\,\hat\eta_X^{(i;j)}(\delta;\eps)$ are in one-to-one correspondence with these regions.
The correspondence is explicitly given by
\beq
\bsp
\phi_\Gamma^{(2;1)}(\eps)\,\hat\eta_X^{(2;1)}(\delta;\eps) &\,\leftrightarrow (h,h)\,,\\
\phi_\Gamma^{(3;1)}(\eps)\,\hat\eta_X^{(3;1)}(\delta;\eps) &\,\leftrightarrow (c_1+c_2,h)+(h,c_1+c_2)  \,,\\
\phi_\Gamma^{(4;1)}(\eps)\,\hat\eta_X^{(4;1)}(\delta;\eps) &\,\leftrightarrow (c_1,c_2)+(c_2,c_1)\,,\\
\phi_\Gamma^{(4;2)}(\eps)\,\hat\eta_X^{(4;2)}(\delta;\eps) &\,\leftrightarrow (s,h)+(h,s)\,,\\
\phi_\Gamma^{(4;3)}(\eps)\,\hat\eta_X^{(4;3)}(\delta;\eps) &\,\leftrightarrow (c_1,c_1)+(c_2,c_2)\,,\\
\phi_\Gamma^{(5;1)}(\eps)\,\hat\eta_X^{(5;1)}(\delta;\eps) &\,\leftrightarrow (c_1+c_2,s)+(s,c_1+c_2)\,,\\
\phi_\Gamma^{(6;1)}(\eps)\,\hat\eta_X^{(6;1)}(\delta;\eps) &\,\leftrightarrow (s,s)\,.
\esp
\eeq
\\
A derivation of the above decomposition will be given in Section~\ref{sec:methods}.  The analytic form of the double soft $(s,s)$
terms $ \hat\eta_X^{(6;1)}(\delta;\eps) $ is rather simple.  We find: 
\bea
\hat{\eta}_{q\bar{q}}^{(6;1)}(\delta;\eps)=&&\frac{1}{N^2}\Bigg[\frac{\delta}{24 \epsilon ^2-16 \epsilon +2}+\frac{\delta^2 \left(-2 \epsilon ^2-2 \epsilon
   +1\right)}{24 \epsilon ^4-16 \epsilon ^3+2 \epsilon
   ^2}\nonumber\\
   &&+\frac{\delta^3 \left(3 \epsilon ^4+6 \epsilon ^3-3 \epsilon
   +1\right)}{72 \epsilon ^6-48 \epsilon ^5+6 \epsilon
   ^4}\Bigg],
\eea
\bea
\hat{\eta}_{qg}^{(6;1)}(\delta;\eps)=&&N^2 \Bigg[\frac{\epsilon -1}{12 \epsilon ^5}+\frac{\delta}{6 \epsilon ^4 (6 \epsilon
   -1)}-\frac{\delta^2 (\epsilon +1)
   \left(3 \epsilon ^3+12 \epsilon ^2-11 \epsilon +2\right)}{24
   \epsilon ^5 \left(12 \epsilon ^2-8 \epsilon
   +1\right)}\nonumber\\
      &&+\frac{\delta^3 (\epsilon -1)}{4 \epsilon ^2 \left(12
   \epsilon ^2-8 \epsilon +1\right)}-\frac{\delta^4 (\epsilon -1)}{8 \epsilon ^2 \left(12 \epsilon
   ^2-8 \epsilon +1\right)}\Bigg],
\eea
\bea
\hat{\eta}_{gg}^{(6;1)}(\delta;\eps)=&&N^2 \Bigg[+\frac{2 (\epsilon -1)}{3
   \delta \epsilon ^5}-\frac{4 (\epsilon -1)}{3 \epsilon
   ^5}+\frac{2 \delta \left(8 \epsilon ^2-7 \epsilon
   +1\right)}{\epsilon ^5 (6 \epsilon -1)}
\nonumber\\
      &&
-\frac{4 \delta^2 \left(9 \epsilon ^2-7 \epsilon
   +1\right)}{3 \epsilon ^5 (6 \epsilon -1)}-\frac{\delta^3 (3 \epsilon -1) \left(\epsilon ^4-20 \epsilon ^3+35
   \epsilon ^2-21 \epsilon +4\right)}{6 (1-2 \epsilon )^2 \epsilon ^5 (6
   \epsilon -1)}\nonumber\\
      &&+\frac{\delta^4 (\epsilon -2) (3
   \epsilon -1)}{3 (1-2 \epsilon )^2 \epsilon ^2 (6 \epsilon
   -1)}-\frac{\delta^5 (\epsilon -2) (3 \epsilon -1)}{6 (1-2 \epsilon
   )^2 \epsilon ^2 (6 \epsilon -1)}\Bigg].
\eea
The remaining $\hat{\eta}_{X}^{a;i}$ terms are more complicated combinations of generalized hypergeometric functions, 
which can be readily cast as a Laurent series in $\delta$
\[ 
\sum_{n=-1}^{\infty} c_n(\epsilon) \delta^n
\] 
Due to the size and complexity of the expressions in terms of hypergeometric functions, we provide in Appendix~\ref{sec:thresholdexp} and in the arXiv submission 
file \textsf{results.txt}~\cite{xsdownload} the terms of the series in the $\delta$ expansion up to ${\cal O}\left(\delta^4\right)$.  
With our computer programs, we have explicitly generated the terms of the series up to order ${\cal O}\left( \delta^{10} \right)$. 
In addition, as explained earlier,  we have computed the $\hat \eta_{X}^{(a;j)}$ functions for arbitrary values of $\delta$ as 
an expansion in $\epsilon$ through order ${\cal O}\left(\epsilon^0\right)$.

\section{Methods}
\label{sec:methods}
In this section,  we discuss how we evaluated the loop and 
phase-space integrals that contribute to the real-virtual squared cross-sections~\eqref{eq:xsections_master}.
We employed various methods that each have their own strengths and weaknesses. 
We checked that we obtain consistent results when comparing the
different approaches. These methods are:
\begin{enumerate}
\item {\bf Threshold expansion of the cross section:} In this 
 approach, we derive a representation of the cross-section
 as an expansion close to threshold where $\delta \to 0$. 
We first expand the loop amplitude in the
limit where the final state parton is soft, and then perform the
phase-space integration order-by-order in the expansion. 
The threshold expansion of the loop amplitude is obtained in two
different ways:  first by finding a suitable representation in terms of convergent 
hypergeometric functions within the entire phase-space, and, second,
by expanding around the relevant soft, collinear and hard regions of
the loop momentum.  We will detail our threshold expansion techniques 
in Section~\ref{sec:sigma_expansion}.
\item {\bf Differential equations for master integrals:} Using a
 duality of loop and phase-space integrals  we reduce them
 simultaneously to a minimal set of master integrals.  The master integrals 
satisfy differential equations that can be solved in two ways: 
either order-by-order in dimensional regularization in terms of
harmonic polylogarithms, or in terms of generalized hypergeometric functions. 
The boundary conditions for the differential equations are obtained by
matching to the  leading term of the threshold expansion, which we compute
with one of our threshold expansion methods mentioned above. 
We present our approach based on differential equations in Section~\ref{sec:revunit}.
\item {\bf Direct integration using multiple polylogarithms:} It is
  possible to derive analytic results of the loop integrals entering
  our amplitudes in terms  of polylogarithmic functions.  These
  expressions are singular in soft and collinear limits of the
  phase-space and we render all integrals convergent by constructing
  appropriate counterterms.  Then we perform the two-body phase-space
  integration by embedding the polylogarithmic functions into a larger
  class of multiple polylogarithms for which the integration is
  trivial. In the end, we recast  the final result in terms of
  harmonic polylogarithms only. This method is explained 
  in Section~\ref{sec:sub}.
\end{enumerate}

\subsection{Expansion of the cross-section around threshold} 
\label{sec:sigma_expansion}
We start by computing the one-loop amplitudes $A^{1}_X$ with $X
\in \{gg \to Hg, q \bar q \to H g, q g \to H q\}$ in dimensional
regularization and for arbitrary values of the regulator $\epsilon$. 
We generate the Feynman diagrams with QGRAF~\cite{Nogueira:1991ex} and 
perform the spin and color algebra using FORM~\cite{Vermaseren:2000nd}. 
Using the methods of refs.~\cite{Anastasiou:1999bn,Anastasiou:2000mf}
for tensor integrals and well established reduction techniques for
scalar integrals~\cite{Laporta:2001dd,Anastasiou:2004vj},
the amplitudes are reduced to the one-loop scalar bubble and box master integrals,
\beq\bsp\label{eq:one-loop-masters}
{\rm Bub}(s_{12}) &\,= 
 \int \frac{d^dk}{i \pi^{\dhalf}} \frac  1  { 
k^{2}\, (k+q_1+q_2)^2
}\,,\\
{\rm Box}(s_{12}, s_{23}, s_{31}) &\,= 
 \int \frac{d^dk}{i \pi^{\dhalf}} \frac  1  { 
k^{2}\, (k+q_1)^2\, (k+q_1+q_2)^2\, (k+q_1+q_2+q_3)^2
}\,,
\esp\eeq
where the $q_i$ are considered light-like and ingoing and $s_{ij}=(q_i+q_j)^2$.

In a next step, we construct the squared
one-loop amplitudes and cast them in terms of three functions, in the form:
\begin{align}
\label{eq:Sggg}
\sum \left| {\cal A}_{g g  \to H g}^{(1)}\right|^2 & = 
\frac{N\, V\, \left| C \right|^2\, g_s^6}{ (4 \pi)^d\,s\,t\,u}
\Bigg\{ 
\left| A_{ggg}(s, -t, -u) \right|^2 
\nonumber \\
& \hspace{-2cm}
+(1-\epsilon) \left[ 
\left| B_{ggg}(s, -t, -u) \right|^2 
+
\left| B_{ggg}(-t, -u, s) \right|^2 
+
\left| B_{ggg}(-u, s, -t) \right|^2 
\right] 
\Bigg\}\,,
\end{align}
\begin{align}
\label{eq:Sqgq}
\sum \left| {\cal A}_{q g  \to H q}^{(1)}\right|^2 & = 
\frac{V\, \left| C \right|^2\, g_s^6\,u}{2\, (4 \pi)^d}
\Bigg\{  
(1-\epsilon)\,\left[  
\left| A_{q\bar q g}(-u, -t, s) \right|^2
+
\left| A_{q\bar q g}(-u, s, -t) \right|^2 \right] 
\nonumber \\ 
& \quad -2 \epsilon\, \Re \left[ A_{q\bar q g}(-u, -t, s)\,  A_{q\bar q g}^*(-u, s, -t) \right]
\Bigg\} \,,
\end{align}
\begin{align}
\label{eq:Sqqbg}
\sum \left| {\cal A}_{q \bar q  \to H g}^{(1)}\right|^2 & = 
\frac{V\, \left| C \right|^2\, g_s^6\,s}{2\, (4 \pi)^d}  
\Bigg\{  
(1-\epsilon)\,\left[  
\left| A_{q\bar q g}(s, -t, -u) \right|^2
+
\left| A_{q\bar q g}(s, -u, -t) \right|^2 \right] 
\nonumber \\ 
& \quad -2 \epsilon\, \Re \left[ A_{q\bar q g}(s, -t, -u) \, A_{q\bar q g}^*(s, -u, -t) \right]
\Bigg\}\,.
\end{align}
Explicit expressions for the functions $A_{q\bar q g}, A_{ggg}, B_{ggg}$ are given in 
Appendix~\ref{app:AB_funcs}.

\paragraph{Threshold expansion using hypergeometric functions.}
Loop integrals in dimensional regularization can be expressed, to all orders in the dimensional regulator, as (generalized) hypergeometric functions. For example, the box integral\footnote{The bubble integral is trivial, and will not be discussed any further.} defined in eq.~\eqref{eq:one-loop-masters} admits the representation (see, e.g., ref.~\cite{Anastasiou:1999cx}),
\begin{align}
\label{eq:boxsymmetric}
{\rm Box}(s_{12}, s_{23}, s_{31}) & = \frac{2 c_\Gamma}{\epsilon^2}
\frac{1}{s_{12} s_{23}} \Bigg\{  
\left( -s_{23}\right)^{-\epsilon}
{_2F_1}\left( 1, -\epsilon; 1-\epsilon; -\frac{s_{31}}{s_{12}} \right)
\nonumber \\
& \hspace{-3cm}
+\left( -s_{12}\right)^{-\epsilon}
{_2F_1}
\left( 1, -\epsilon; 1-\epsilon; -\frac{s_{31}}{s_{23}} \right)
-  \left(-M_h^2 \right)^{-\epsilon}
{_2F_1}\left( 1, -\epsilon; 1-\epsilon; 
-\frac{M_h^2 s_{31}}{s_{12} s_{23}} \right)
\Bigg\}\,.
\end{align}
Our goal is to insert the parametrization~\eqref{eq:mandelstam} and
then to perform the integration over $\lambda$ term-by-term in the
series representation of the hypergeometric functions. The result is a
power series in $\delta$, i.e., the desired expansion of the cross-section close to threshold.

While all the hypergeometric series in eq.~\eqref{eq:boxsymmetric} are
convergent in the Euclidean region 
where $s_{ij}<0$, the one-loop amplitude in the physical region
involves the functions ${\rm Box}(-t, -u, s)$,  ${\rm Box}(s, -t, -u)$
and $ {\rm Box}(s, -u, -t)$. 
It is easy to check that the corresponding hypergeometric series are no longer convergent in the physical scattering region. 
It is, however, always possible to analytically continue the ${}_2F_1$ function such that arguments  lie inside the unit disc, yielding
another representation in terms of ${}_2F_1$ functions. 
While this approach is adequate to find a meaningful expansion around $\epsilon$ in terms of polylogarithms, 
it does not allow one to find (convergent) hypergeometric series expansions around
$\delta = 0$. 
Instead, one needs at least a double sum representation to achieve this task. 
It turns out that such a representation is known in the literature~\cite{Anastasiou:1999cx},
\beq\bsp
\label{eq:boxtu}
{\rm Box}(-t, -u, s) &\, =  
\frac{2 c_\Gamma}{\epsilon^2} \Gamma(1+\epsilon) \Gamma(1-\epsilon)\,e^{-i\pi\eps}\,
\frac{\left(\frac{t u}{s} \right)^{-\epsilon} }{tu} 
-\frac{2 c_\Gamma}{\epsilon (1+\epsilon)}
\frac{t^{-\epsilon-1}}{s} {_2F_1}\left( 1, 1+\epsilon ; 2+\epsilon ;
  \frac u s \right) \\
& \,
-\frac{2 c_\Gamma}{\epsilon (1+\epsilon)}
\frac{u^{-\epsilon-1}}{s} {_2F_1}\left( 1, 1+\epsilon ; 2+\epsilon ;
  \frac t s \right) \\
&\, -\frac{2 c_\Gamma}{\epsilon (1+\epsilon)}\,e^{i\pi\eps}\,
s^{-2-\epsilon} \, F_2\left(  
2+\epsilon ; 1+\epsilon, 1+\epsilon ; 2+\epsilon, 2+\epsilon ;  \frac
u s ,  \frac t s
\right)\,, \\
\esp\eeq
\beq\bsp
\label{eq:boxst}
 {\rm Box}(s, -t, -u) &\, = 
\frac{2 c_\Gamma}{\epsilon^2} \frac{t^{-\epsilon}}{s (-t)}
{_2F_1}\left( 1 , -\epsilon ; 1-\epsilon ; \frac{u}{s} \right)   \\
& \,
+ \frac{2 c_\Gamma}{\epsilon} 
s^{-2-\epsilon} \,e^{i\pi\eps}\,
S_1\left( 
2+\epsilon; 1, 1+\epsilon; 2 , 2+\epsilon, \frac t s, \frac u s
\right)\,.  
\esp\eeq
The corresponding result for ${\rm Box}(s, -u, -t)$ is obtained from
${\rm Box}(s, -t, -u)$ by exchanging $t$ and $u$. 
The generalized Kamp\'e de F\'eriet function $S_1$ is defined as 
\begin{equation}
S_1(a_1; a_2, b_1; a_3, b_2; x_1, x_2) = 
\sum_{n,m=0}^\infty 
\frac{\left(a_1\right)_{m+n} \left(a_2\right)_{m+n} \left(b_1\right)_m
}{\left(a_3\right)_{m+n} \left(b_2\right)_m } \frac{x_1^m\, x_2^n}{m!\, n!},
\end{equation}
and the Appel function $F_2$ is defined as 
\begin{equation}
F_2(a; b_1, b_2; c_1, c_2; x_1, x_2) = 
\sum_{n,m=0}^\infty 
\frac{\left(a\right)_{m+n} \left(b_1\right)_m \left(b_2\right)_n
}{\left(c_1\right)_m \left(c_2\right)_n } \frac{x_1^m\, x_2^n}{m!\,n!}.
\end{equation}
Using these expressions for the box functions, we can easily exchange
the phase-space integration and the infinite 
summations, and all the integrals can be performed in terms of Euler's Beta function,
\beq
B(\alpha,\beta) = \int_0^1d\lambda\,\lambda^{\alpha-1}\,(1-\lambda)^{\beta-1} = \frac{\Gamma(\alpha)\,\Gamma(\beta)}{\Gamma(\alpha+\beta)}\,.
\eeq

\paragraph{Threshold expansion of hard, soft and collinear regions.}
\label{sec:regions}


It is possible to derive representations such as the ones of
\eqref{eq:boxtu} and \eqref{eq:boxst} with a more physical method, 
performing Taylor expansions around soft, collinear
and hard regions of the integrand of loop integrals in momentum
space. The method of expansions by
regions~\cite{Beneke:1997zp} promises 
to hold in general, although its generality has only been stated as a
conjecture and  a verification of the validity of the approach is necessary in specific 
cases~\footnote{Full proofs of the validity of 
asymptotic expansions by regions are hard to derive or unknown. 
For efforts in this direction we refer the reader 
to ref.~\cite{Pak:2010pt, Jantzen:2011nz} and references therein.}. 

For the production of the Higgs boson near threshold, the partonic
center of mass energy is close in value to the Higgs boson mass, and
thus we have a kinematic 
variable which is small, $\delta = 1-z \sim 0$. 
From eq.~\eqref{eq:mandelstam} we infer that the external momenta scale as 
\begin{equation}
p_1 \sim p_2 \sim \sqrt s, \quad p_3 \sim \sqrt s  \, \delta\,.
\end{equation}
For a particle propagating in the loop, we find four types of non-trivial scalings of its momentum $k$: 
\begin{itemize}
\item {\it Hard} $(h)$
\[ 
k^\mu \sim \sqrt s\,,
\] where all propagators in the loop are off-shell, 
\item {\it Soft} $(s)$
\[ 
k^\mu  \sim \sqrt s  \,  \delta\,, 
\] where the loop integrand is singular at the point $k^\mu = 0$, 
\item {\it Collinear to $p_1$} $(c_1)$ 
\[ 
\frac {2 k \cdot p_1}{ s} \sim \delta, \quad    \frac {2 k \cdot p_2}{ s} \sim 1, \quad k_\perp \sim \sqrt{s \, \delta}\,,  
\] where the integrand has a singular surface as $k^\mu \propto p_1^\mu $,
\item {\it Collinear to $p_2$} $(c_2)$ 
\[ 
\frac {2 k \cdot p_2}{ s} \sim \delta, \quad    \frac {2 k \cdot p_1}{ s} \sim 1, \quad k_\perp \sim \sqrt{s \, \delta}\,,  
\]
where the integrand has a singular surface as $k^\mu \propto p_2^\mu$.
\end{itemize}
In the above, the transverse momentum $k_\perp$ of the particle is defined via: 
\begin{equation}
k =  p_1 \frac {2 k \cdot p_2}{ s} + p_2 \frac {2 k \cdot p_1}{ s} + k_\perp \,.
\end{equation}
A scaling of the loop momentum is called a {\it region}. In a given region, we
can perform a systematic  expansion of the integrand around $\delta
=0$. This yields multiple new integrals which are simpler than the
unexpanded integral. For some regions, we are able to compute
analytically all (i.e., an infinite number of) terms of the expansion. For the
remaining regions, we limit ourselves to a finite number of terms in
the expansion and perform an algebraic reduction~\cite{Laporta:2001dd,Anastasiou:2004vj} (after expansion) to master
integrals. The soft and collinear regions of our loop integrals 
correspond to the singular surfaces which solve the Landau equations~\cite{Coleman:1965xm} while the 
loop momentum scalings can be identified with the scalings of the coordinates which are 
normal to the singular surfaces~\cite{Sterman:1978bi}.
In the following we discuss how we can reproduce the hypergeometric function representations given in eq.~\eqref{eq:boxtu}.

We start by discussing the asymptotic expansion of ${\rm Box}(s, -t, -u)$, which we find convenient to parametrize as
\begin{equation}
{\rm Box}(s, -t, -u) = \int \frac{d^dk}{i \pi^{\dhalf}} \frac  1  { 
A_1 A_2 A_3 A_4
}\,,
\end{equation}
with 
\beq\bsp
A_1  &\,=    \left ( k-p_{12} \right)^2\,, \\
A_2  &\,=    \left ( k-p_2 \right)^2\,, \\
A_3  &\,=    k^2\,, \\
A_4  &\,=    \left ( k-p_3 \right)^2\,.
\esp\eeq
We find that the full integral is reconstructed by two regions:
\begin{enumerate}
\item $(c_2)$-region, where $k$ is collinear to $p_2$. 
\item $(h)-$region, where $k$ is hard. 
\end{enumerate}
After Taylor expanding the loop integrand in every region in the small variable $\delta$, we can use integration-by-parts identities to reduce the coefficients of the Taylor expansion to a small set of master integrals. In the $(c_2)$-region we find that all the coefficients are proportional to
the one-loop bubble integral ${\rm Bub}\left(-t\right)$, and this region reconstructs, order by order in $\delta$, the first term of the hypergeometric representation for ${\rm Box}(s, -t, -u)$ given eq.~\ref{eq:boxst}, and we have verified this statement explicitly up to ${\cal
  O}\left(\delta^{10}\right)$. 
The $(h)$-region yields the second and last term of eq.~\eqref{eq:boxst}. In this
region, we have been able to calculate all terms in the expansion
around $\delta=0$ with an analytic integration. We see that 
the sum  of the $(h)$ and $(c_2)$ regions is equal to the correct
expression for the one-loop box.  All other
soft and collinear regions are zero, as we can readily verify.

Next, we turn to the asymptotic expansion of the ${\rm Box}(-t, -u, s)$, given by
\begin{equation}
{\rm Box}(-t, -u, s) = \int \frac{d^dk}{i \pi^{\dhalf}} \frac  1  { 
A_1 A_2 A_3 A_4
}\,,
\end{equation}
with 
\beq\bsp
A_1  &\,=    \left ( k-p_1 \right)^2\,, \\
A_2  &\,=    \left ( k\right)^2\,, \\
A_3  &\,=    \left ( k-p_3 \right)^2\,, \\
A_4  &\,=    \left ( k-p_3+p_2 \right)^2 \,.
\esp\eeq
We find that the expression for ${\rm Box}(-t, -u, s)$ given in eq.~\eqref{eq:boxtu} is reconstructed
entirely from the following regions: 
\begin{enumerate}
\item  $(s)-$region, where the $k \sim \delta$, yielding the first
  term of eq.~\eqref{eq:boxtu}.  It is interesting that the $(s)-$region
  consists of a single term without any subleading terms in the
   expansion in $\delta$.
\item  $(c_1)$-region where the momentum $k$ is collinear to
  $p_1$. This region reconstructs the second term of
  eq.~\eqref{eq:boxtu} as we have verified explicitly up to ${\cal
  O}\left( \delta^{10}\right)$. 
\item  $(c_2)$-region where the momentum $k-p_3$ is collinear to
  $p_2$. This region reconstructs the third term of
  eq.~\eqref{eq:boxtu} as we have verified explicitly up to ${\cal
  O}\left( \delta^{10}\right)$. 
\item $(h)$-region, where $k$  is hard.  This region reconstructs the
  last term of eq.~\eqref{eq:boxtu} as we have verified explicitly up to ${\cal
  O}\left( \delta^{10}\right)$. 
\end{enumerate}
All other soft and collinear regions are zero.

We have seen that an expansion in hard, soft and collinear regions
yields series representations for the one-loop master integrals of
the required amplitudes which converge in the entire phase-space, and thus we can immediately
perform the phase-space integration in terms of Beta functions order-by-order in the expansion.
While in our case the strategy of expansion by regions is only an alternative
method for deriving the threshold expansions of
Section~\ref{sec:sigma_expansion}, it can be the method of
choice for the phase-space integration of more complicated one-loop
amplitudes.  Here we have presented expansions by regions at the level
of master integrals. We would like to remark that such expansions can also be 
performed at the integrand of loop-amplitudes before any reduction to
master integrals has taken place.  Combined with the method of
reverse-unitarity~\cite{Anastasiou:2013srw} we have a powerful algebraic technique for
the simultaneous threshold expansion of integrals over loop and
external momenta.


\subsection{Reverse unitarity and differential equations}	
\label{sec:revunit}

In this section we evaluate the real-virtual squared cross-sections
using the reverse-unitarity
approach~\cite{Anastasiou:2002yz,Anastasiou:2002qz,Anastasiou:2003yy,Anastasiou:2003ds}. Reverse
unitarity establishes a duality between phase-space integrals and loop
integrals. Specifically, on-shell and other phase-space constraints
are dual to ``cut'' propagators
\beq
\delta_+(q^2) \rightarrow \left[\frac{1}{q^2}\right]_c=\frac{1}{2\pi i}\,\textrm{Disc}\, \frac{1}{q^2}=\frac{1}{2\pi i}\left[\frac{1}{q^2+i0}-\frac{1}{q^2-i0}\right].
\eeq
A cut-propagator can be differentiated similarly to an ordinary
propagator with respect to its momenta. It is therefore possible to
derive integration-by-parts (IBP)
identities~\cite{Chetyrkin:1981qh,Tkachov:1981wb} for phase-space
integrals in the same way as for loop integrals. The only difference
is an additional simplifying constraint that a cut-propagator raised to a 
negative power vanishes:
\beq
\label{eq:zerooncut}
\left[\frac{1}{q^2}\right]_c^{-\nu}=0,\hspace{1cm }\nu \geq 0\,.
\eeq
In this approach, we are not obliged to perform a strictly sequential 
evaluation of the loop integrals in the amplitude followed by
the nested phase-space integrals. Rather, we combine the two types of
integrals into a single multiloop-like type of integration by
introducing cut-propagators and then derive and solve IBP 
identities for the combined integrals. 
We solve the large system of IBP identities which are relevant for our
calculation with the Gauss elimination algorithm of
Laporta~\cite{Laporta:2001dd}. 
We have made an independent implementation of the algorithm in
\textsc{C++} using also the \textsc{GiNaC} library~\cite{Bauer:2000cp}. In comparison to
AIR~\cite{Anastasiou:2004vj}, which is a second reduction program used
in this work, the \textsc{C++} implementation is faster and more
powerful, storing all identities in virtual memory rather than in the
file system.  All integrals
that appear in the real-virtual squared cross section are reduced to
linear combinations of 19 master integrals, which we choose as follows:
\beq
\mathcal{M}_1=\scalebox{0.25}{\Mone{120}}=\int d\Phi_{2}\, \text{Bub}(s_{23})\,\text{Bub}^*(s_{13}).
\eeq
\beq
\mathcal{M}_2=\scalebox{0.3}{\Mthree{85}} =\int d\Phi_{2}\, \text{Bub}(s_{12})\,\text{Bub}^*(s_{12}).
\eeq
\beq
\mathcal{M}_3=\scalebox{0.3}{\Mseven{85}}=\int d\Phi_{2}\, \text{Bub}(s_{13})\,\text{Bub}^*(s_{12}).
\eeq
\beq
\mathcal{M}_4=\scalebox{0.3}{\Mnine{90}}=\int d\Phi_{2}\, \text{Bub}(s_{13})\,\text{Bub}^*(s_{13}).
\eeq
\beq
\mathcal{M}_{5}=\scalebox{0.3}{\Mten{90}} =\int d\Phi_{2}\, \text{Tri}(s_{12}+s_{23})\,\text{Bub}^*(s_{23}).
\eeq
\beq
\mathcal{M}_{6}=\scalebox{0.3}{\Mtwenty{90}}=\int d\Phi_{2}\, \text{Tri}(s_{12}+s_{13})\,\text{Bub}^*(s_{23}).
\eeq
\beq
\mathcal{M}_{7}=\scalebox{0.3}{\Mtwentytwo{90}}=\int d\Phi_{2}\,\text{Tri}(s_{12}+s_{23})\, \text{Bub}^*(s_{12}).
\eeq
\beq
\mathcal{M}_{8}=\scalebox{0.3}{\Mtwentythree{90}}=\int d\Phi_{2}\, \text{Bub}(s_{23})\, \text{Box}^*(s_{12},s_{23},s_{13}).
\eeq
\beq
\mathcal{M}_{9}=\scalebox{0.3}{\Mtwentynine{90}}=\int d\Phi_{2}\, \text{Bub}(s_{12})\, \text{Box}^*(s_{13},s_{23},s_{12}).
\eeq
\beq
\mathcal{M}_{10}=\scalebox{0.3}{\Mfortyone{90}}=\int d\Phi_{2}\, \text{Bub}(s_{23})\, \text{Box}^*(s_{12},s_{13},s_{23}).
\eeq
\beq
\mathcal{M}_{11}=\scalebox{0.3}{\Mfortytwo{90}}=\int d\Phi_{2}\, \text{Bub}(s_{13})\, \text{Box}^*(s_{13},s_{23},s_{12}).
\eeq
\beq
\mathcal{M}_{12}=\scalebox{0.3}{\Mnineteen{100}}=\int d\Phi_{2}\, \text{Tri}(s_{12}+s_{13})\, \text{Box}^*(s_{13},s_{23},s_{12}).
\eeq
\beq
\mathcal{M}_{13}=\scalebox{0.3}{\Mthirtythree{90}}=\int d\Phi_{2}\, \text{Tri}(s_{12}+s_{13})\, \text{Box}^*(s_{12},s_{13},s_{23}).
\eeq
\beq
\mathcal{M}_{14}=\scalebox{0.3}{\Mthirtyfour{90}}=\int d\Phi_{2}\, \text{Tri}(s_{12}+s_{13})\, \text{Box}^*(s_{12},s_{23},s_{13}).
\eeq
\beq
\mathcal{M}_{15}=\scalebox{0.3}{\Mfive{90}}=\int d\Phi_{2}\, \text{Box}(s_{12},s_{13},s_{23})\, \text{Box}^*(s_{12},s_{23},s_{13}).
\eeq
\beq
\mathcal{M}_{16}=\scalebox{0.3}{\Msix{90}}=\int d\Phi_{2}\, \text{Box}(s_{12},s_{13},s_{23})\, \text{Box}^*(s_{13},s_{23},s_{12}).
\eeq
\beq
\mathcal{M}_{17}=\scalebox{0.3}{\Mtwentyeight{90}}=\int d\Phi_{2}\, \text{Box}(s_{12},s_{23},s_{13})\,\text{Box}^*(s_{12},s_{23},s_{13}).
\eeq
\beq
\mathcal{M}_{18}=\scalebox{0.3}{\Msixtysix{85}}=\int d\Phi_{2}\, \text{Box}(s_{13},s_{23},s_{12}) \,\text{Box}^*(s_{13},s_{23},s_{12}).
\eeq
\beq
\mathcal{M}_{19}=\scalebox{0.3}{\Mthirtysix{90}}=\int d\Phi_{2}\, \text{Bub}(M_h^2) \,\text{Box}^*(s_{12},s_{23},s_{13})\frac{1}{s_{23}}.
\eeq
\\
\vspace{1cm}
\\

Single solid lines represent scalar massless propagators. The
phase-space integration is represented by the dashed line and the
cut-propagators are the lines cut by the dashed line.  The cut
propagator of the Higgs boson is depicted by the double-line. Every
master integral has a one-loop integral on the left- and a
complex-conjugated one-loop integral on the right-hand side of the
cut. In each side of the cut, we find scalar bubble, box 
or triangle integrals, where the latter is defined by
\beq\bsp\label{eq:triangle_def}
\text{Tri}(s_{12}) &\,=\int\frac{d^Dk}{i(\pi)^{D/2}}\frac{1}{k^2 (k+q_{1})^2(k+q_1+q_2)^2}\,,\\
\text{Tri}(p_1^2,p_2^2) &\,=\int\frac{d^Dk}{i(\pi)^{D/2}}\frac{1}{k^2 (k+p_{1})^2(k+p_1+p_2)^2}\,,
\esp\eeq
with $q_i^2=0$, $p_i^2\neq 0$ and $(p_1+p_2)^2=0$.
The scalar bubble and box integrals have been defined in \eqref{eq:one-loop-masters}. A comment is in order about the appearance of the triangle integrals in this approach, which seems to be at odds 
with the fact that in the expression of the one-loop amplitude presented Section~\ref{sec:sigma_expansion} only bubble and box integrals appeared. Indeed, it is well-known that eq.~\eqref{eq:triangle_def} can be expressed as a linear combination of bubble integrals,
\beq\bsp\label{eq:tri_to_bub}
\text{Tri}(s_{12}) &\,= \frac{1-2\eps}{\eps\,s_{12}}\,\textrm{Bub}(s_{12})\,,\\
\text{Tri}(p_1^2,p_2^2) &\,= \frac{1-2\eps}{\eps\,(p_1^2-p_2^2)}\,[\textrm{Bub}(p_1^2)-\textrm{Bub}(p_1^2)]\,.\\
\esp\eeq
These relations however introduce new denominators which need to be taken into account in the reduction of the phase-space integrals. We therefore prefer not to use eq.~\eqref{eq:tri_to_bub}, but we work directly with the triangle integrals instead.


To evaluate the master integrals we employ the method of differential 
equations~\cite{Kotikov:1990kg,Gehrmann:1999as,Anastasiou:2002yz}. 
Differentiating the corresponding cut-propagator with respect to the
square of the Higgs mass,
\beq
\frac{\partial}{\partial  {M_h^2}}\left(\frac{1}{p_h^2-M_h^2 }\right)_c= \left(\frac{1}{p_H^2-M_h^2}\right)_c^2,
\eeq
results in another phase-space integral. This new integral can again
be reduced by IBP identities to our basis of master
integrals. Proceeding in this way we obtain a system of linear first
order differential equations for the master integrals, 
\beq
\frac{\partial}{\partial  \delta}\mathcal{M}_i( \delta)=A_{ij}( \delta)\, \mathcal{M}_j( \delta)\,.
\eeq
The system is triangular
\beq\label{eq:triangular_system}
\frac{\partial}{\partial \delta} \mathcal{M}_i(\delta)=A_{ii}(\delta)\, \mathcal{M}_i(\delta) +y_i(\delta)\,,
\eeq
where $y_i(\delta)$ only depends on master integrals that can be solved for independently of $\mathcal{M}_i(\delta)$.
In other words, the system can be solved hierarchically, starting from the differential
equations with vanishing or known functions $y$. Every time we solve
such an equation, its solution  serves to determine the  $y$ function
of a next equation. In this way, at any stage of this procedure  the
$y$ function is a linear 
combination of already evaluated master integrals
\beq
y_i(\delta)=\sum\limits_{j\neq i}A_{ij} (\delta)\,\mathcal{M}_j(\delta)\,,
\eeq
 that can be integrated in order to determine the integral $\mathcal{M}_j(\delta)$. 
The coefficients $A_{ij}(\delta)$ are rational functions in $\delta$ and $\epsilon$ and have isolated singularities in $\delta$ only at $\delta=0,1,2$.
The first step to solving this type of differential equation is to
find a solution for the homogeneous part. 
The general homogeneous solution associated to the differential equation~\eqref{eq:triangular_system} is given by
\beq
\mathcal{M}^h_{i}(\delta)=\mathcal{M}_{i}(0) \,\exp\left[{\int\limits^{\delta}_0 d\delta^\prime A_{ii}(\delta ^\prime)}\right]\,.
\eeq
and is determined up to an integration constant $\mathcal{M}_{i}(0)$. We determine this integration constant by calculating the soft limit of the master integral explicitly following the methods discussed in Section \ref{sec:regions}.
We find that only $7$ of our $19$ master integrals have non-trivial
boundary conditions. Interestingly, with our choice of basis of master
integrals, the non-trivial boundary conditions are in one-to-one
correspondence to the leading terms of the $7$ regions of the soft
expansion of the squared amplitude of eqs.~\eqref{eq:eta_not_4}-\eqref{eq:eta_4}.
The non-trivial boundary conditions are: 
\bea
   \mathcal{M}^S_1&=&(4\pi)^{-1+\epsilon}\,\omega_\Gamma\,\delta^{1-4\epsilon}\, \frac{\phi_\Gamma^{(4;1)}}{2 \epsilon^2(1-2 \epsilon )^2(1-4\epsilon)}\,,\nonumber\\
   \mathcal{M}^S_2&=&(4\pi)^{-1+\epsilon}\,\omega_\Gamma\, \delta^{1-2\epsilon}\,\frac{\phi_\Gamma^{(2;1)}}{2 \epsilon^2(1-2 \epsilon )^3}\,,\nonumber\\
   \mathcal{M}^S_3&=&(4\pi)^{-1+\epsilon}\,\omega_\Gamma\,\delta^{1-3\epsilon} \,\frac{\phi_\Gamma^{(3;1)}}{2 \epsilon^2(1-2 \epsilon )^2(1-3\epsilon)}\,,\nonumber\\
   \mathcal{M}^S_4&=&(4\pi)^{-1+\epsilon}\,\omega_\Gamma\delta^{1-4\epsilon} \,\frac{\phi_\Gamma^{(4;3)}}{2 \epsilon^2(1-2 \epsilon )^2(1-4\epsilon)}\,,\\
   \mathcal{M}^S_{9}&=&-(4\pi)^{-1+\epsilon}\,\omega_\Gamma\, \delta^{-1-4\epsilon}\,\frac{\phi_\Gamma^{(4;2)}}{ \epsilon^4(1-2 \epsilon )}\,,\nonumber\\
   \mathcal{M}^S_{11}&=&-(4\pi)^{-1+\epsilon}\,\omega_\Gamma\, \delta^{-1-5\epsilon}\,\frac{5\phi_\Gamma^{(5;1)}}{6 \epsilon^4(1-2 \epsilon )}\,,\nonumber\\
   \mathcal{M}^S_{18}&=&-(4\pi)^{-1+\epsilon}\,\omega_\Gamma\,\delta^{-3-6\epsilon }\,\frac{8(1+6\epsilon)\phi_\Gamma^{(6;1)}}{3\epsilon^5(1+3\epsilon)}\,.\nonumber
\eea
Only the real part of the boundary conditions is presented here, given
that the imaginary part does not contribute to the cross-section.

Once the homogeneous solution is found we can compute a particular solution to the inhomogeneous equation by
\beq
\label{eq:inhom}
\mathcal{M}_i^p(\delta)= \mathcal{M}_i^h(\delta) \int\limits_0 ^{\delta} d\delta ^\prime \frac{y(\delta^\prime)}{\mathcal{M}_i^h(\delta^\prime)},
\eeq
The full solution for the master integral is then given by
\beq
\mathcal{M}_i(\delta)=\mathcal{M}_i^h(\delta)+\mathcal{M}_i^p(\delta).
\eeq
We perform the integration in the equation above with two different approaches.

\paragraph{Solving differential equations as an expansion in $\epsilon$.}
One well established strategy is to expand the differential equations
in powers of the dimensional
regulator~\cite{Kotikov:1990kg,Gehrmann:1999as}. After expanding the
integral of \eqref{eq:inhom} a solution is naturally given 
by iterated integrals leading to multiple
polylogarithms~\cite{Goncharov:1998kja} of the form $G(a_1,\ldots,a_n;\delta)$, 
with $a_i\in\{0,1,2\}$. Expressing the functions in terms of the variable
$z=1-\delta$ recasts the solutions in terms of more familiar harmonic
polylogarithms~\cite{Gehrmann:1999as}. 

\paragraph{Solving differential equations in terms of hypergeometric functions.}
The integrand of eq.~\eqref{eq:inhom} takes the form 
\beq
\label{eq:inhomtype}
\mathcal{M}^s_i(\delta )\sim \int_0^{\delta}d \delta^\prime (\delta^\prime)^{c_{1}} (1-\delta^\prime)^{c_2} (2-\delta^\prime)^{c_3} \mathcal{M}_{j\neq i}(\delta^{\prime}),
\eeq
where $c_1,c_2$ and $c_3$ are linear polynomials in $\epsilon$. This
structure reminds of the Euler-type integral representation of hypergeometric functions. Inspired by the large variety of techniques available for the solution of iterated integrals in terms of multiple polylogarithms~\cite{Goncharov:1998kja} we define an iterated integral with integration kernel $\delta^{a-1}(1-x \delta)^{-b}$. The $n^{th}$ iterated integral is then recursively defined by
\beq
\mathcal{F}_{\vec{a}_n,\dots,\vec{a}_1} (x_n,\dots,x_1;\delta)=\int_0^\delta d \delta^\prime (\delta^\prime )^{a_n-1}(1-x_n \delta^\prime)^{-b_n}\mathcal{F}_{\vec{a}_{n-1},\dots,\vec{a}_1} (x_{n-1},\dots,x_1;\delta^\prime)\,.
\eeq
where we have abbreviated $\tiny{\vec{a}_i=\left(\begin{array}{c} a_i \\ b_i \end{array}\right)}$. 

We find that these iterated integrals interpolate between multiple polylogarithms and hypergeometric functions. For example, in this framework the multiple polylogarithm is given by
\beq\bsp
\mathrm{Li}&_{m_1,\dots,m_k}(x_1,\dots,x_k)\\
&\,=\left(\prod_{i=1}^kx_i^{k-i+1}\right)\mathcal{F}_{\scriptsize\underbrace{\vec{0},\dots,\vec{0}}_{m_1-1},\vec{1},\dots,\underbrace{\vec{0},\dots,\vec{0}}_{m_k-1},\vec{1}}(\underbrace{0,\dots,0}_{m_1-1},x_1,\dots,\underbrace{0,\dots,0}_{m_k-1},x_k\dots x_1;1)\,.
\esp\eeq
where $\tiny{\vec{0}=\left(\begin{array}{c}0\\0\end{array}\right)}$ and $\tiny{\vec{1}=\left(\begin{array}{c}1\\1\end{array}\right)}$.
The Gauss hypergeometric function is given by
\beq
_{2}F_1(a_2,a_1;b_1;z)=a_1\frac{\Gamma(b_1)}{\Gamma(a_2)\Gamma(b_1-a_2)} \mathcal{F}_{\vec{v}\vec{u}}(1,z;1),
\eeq
with $\tiny{\vec{v}=\left(\begin{array}{c}a_{2}-a_1\\ 1-a_{2}-b_1\end{array}\right)}$ and  $\tiny{\vec {u}=\left(\begin{array}{c}a_1\\a_1+1\end{array}\right)}$.
A large variety of hypergeometric functions can be expressed in terms of these iterated integrals. 
With the definition
\beq
A_n=\sum_{i=1}^n a_i, \hspace{1cm} K_n=\sum_{i=1}^n k_i\,,
\eeq
we find an explicit sum representation for this type of iterated integrals.
\bea
\label{eq:sumrep}
\mathcal{F}_{\vec{a}_n,\dots,\vec{a}_1} (x_n,\dots,x_1;\delta)&=&\frac{\delta^{A_n}}{\prod\limits_{i=1}^n A_i}\,\sum\limits_{k_1,\dots,k_n=0}^\infty \, \prod\limits_{i=1}^n \left(\frac{(A_i)_{K_i}}{(A_i+1)_{K_i}} (b_i)_{k_i} \frac{(x_i \delta)^{k_i}}{k_i !}\right)\,.
\eea
Equation~\eqref{eq:sumrep} is valid whenever the sums are convergent. Further properties and derivations are discussed in more detail in Appendix~\ref{sec:itint}.
The solution of differential equations using iterated integrals is
illustrated with an example in Appendix~\ref{sec:NNMasters}.

The iterated integrals defined in this section are a powerful tool and
enable us to solve all 19 master integrals in terms of hypergeometric functions. 
The results is valid to all orders in $\epsilon$. The iterated
integrals can be written as multiple sums eq.~\eqref{eq:sumrep} from
which it is very convenient to extract a threshold expansion in
$\delta$. 

\paragraph{Results for the master integrals.}
The master integrals that we have computed in this section are useful
for the evaluation of any cross-section for a $2 \to 1$ process at
N$^3$LO. 
Due to the length of their expressions,  we provide them in terms of
harmonic polylogarithms in the form of the text file
\textsf{masters.txt}  enclosed with the arXiv submission of this work
and on the web-page \cite{xsdownload}. 
We have computed all 19 master integrals to all orders in $\epsilon$ in terms of hypergeometric functions and as an expansion in $\epsilon$ in terms of harmonic polylogarithms as described above.


\subsection{Direct integration using multiple polylogarithms}
\label{sec:sub}

We present here an alternate method to compute the ${\rm (RV)}^2$
Higgs boson cross-section, based on subtraction terms. 
The phase-space integral over the squared amplitude can be written schematically as,
\beq
\int d\Phi_2\, |\mathcal{A}|^2 = \int d\Phi_2\, \sum_{i,j}M_{i}(s_{12},s_{23},s_{13})\,M_{j}(s_{12},s_{23},s_{13})\,N_{i,j}(s_{12},s_{23},s_{13}).
\eeq
In this expression $M_{i}$ denote the one-loop master integrals and $N_{i,j}$ are rational functions, all of which depend on the invariants $s_{12},s_{23}$ and $s_{13}$.
Since the results for the required one-loop master integrals are known to all orders in $\epsilon$ \cite{Anastasiou:1999cx}, the integrals are well defined in dimensional regularization.
Our goal is to expand the integrals in $\eps$ under the integration sign and to perform the integration order by order in $\eps$. After expansion, however, the integrals may develop soft and collinear divergencies. The strategy is to subtract the singular limits of the integrand \emph{before} expansion, and to perform the remaining (finite)
integration in terms of multiple
polylogarithms.

The construction of the counterterms that render the integration finite proceeds in two steps. First, we analytically continue all the hypergeometric functions that appear in the all order expressions of the one-loop master integrals such that they are convergent in the whole phase-space. This is achieved by using the well-known identities
\beq\bsp
\label{eq:flip}
\f21(a,b;c;z) &\,= (1-z)^{-b}\,\f21\left(b,c-a;c;\frac{z}{z-1}\right)\,,\\
\f21(a,b;c;z) &\,= \frac{\Gamma(b-a)\Gamma(c)}{\Gamma(b)\Gamma(c-a)}\,(-z)^{-a}\,\f21(a,a-c+1;a-b+1;z^{-1}) \\
&\,+ \frac{\Gamma(a-b)\Gamma(c)}{\Gamma(a)\Gamma(c-b)}\,(-z)^{-b}\,\f21(b,b-c+1;b-a+1;z^{-1})\,.
\esp\eeq
Second, the soft and collinear counterterms are easily constructed by expanding the integrand around the collinear limits,
i.e., $s_{13} \to 0$ or $s_{23} \to 0$.
The counterterms can be trivially integrated to all orders in the dimensional regulator in terms of $\Gamma$ functions.


At the end of this procedure we are left with finite one-dimensional integrals. 
We expand the hypergeometric functions appearing in the integrand in $\epsilon$ using
\texttt{HypExp}~\cite{Huber:2007dx}, resulting in a representation for the integrand in terms of classical polylogarithms up to weight four. More specifically, we are left with integrals of the form
\beq
\int_0^1d\lambda\,\sum_i \frac{P_i(\lambda,z)}{\lambda(1-\lambda)} \text{Li}_n\left(R_{i,1}(\lambda,z)\right)\,\text{Li}_m\left(R_{i,2}(\lambda,z)\right)\,,
\eeq
with $n+m \leq 4$ and where $P_i$ is a polynomial and $R_{i,k}$ are rational functions. Note that, while individual terms in the sum are singular for $\lambda\to0,1$, the sum is finite by construction, and so the integral is well defined.
In order to perform the integration over $\lambda$, we rewrite the classical polylogarithms in terms of multiple polylogarithms~\cite{Goncharov:1998kja} of the form $G(a_1(z),\ldots,a_n(z);\lambda)$, where $a_i(z)$ are rational functions of $z$ using symbols~\cite{symbolsC,symbolsB,symbols,Goncharov2010jf,Duhr2011zq} and the Hopf algebra structure of multiple polylogarithms~\cite{GoncharovGalois,Brown2011ik,Duhr2012fh,Anastasiou:2013srw}. All the integrals can then easily be performed using the recursive definition of the multiple polylogarithms,
\beq
\int_{0}^1\frac{d\lambda}{\lambda-a_1}\,G(a_2,\ldots,a_n;\lambda) = G(a_1,\ldots,a_n;1)\,.
\eeq
Finally, we observe that the results of the integration can also be expressed in terms of harmonic polylogarithms~\cite{Remiddi:1999ew} of weight up to five, and we checked that the results are in agreement with the differential equation approach.

\section{Conclusions}
\label{sec:conclusions}

In this article, we have taken one more step towards the evaluation of 
the Higgs boson cross-section at N$^3$LO in perturbative 
QCD. Specifically, we have analytically evaluated the one-loop contributions to the partonic
cross-sections from $2 \to 2$ processes. This has required us to perform 
single real emission phase space integrals over the square of one-loop amplitudes. 
We performed these integrations using various independent methods.
The motivation for this was, besides having a number of cross checks for the correctness of our
results, to try out the most promising state of the art techniques for the evaluation of multi-loop integrals. 
The (RV)$^2$-corrections are simpler than other, yet unknown, contributions to the N$^3$LO Higgs boson 
cross-section. However,  they share many of the complexities that
appear in corrections with mixed real and virtual
radiation. Therefore, our calculations in this paper serve as a perfect testing ground for our 
computational techniques. 

In this publication,  we have achieved two goals. 
We have obtained analytic expressions for the (RV)$^2$ partonic cross-sections which 
are valid for arbitrary values of the Higgs mass and energy  as an expansion in $\epsilon$ 
through the finite part. Given that a calculation of the full hadronic Higgs boson cross-section at N$^3$LO 
will most likely be achieved first as an expansion around threshold, we have obtained here a threshold series 
expansion for the (RV)$^2$ partonic cross-sections. The coefficients of the threshold expansion are valid for 
arbitrary values of $\epsilon$. We performed our calculations with various methods and techniques.

Our first method is based on reverse unitarity and integration by part identities 
in order to reduce the integrals of the partonic cross-sections to $19$ master integrals. For these master integrals 
we derived  a system of differential equations which we could solve to all orders in the dimensional regulator in terms
of iterative functions of the kind of hypergeometric functions which we defined explicitly for this purpose. 
Their series representations yield a threshold expansion for the master integrals and the partonic cross-sections. 
The newly introduced hypergeometric functions share many properties 
with multiple polylogarithms, for example they satisfy a shuffle algebra. 
They also appear to cover a wide range of known multi-dimensional hypergeometric functions. 
It will be an exciting new direction for further research to establish further properties of these functions and their usefulness in higher order perturbative corrections. 
We have also solved the differential equations for the master integrals order by order in $\epsilon$ in terms of 
harmonic polylogarithms.  This yields the main result of this publication, which is the (RV)$^2$ partonic 
cross-sections for arbitrary values of the Higgs mass and the partonic  center of mass energy as an expansion in $\epsilon$ 
through ${\cal O}\left( \epsilon^0\right)$.

We have also followed a direct integration approach to obtain the same results. 
For this task, we introduced counter-terms at the level of the squared 
matrix element in order to subtract its collinear and soft divergences. 
Having rendered the phase-space integrand finite, we then 
expand in the dimensional regulator and perform a direct integration in terms of harmonic polylogarithms.  The integration is made possible by embedding the classical polylogarithms that result form the expansion of the hypergeometric function in a larger space of multiple polylogarithms and exploiting their Hopf algebra structure.

We have also followed a different strategy, with a more restricted objective to obtain a threshold expansion 
of the cross-sections. 
Experience from NNLO has shown that very good approximations to the inclusive Higgs boson cross-section 
can be obtained with the first few terms of this expansion. We obtain the series around threshold by expanding the integrand of the loop amplitudes in 
soft, collinear and hard regions and perform the phase-space integration term by term.
%
We have successfully applied this method to compute a sufficiently 
large number of terms in the threshold expansion to all orders in the dimensional regulator. 
Unlike some of the other methods we have used, the strategy of regions is generic,  
which makes it a  particularly attractive option  for computing the (RV)  and (RV)$^2$ contributions in more complicated processes, where a direct evaluation in terms of polylogarithms may be unfeasible 
or just extremely difficult. 

Our results are rather lengthy and we have only typeset parts of them in this document. Instead, 
we provide in the source of the arXiv submission two text files {\tt results.txt} and {\tt masters.txt} with the expressions for the partonic cross-sections and the master integrals respectively. The two files can also be downloaded from \cite{xsdownload}.

We believe that with our calculation and the methods that we have developed in this paper to be closer in our objective to compute the Higgs boson cross-section at N$^3$LO.   We look forward to applying the techniques presented here 
towards this objective. 

\section*{Acknowledgements}
We thank Thomas Becher, Aneesh Manohar, Eric Laanen, George
Sterman and Leonardo Vernazza for useful discussions. 
This research is supported by the ERC Starting Grant for the project
``IterQCD'', the Swiss National Foundation under contract SNF
200021\_143781, by the Research Executive Agency (REA) of the European Union under the Grant Agreement number PITN-GA-2010-264564 (LHCPhenoNet) and the FP7 Marie
Curie Initial Training Network MCnetITN under contract PITN-GA-2012-315877. The work of F.H. was supported by ERC grant 291377.

\newpage
\appendix 
\input{APP_IteratedIntegrals.tex}

\input{APP_NNMasters.tex}

\input{APP_MatrixElements.tex}
\newpage
\section{Results: Threshold Expansion}
\label{sec:thresholdexp}
In this  section, we present the results of  our calculation for the functions $\eta_X^{a;j}$ of Section~\ref{sec:setup}.

\input{XSthreshold.tex}

\end{document}

%% file: masterpics.tex
\def\Mone#1{
\begin{picture}(400,180)(0,#1)
  \begin{picture}(432,274) (19,-14)
    \SetWidth{4.0}
    \SetColor{Black}
    \Line[double,sep=8](96,33)(336,34)
    \SetWidth{6.0}
    \Line[dash,dashsize=10](208,258)(208,-14)
    \SetWidth{4.0}
    \Line[arrow,arrowpos=0.5,arrowlength=6.667,arrowwidth=4,arrowinset=0.2](32,210)(96,210)
    \Line(96,210)(336,210)
    \Line[arrow,arrowpos=0.5,arrowlength=6.667,arrowwidth=4,arrowinset=0.2](336,210)(400,210)
    \Line[arrow,arrowpos=0.5,arrowlength=6.667,arrowwidth=4,arrowinset=0.2](32,34)(97,33)
    \Line[arrow,arrowpos=0.5,arrowlength=6.667,arrowwidth=4,arrowinset=0.2](336,34)(400,34)
    \Arc(153,121.5)(105.268,122.784,237.216)
    \Arc(39,121.5)(105.268,-57.216,57.216)
    \Arc(392.369,121.816)(104.661,122.587,238.06)
    \Arc(280,122)(104.307,-57.529,57.529)
    \Text(416,210)[lb]{\Huge{\Black{$1$}}}
    \Text(416,34)[lb]{\Huge{\Black{$2$}}}
    \Text(16,210)[lb]{\Huge{\Black{$2$}}}
    \Text(16,34)[lb]{\Huge{\Black{$1$}}}
  \end{picture}
\end{picture}
}

\def\Mthree#1{
\begin{picture}(400,180)(0,#1)
  \begin{picture}(416,174) (3,-78)
    \SetWidth{6.0}
    \SetColor{Black}
    \Line[dash,dashsize=10](192,94)(192,-82)
    \SetWidth{4.0}
    \Line[arrow,arrowpos=0.5,arrowlength=6.667,arrowwidth=4,arrowinset=0.2](320,14)(352,46)
    \Line[arrow,arrowpos=0.5,arrowlength=6.667,arrowwidth=4,arrowinset=0.2,flip](352,-18)(320,14)
    \Arc(192,-4.667)(66.667,16.26,163.74)
    \Arc[double,sep=6,clock](192,32.667)(66.667,-16.26,-163.74)
    \Line[arrow,arrowpos=0.5,arrowlength=6.667,arrowwidth=4,arrowinset=0.2](32,46)(64,14)
    \Line[arrow,arrowpos=0.5,arrowlength=6.667,arrowwidth=4,arrowinset=0.2](32,-18)(64,14)
    \Arc(96,14)(32,90,450)
    \Arc(288,14)(32,90,450)
    \Text(10,46)[lb]{\Huge{\Black{$1$}}}
    \Text(10,-18)[lb]{\Huge{\Black{$2$}}}
    \Text(365,46)[lb]{\Huge{\Black{$1$}}}
    \Text(365,-18)[lb]{\Huge{\Black{$2$}}}
  \end{picture}
  \end{picture}
 }

  \def\Mseven#1{
\begin{picture}(400,180)(0,#1)
  \begin{picture}(400,173) (35,-75)
    \SetWidth{6.0}
    \SetColor{Black}
    \Line[dash,dashsize=10](192,77)(192,-83)
    \SetWidth{4.0}
    \Line[arrow,arrowpos=0.5,arrowlength=6.667,arrowwidth=4,arrowinset=0.2](64,77)(112,77)
    \Line(112,77)(256,13)
    \Line[arrow,arrowpos=0.5,arrowlength=6.667,arrowwidth=4,arrowinset=0.2](64,-51)(112,-51)
    \Line[double,sep=8](112,-51)(256,13)
    \Line[arrow,arrowpos=0.5,arrowlength=6.667,arrowwidth=4,arrowinset=0.2](320,13)(368,45)
    \Arc(162.181,14.091)(82.188,132.175,232.37)
    \Arc(64,13)(80,-53.13,53.13)
    \Line[arrow,arrowpos=0.5,arrowlength=6.667,arrowwidth=4,arrowinset=0.2,flip](368,-19)(320,13)
    \Arc(288,13)(32,90,450)
    \Text(42,77)[lb]{\Huge{\Black{$1$}}}
    \Text(42,-51)[lb]{\Huge{\Black{$2$}}}
    \Text(380,45)[lb]{\Huge{\Black{$1$}}}
    \Text(380,-19)[lb]{\Huge{\Black{$2$}}}
  \end{picture}
  \end{picture}
 }
 
   \def\Mnine#1{
\begin{picture}(400,180)(0,#1)
  \begin{picture}(352,194) (35,-62)
    \SetWidth{6.0}
    \SetColor{Black}
    \Line[dash,dashsize=10](192,130)(192,-62)
    \SetWidth{4.0}
    \Line[arrow,arrowpos=0.5,arrowlength=6.667,arrowwidth=4,arrowinset=0.2](64,98)(112,98)
    \Line(112,98)(288,98)
    \Line[arrow,arrowpos=0.5,arrowlength=6.667,arrowwidth=4,arrowinset=0.2](288,98)(320,98)
    \Line[arrow,arrowpos=0.5,arrowlength=6.667,arrowwidth=4,arrowinset=0.2](64,-30)(112,-30)
    \Line[double,sep=8](112,-30)(272,-30)
    \Line[arrow,arrowpos=0.5,arrowlength=6.667,arrowwidth=4,arrowinset=0.2](272,-30)(320,-30)
    \Arc(162.181,35.091)(82.188,132.175,232.37)
    \Arc(64,34)(80,-53.13,53.13)
    \Arc(320,34)(80,126.87,233.13)
    \Arc(224,34)(80,-53.13,53.13)
    \Text(42,98)[lb]{\Huge{\Black{$1$}}}
    \Text(42,-30)[lb]{\Huge{\Black{$2$}}}
    \Text(342,98)[lb]{\Huge{\Black{$1$}}}
    \Text(342,-30)[lb]{\Huge{\Black{$2$}}}
  \end{picture}
  \end{picture}
 }
 
    \def\Mten#1{
\begin{picture}(400,180)(-25,#1)
  \begin{picture}(352,205) (19,-33)
    \SetWidth{4.0}
    \SetColor{Black}
    \Line(272,141)(80,141)
    \SetWidth{6.0}
    \Line[dash,dashsize=10](182,170)(182,-31)
    \SetWidth{4.0}
    \Line[arrow,arrowpos=0.5,arrowlength=6.667,arrowwidth=4,arrowinset=0.2](32,141)(80,141)
    \Line[arrow,arrowpos=0.5,arrowlength=6.667,arrowwidth=4,arrowinset=0.2](32,-3)(80,-3)
    \Line(80,141)(80,-3)
    \Line(80,-3)(160,-3)
    \Line[double,sep=8](160,-3)(272,-3)
    \Arc[clock](208,69)(96.333,48.366,-48.366)
    \Arc[clock](336,69)(96.333,-131.634,-228.366)
    \Line[arrow,arrowpos=0.5,arrowlength=6.667,arrowwidth=4,arrowinset=0.2,flip](320,141)(272,141)
    \Line[arrow,arrowpos=0.5,arrowlength=6.667,arrowwidth=4,arrowinset=0.2,flip](320,-3)(272,-3)
    \Line(160,-3)(80,141)
    \Text(336,141)[lb]{\Huge{\Black{$2$}}}
    \Text(336,-3)[lb]{\Huge{\Black{$1$}}}
    \Text(16,-3)[lb]{\Huge{\Black{$2$}}}
    \Text(16,141)[lb]{\Huge{\Black{$1$}}}
  \end{picture}
  \end{picture}
 }
 
     \def\Mtwenty#1{
\begin{picture}(400,180)(0,#1)
  \begin{picture}(387,208) (0,0)
    \SetWidth{4.0}
    \SetColor{Black}
    \Line(274,160)(96,160)
    \SetWidth{1.0}
    \SetWidth{6.0}
    \Line[dash,dashsize=10](192,192)(192,0)
    \SetWidth{4.0}
    \Line[arrow,arrowpos=0.5,arrowlength=6.667,arrowwidth=4,arrowinset=0.2](48,160)(96,160)
    \Line[arrow,arrowpos=0.5,arrowlength=6.667,arrowwidth=4,arrowinset=0.2](48,32)(96,32)
    \Line(96,160)(96,32)
    \Line(96,32)(160,32)
    \Line[double,sep=8](160,32)(272,32)
    \Arc[clock](224,96)(80,53.13,-53.13)
    \Arc[clock](320,96)(80,-126.87,-233.13)
    \Line[arrow,arrowpos=0.5,arrowlength=6.667,arrowwidth=4,arrowinset=0.2,flip](320,160)(272,160)
    \Line[arrow,arrowpos=0.5,arrowlength=6.667,arrowwidth=4,arrowinset=0.2,flip](320,32)(272,32)
    \Line(160,32)(96,160)
    \Text(26,160)[lb]{\Huge{\Black{$2$}}}
    \Text(26,32)[lb]{\Huge{\Black{$1$}}}
    \Text(342,160)[lb]{\Huge{\Black{$2$}}}
    \Text(342,32)[lb]{\Huge{\Black{$1$}}}
  \end{picture}
  \end{picture}
 }
 
      \def\Mtwentytwo#1{
\begin{picture}(400,180)(0,#1)
  \begin{picture}(417,194) (34,-62)
    \SetWidth{6.0}
    \SetColor{Black}
    \Line[dash,dashsize=10](208,130)(208,-62)
    \SetWidth{4.0}
    \Line[arrow,arrowpos=0.5,arrowlength=6.667,arrowwidth=4,arrowinset=0.2](64,98)(112,98)
    \Line[arrow,arrowpos=0.5,arrowlength=6.667,arrowwidth=4,arrowinset=0.2](64,-30)(112,-30)
    \Line(112,98)(112,-30)
    \Line(112,-30)(176,-30)
    \Line[double,sep=8](176,-30)(288,34)
    \Arc(320,34)(32,90,450)
    \Line[arrow,arrowpos=0.5,arrowlength=6.667,arrowwidth=4,arrowinset=0.2](352,34)(384,66)
    \Line(176,-30)(112,98)
    \Line[arrow,arrowpos=0.5,arrowlength=6.667,arrowwidth=4,arrowinset=0.2](352,34)(384,2)
    \Line(112,98)(288,34)
    \Text(42,96)[lb]{\Huge{\Black{$1$}}}
    \Text(42,-31)[lb]{\Huge{\Black{$2$}}}
    \Text(401,66)[lb]{\Huge{\Black{$1$}}}
    \Text(401,2)[lb]{\Huge{\Black{$2$}}}
  \end{picture}
  \end{picture}
 }

\def\Mtwentythree#1{
\begin{picture}(400,180)(0,#1)
  \begin{picture}(403,208) (0,0)
    \SetWidth{4.0}
    \SetColor{Black}
    \Line[double,sep=8](112,32)(240,32)
    \Line(304,160)(304,32)
    \Line(240,32)(304,32)
    \SetWidth{6.0}
    \Line[dash,dashsize=10](208,192)(208,0)
    \SetWidth{4.0}
    \Line[arrow,arrowpos=0.5,arrowlength=6.667,arrowwidth=4,arrowinset=0.2](64,160)(112,160)
    \Line[arrow,arrowpos=0.5,arrowlength=6.667,arrowwidth=4,arrowinset=0.2](64,32)(112,32)
    \SetWidth{1.0}

    \SetWidth{4.0}
    \Line[arrow,arrowpos=0.5,arrowlength=6.667,arrowwidth=4,arrowinset=0.2,flip](336,160)(304,160)
    \Line[arrow,arrowpos=0.5,arrowlength=6.667,arrowwidth=4,arrowinset=0.2,flip](336,32)(304,32)
    \Line(112,160)(304,160)
    \Line(240,160)(240,32)
    \Arc(160,96)(80,126.87,233.13)
    \Arc(64,96)(80,-53.13,53.13)
    \Text(42,32)[lb]{\Huge{\Black{$1$}}}
    \Text(42,160)[lb]{\Huge{\Black{$2$}}}
    \Text(353,160)[lb]{\Huge{\Black{$1$}}}
    \Text(353,32)[lb]{\Huge{\Black{$2$}}}
  \end{picture}
  \end{picture}
 }
 
 \def\Mtwentynine#1{
\begin{picture}(400,180)(0,#1)
  \begin{picture}(384,194) (19,-62)
    \SetWidth{4.0}
    \SetColor{Black}
    \Line(240,-30)(304,-30)
    \SetWidth{6.0}
    \Line[dash,dashsize=10](208,130)(208,-62)
    \SetWidth{4.0}
    \Line[arrow,arrowpos=0.5,arrowlength=6.667,arrowwidth=4,arrowinset=0.2](32,66)(64,34)
    \Line[arrow,arrowpos=0.5,arrowlength=6.667,arrowwidth=4,arrowinset=0.2](32,2)(64,34)
    \Line[double,sep=8](128,34)(240,-30)
    \Line[arrow,arrowpos=0.5,arrowlength=6.667,arrowwidth=4,arrowinset=0.2,flip](336,98)(304,98)
    \Line[arrow,arrowpos=0.5,arrowlength=6.667,arrowwidth=4,arrowinset=0.2,flip](336,-30)(304,-30)
    \Line(128,34)(240,98)
    \Line(304,98)(240,-30)
    \SetWidth{1.0}
    \SetColor{White}
    \Vertex(272,34){16}
    \SetWidth{4.0}
    \SetColor{Black}
    \Line(240,98)(304,-30)
    \Arc(96,34)(32,180,540)
    \Line(240,98)(304,98)
    \Text(16,66)[lb]{\Huge{\Black{$1$}}}
    \Text(16,2)[lb]{\Huge{\Black{$2$}}}
    \Text(353,98)[lb]{\Huge{\Black{$1$}}}
    \Text(353,-30)[lb]{\Huge{\Black{$2$}}}
  \end{picture}
  \end{picture}
 }
 
  \def\Mfortyone#1{
\begin{picture}(400,180)(0,#1)
  \begin{picture}(387,208) (0,0)
    \SetWidth{4.0}
    \SetColor{Black}
    \Line[double,sep=8](112,32)(240,32)
    \Line(304,160)(304,32)
    \Line(240,32)(304,32)
    \SetWidth{6.0}
    \Line[dash,dashsize=10](208,192)(208,0)
    \SetWidth{4.0}
    \Line[arrow,arrowpos=0.5,arrowlength=6.667,arrowwidth=4,arrowinset=0.2](64,160)(112,160)
    \Line[arrow,arrowpos=0.5,arrowlength=6.667,arrowwidth=4,arrowinset=0.2](64,32)(112,32)
    \SetWidth{1.0}
    \SetWidth{4.0}
    \Line[arrow,arrowpos=0.5,arrowlength=6.667,arrowwidth=4,arrowinset=0.2,flip](336,160)(304,160)
    \Line[arrow,arrowpos=0.5,arrowlength=6.667,arrowwidth=4,arrowinset=0.2,flip](336,32)(304,32)
    \Line(112,160)(304,160)
    \Line(240,160)(240,32)
    \Arc(160,96)(80,126.87,233.13)
    \Arc(64,96)(80,-53.13,53.13)
    \Text(42,160)[lb]{\Huge{\Black{$2$}}}
    \Text(42,32)[lb]{\Huge{\Black{$1$}}}
    \Text(342,160)[lb]{\Huge{\Black{$2$}}}
    \Text(342,32)[lb]{\Huge{\Black{$1$}}}
  \end{picture}
  \end{picture}
 }
 
 \def\Mfortytwo#1{
\begin{picture}(400,180)(0,#1)
  \begin{picture}(403,208) (0,0)
    \SetWidth{4.0}
    \SetColor{Black}
    \Line(240,160)(304,32)
    \Line(240,32)(304,32)
    \Line[arrow,arrowpos=0.5,arrowlength=6.667,arrowwidth=4,arrowinset=0.2,flip](336,32)(304,32)
    \SetWidth{1.0}
    \SetColor{White}
    \Vertex(272,96){16}
    \SetWidth{4.0}
    \SetColor{Black}
    \Line[double,sep=8](112,32)(240,32)
    \SetWidth{6.0}
    \Line[dash,dashsize=10](208,192)(208,0)
    \SetWidth{4.0}
    \Line[arrow,arrowpos=0.5,arrowlength=6.667,arrowwidth=4,arrowinset=0.2](64,160)(112,160)
    \Line[arrow,arrowpos=0.5,arrowlength=6.667,arrowwidth=4,arrowinset=0.2](64,32)(112,32)
    \SetWidth{1.0}
    \SetWidth{4.0}
    \Line[arrow,arrowpos=0.5,arrowlength=6.667,arrowwidth=4,arrowinset=0.2,flip](336,160)(304,160)
    \Line(112,160)(304,160)
    \Line(304,160)(240,32)
    \Arc(160,96)(80,126.87,233.13)
    \Arc(64,96)(80,-53.13,53.13)
    \Text(42,160)[lb]{\Huge{\Black{$1$}}}
    \Text(42,32)[lb]{\Huge{\Black{$2$}}}
    \Text(358,160)[lb]{\Huge{\Black{$2$}}}
    \Text(358,32)[lb]{\Huge{\Black{$1$}}}
  \end{picture}
  \end{picture}
 }
 
  \def\Mnineteen#1{
\begin{picture}(400,180)(0,#1)
  \begin{picture}(371,215) (-35,1)
    \SetWidth{4.0}
    \SetColor{Black}
    \Line(288,183)(208,39)
    \Line(208,39)(288,39)
    \SetWidth{6.0}
    \Line[dash,dashsize=10](178,214)(178,1)
    \SetWidth{4.0}
    \Line[arrow,arrowpos=0.5,arrowlength=6.667,arrowwidth=4,arrowinset=0.2](16,183)(80,183)
    \Line[arrow,arrowpos=0.5,arrowlength=6.667,arrowwidth=4,arrowinset=0.2](16,39)(64,39)
    \Line(64,183)(64,39)
    \Line(64,39)(144,39)
    \Line[double,sep=8](144,39)(208,39)
    \SetWidth{1.0}

    \SetWidth{4.0}
    \Line[arrow,arrowpos=0.5,arrowlength=6.667,arrowwidth=4,arrowinset=0.2,flip](320,183)(288,183)
    \Line[arrow,arrowpos=0.5,arrowlength=6.667,arrowwidth=4,arrowinset=0.2,flip](320,39)(288,39)
    \Line(64,183)(144,39)
    \Line(80,183)(288,183)
    \SetWidth{1.0}
    \SetColor{White}
    \Vertex(247,108){19}
    \SetWidth{4.0}
    \SetColor{Black}
    \Line(208,183)(288,39)
    \Text(0,39)[lb]{\Huge{\Black{$1$}}}
    \Text(0,183)[lb]{\Huge{\Black{$2$}}}
    \Text(336,39)[lb]{\Huge{\Black{$2$}}}
    \Text(336,183)[lb]{\Huge{\Black{$1$}}}
  \end{picture}
  \end{picture}
 }
 
   \def\Mthirtythree#1{
\begin{picture}(400,180)(0,#1)
  \begin{picture}(387,208) (0,0)
    \SetWidth{4.0}
    \SetColor{Black}
    \Line(224,160)(224,32)
    \Line(224,32)(288,32)
    \SetWidth{6.0}
    \Line[dash,dashsize=10](192,192)(192,0)
    \SetWidth{4.0}
    \Line[arrow,arrowpos=0.5,arrowlength=6.667,arrowwidth=4,arrowinset=0.2](48,160)(96,160)
    \Line[arrow,arrowpos=0.5,arrowlength=6.667,arrowwidth=4,arrowinset=0.2](48,32)(96,32)
    \Line(96,160)(96,32)
    \Line(96,32)(160,32)
    \Line[double,sep=8](160,32)(224,32)
    \SetWidth{1.0}

    \SetWidth{4.0}
    \Line[arrow,arrowpos=0.5,arrowlength=6.667,arrowwidth=4,arrowinset=0.2,flip](320,160)(288,160)
    \Line[arrow,arrowpos=0.5,arrowlength=6.667,arrowwidth=4,arrowinset=0.2,flip](320,32)(288,32)
    \Line(96,160)(160,32)
    \Line(96,160)(288,160)
    \Line(288,160)(288,32)
    \Text(26,160)[lb]{\Huge{\Black{$2$}}}
    \Text(26,32)[lb]{\Huge{\Black{$1$}}}
    \Text(342,160)[lb]{\Huge{\Black{$2$}}}
    \Text(342,32)[lb]{\Huge{\Black{$1$}}}
  \end{picture}
  \end{picture}
 }

    \def\Mthirtyfour#1{
\begin{picture}(400,180)(0,#1)
  \begin{picture}(387,208) (0,0)
    \SetWidth{4.0}
    \SetColor{Black}
    \Line(224,160)(224,32)
    \Line(224,32)(288,32)
    \SetWidth{6.0}
    \Line[dash,dashsize=10](192,192)(192,0)
    \SetWidth{4.0}
    \Line[arrow,arrowpos=0.5,arrowlength=6.667,arrowwidth=4,arrowinset=0.2](48,160)(96,160)
    \Line[arrow,arrowpos=0.5,arrowlength=6.667,arrowwidth=4,arrowinset=0.2](48,32)(96,32)
    \Line(96,160)(96,32)
    \Line(96,32)(160,32)
    \Line[double,sep=8](160,32)(224,32)
    \SetWidth{1.0}

    \SetWidth{4.0}
    \Line[arrow,arrowpos=0.5,arrowlength=6.667,arrowwidth=4,arrowinset=0.2,flip](320,160)(288,160)
    \Line[arrow,arrowpos=0.5,arrowlength=6.667,arrowwidth=4,arrowinset=0.2,flip](320,32)(288,32)
    \Line(96,160)(160,32)
    \Line(96,160)(288,160)
    \Line(288,160)(288,32)
    \Text(26,160)[lb]{\Huge{\Black{$2$}}}
    \Text(26,32)[lb]{\Huge{\Black{$1$}}}
    \Text(342,160)[lb]{\Huge{\Black{$1$}}}
    \Text(342,32)[lb]{\Huge{\Black{$2$}}}
  \end{picture}
  \end{picture}
 }
 
    \def\Mfive#1{
\begin{picture}(400,180)(0,#1)
  \begin{picture}(387,208) (0,0)
    \SetWidth{4.0}
    \SetColor{Black}
    \Line(224,160)(224,32)
    \Line(224,32)(288,32)
    \SetWidth{6.0}
    \Line[dash,dashsize=10](192,192)(192,0)
    \SetWidth{4.0}
    \Line[arrow,arrowpos=0.5,arrowlength=6.667,arrowwidth=4,arrowinset=0.2](48,160)(96,160)
    \Line[arrow,arrowpos=0.5,arrowlength=6.667,arrowwidth=4,arrowinset=0.2](48,32)(96,32)
    \Line(96,160)(96,32)
    \Line(96,32)(160,32)
    \Line[double,sep=8](160,32)(224,32)
    \SetWidth{1.0}

    \SetWidth{4.0}
    \Line[arrow,arrowpos=0.5,arrowlength=6.667,arrowwidth=4,arrowinset=0.2,flip](320,160)(288,160)
    \Line[arrow,arrowpos=0.5,arrowlength=6.667,arrowwidth=4,arrowinset=0.2,flip](320,32)(288,32)
    \Line(160,160)(160,32)
    \Line(96,160)(288,160)
    \Line(288,160)(288,32)
    \Text(26,160)[lb]{\Huge{\Black{$2$}}}
    \Text(26,32)[lb]{\Huge{\Black{$1$}}}
    \Text(342,160)[lb]{\Huge{\Black{$1$}}}
    \Text(342,32)[lb]{\Huge{\Black{$2$}}}
  \end{picture}
  \end{picture}
 }
 
     \def\Msix#1{
\begin{picture}(400,180)(0,#1)
  \begin{picture}(387,208) (0,0)
    \SetWidth{4.0}
    \SetColor{Black}
    \Line(224,160)(288,32)
    \Line(224,32)(288,32)
    \SetWidth{1.0}
    \SetColor{White}
    \Vertex(256,96){16}
    \SetWidth{4.0}
    \SetColor{Black}
    \Line[arrow,arrowpos=0.5,arrowlength=6.667,arrowwidth=4,arrowinset=0.2,flip](320,32)(288,32)
    \Line(288,160)(224,32)
    \SetWidth{6.0}
    \Line[dash,dashsize=10](192,192)(192,0)
    \SetWidth{4.0}
    \Line[arrow,arrowpos=0.5,arrowlength=6.667,arrowwidth=4,arrowinset=0.2](48,160)(96,160)
    \Line[arrow,arrowpos=0.5,arrowlength=6.667,arrowwidth=4,arrowinset=0.2](48,32)(96,32)
    \Line(96,160)(96,32)
    \Line(96,32)(160,32)
    \Line[double,sep=8](160,32)(224,32)
    \SetWidth{1.0}

    \SetWidth{4.0}
    \Line[arrow,arrowpos=0.5,arrowlength=6.667,arrowwidth=4,arrowinset=0.2,flip](320,160)(288,160)
    \Line(160,160)(160,32)
    \Line(96,160)(288,160)
    \Text(26,160)[lb]{\Huge{\Black{$2$}}}
    \Text(26,32)[lb]{\Huge{\Black{$1$}}}
    \Text(342,160)[lb]{\Huge{\Black{$2$}}}
    \Text(342,32)[lb]{\Huge{\Black{$1$}}}
  \end{picture}
  \end{picture}
 }
     \def\Mtwentyeight#1{
\begin{picture}(400,180)(0,#1)
  \begin{picture}(387,208) (0,0)
    \SetWidth{4.0}
    \SetColor{Black}
    \Line(224,160)(224,32)
    \Line(224,32)(288,32)
    \SetWidth{6.0}
    \Line[dash,dashsize=10](192,192)(192,0)
    \SetWidth{4.0}
    \Line[arrow,arrowpos=0.5,arrowlength=6.667,arrowwidth=4,arrowinset=0.2](48,160)(96,160)
    \Line[arrow,arrowpos=0.5,arrowlength=6.667,arrowwidth=4,arrowinset=0.2](48,32)(96,32)
    \Line(96,160)(96,32)
    \Line(96,32)(160,32)
    \Line[double,sep=8](160,32)(224,32)
    \SetWidth{1.0}

    \SetWidth{4.0}
    \Line[arrow,arrowpos=0.5,arrowlength=6.667,arrowwidth=4,arrowinset=0.2,flip](320,160)(288,160)
    \Line[arrow,arrowpos=0.5,arrowlength=6.667,arrowwidth=4,arrowinset=0.2,flip](320,32)(288,32)
    \Line(160,160)(160,32)
    \Line(96,160)(288,160)
    \Line(288,160)(288,32)
    \Text(26,160)[lb]{\Huge{\Black{$1$}}}
    \Text(26,32)[lb]{\Huge{\Black{$2$}}}
    \Text(342,160)[lb]{\Huge{\Black{$1$}}}
    \Text(342,32)[lb]{\Huge{\Black{$2$}}}
  \end{picture}
  \end{picture}
 }
     \def\Mthirtysix#1{
\begin{picture}(400,180)(0,#1)
  \begin{picture}(387,209) (0,0)
    \SetWidth{4.0}
    \SetColor{Black}
    \Line(224,161)(224,33)
    \Line(224,33)(288,33)
    \SetWidth{6.0}
    \Line[dash,dashsize=10](192,193)(192,1)
    \SetWidth{4.0}
    \Line[arrow,arrowpos=0.5,arrowlength=6.667,arrowwidth=4,arrowinset=0.2](48,161)(96,161)
    \Line[arrow,arrowpos=0.5,arrowlength=6.667,arrowwidth=4,arrowinset=0.2](48,33)(96,33)
    \Line(96,161)(96,33)
    \Line[double,sep=8](160,33)(224,33)
    \SetWidth{1.0}

    \SetWidth{4.0}
    \Line[arrow,arrowpos=0.5,arrowlength=6.667,arrowwidth=4,arrowinset=0.2,flip](320,161)(288,161)
    \Line[arrow,arrowpos=0.5,arrowlength=6.667,arrowwidth=4,arrowinset=0.2,flip](320,33)(288,33)
    \Line(96,161)(288,161)
    \Line(288,161)(288,33)
    \Arc(128,33)(32,90,450)
    \Text(26,161)[lb]{\Huge{\Black{$2$}}}
    \Text(26,33)[lb]{\Huge{\Black{$1$}}}
    \Text(342,161)[lb]{\Huge{\Black{$1$}}}
    \Text(342,33)[lb]{\Huge{\Black{$2$}}}
  \end{picture}
  \end{picture}
 }
     \def\Msixtysix#1{
\begin{picture}(400,180)(0,#1)
  \begin{picture}(387,208) (0,0)
    \SetWidth{4.0}
    \SetColor{Black}
    \Line(224,160)(288,32)
    \Line(224,32)(288,32)
    \SetWidth{1.0}
    \SetColor{White}
    \Vertex(256,96){16}
    \SetWidth{4.0}
    \SetColor{Black}
    \Line[arrow,arrowpos=0.5,arrowlength=6.667,arrowwidth=4,arrowinset=0.2,flip](320,32)(288,32)
    \Line(288,160)(224,32)
    \SetWidth{6.0}
    \Line[dash,dashsize=10](192,192)(192,0)
    \SetWidth{4.0}
    \Line[arrow,arrowpos=0.5,arrowlength=6.667,arrowwidth=4,arrowinset=0.2](48,160)(96,160)
    \Line[arrow,arrowpos=0.5,arrowlength=6.667,arrowwidth=4,arrowinset=0.2](48,32)(96,32)
    \Line(96,32)(160,32)
    \Line[double,sep=8](160,32)(224,32)
    \SetWidth{1.0}

    \SetWidth{4.0}
    \Line[arrow,arrowpos=0.5,arrowlength=6.667,arrowwidth=4,arrowinset=0.2,flip](320,160)(288,160)
    \Line(96,160)(160,32)
    \Line(96,160)(288,160)
    \Text(26,160)[lb]{\Huge{\Black{$2$}}}
    \Text(26,32)[lb]{\Huge{\Black{$1$}}}
    \Text(342,160)[lb]{\Huge{\Black{$2$}}}
    \Text(342,32)[lb]{\Huge{\Black{$1$}}}
    \SetWidth{1.0}
    \SetColor{White}
    \Vertex(128,96){16}
    \SetWidth{4.0}
    \SetColor{Black}
    \Line(160,160)(96,32)
  \end{picture}
  \end{picture}
 }
 
     \def\Yone#1{
  \begin{picture}(224,157) (20,55)
    \SetWidth{0.5}
    \SetColor{Black}
    \Text(224,125)[lb]{\Huge{\Black{$1$}}}
    \Text(224,-3)[lb]{\Huge{\Black{$2$}}}
    \SetWidth{6.0}
    \Line[dash,dashsize=10](112,141)(112,-19)
    \SetWidth{4.0}
    \Line[arrow,arrowpos=0.5,arrowlength=6.667,arrowwidth=4,arrowinset=0.2](48,77)(64,61)
    \Line[arrow,arrowpos=0.5,arrowlength=6.667,arrowwidth=4,arrowinset=0.2](48,45)(64,61)
    \Line[double,sep=8](64,61)(176,-3)
    \Line[arrow,arrowpos=0.5,arrowlength=6.667,arrowwidth=4,arrowinset=0.2,flip](208,125)(176,125)
    \Line[arrow,arrowpos=0.5,arrowlength=6.667,arrowwidth=4,arrowinset=0.2,flip](208,-3)(176,-3)
    \Line(64,61)(176,125)
    \Text(32,77)[lb]{\Huge{\Black{$2$}}}
    \Text(32,45)[lb]{\Huge{\Black{$1$}}}
    \Arc(224,61)(80,126.87,233.13)
    \Arc(128,61)(80,-53.13,53.13)
  \end{picture}
 }
 
     \def\Ythree#1{
  \begin{picture}(224,157) (20,55)
    \SetWidth{0.5}
    \SetColor{Black}
    \Text(224,125)[lb]{\Huge{\Black{$1$}}}
    \Text(224,-3)[lb]{\Huge{\Black{$2$}}}
    \SetWidth{6.0}
    \Line[dash,dashsize=10](112,141)(112,-19)
    \SetWidth{4.0}
    \Line[arrow,arrowpos=0.5,arrowlength=6.667,arrowwidth=4,arrowinset=0.2](48,77)(64,61)
    \Line[arrow,arrowpos=0.5,arrowlength=6.667,arrowwidth=4,arrowinset=0.2](48,45)(64,61)
    \Line[double,sep=8](64,61)(144,13)
    \Line[arrow,arrowpos=0.5,arrowlength=6.667,arrowwidth=4,arrowinset=0.2,flip](208,125)(176,125)
    \Line[arrow,arrowpos=0.5,arrowlength=6.667,arrowwidth=4,arrowinset=0.2,flip](208,-3)(176,-3)
    \Line(64,61)(176,125)
    \Text(32,77)[lb]{\Huge{\Black{$2$}}}
    \Text(32,45)[lb]{\Huge{\Black{$1$}}}
    \Line(176,125)(144,13)
    \Line(144,13)(176,-3)
    \Line(176,-3)(176,125)
  \end{picture}
 }

     \def\Yfour#1{
  \begin{picture}(224,157) (20,90)
    \SetWidth{0.5}
    \SetColor{Black}
    \Text(256,162)[lb]{\Huge{\Black{$1$}}}
    \Text(256,34)[lb]{\Huge{\Black{$2$}}}
    \SetWidth{4.0}
    \Line(145,164)(145,36)
    \Line(145,36)(209,36)
    \SetWidth{6.0}
    \Line[dash,dashsize=10](113,196)(113,4)
    \SetWidth{4.0}
    \Line[arrow,arrowpos=0.5,arrowlength=6.667,arrowwidth=4,arrowinset=0.2](33,164)(81,164)
    \Line[arrow,arrowpos=0.5,arrowlength=6.667,arrowwidth=4,arrowinset=0.2](33,36)(81,36)
    \Line(81,164)(81,36)
    \Line[double,sep=8](81,36)(145,36)
    \SetWidth{4.0}
    \Line[arrow,arrowpos=0.5,arrowlength=6.667,arrowwidth=4,arrowinset=0.2,flip](241,164)(209,164)
    \Line[arrow,arrowpos=0.5,arrowlength=6.667,arrowwidth=4,arrowinset=0.2,flip](241,36)(209,36)
    \Line(81,164)(209,164)
    \Line(209,164)(209,36)
    \Text(17,34)[lb]{\Huge{\Black{$1$}}}
    \Text(17,164)[lb]{\Huge{\Black{$2$}}}
  \end{picture}
 }

     \def\Yfive#1{
  \begin{picture}(300,157) (20,55)
    \SetWidth{0.5}
    \SetColor{Black}
    \Text(288,74)[lb]{\Huge{\Black{$1$}}}
    \Text(288,42)[lb]{\Huge{\Black{$2$}}}
    \SetWidth{6.0}
    \Line[dash,dashsize=10](112,138)(112,-22)
    \SetWidth{4.0}
    \Line[arrow,arrowpos=0.5,arrowlength=6.667,arrowwidth=4,arrowinset=0.2](48,74)(64,58)
    \Line[arrow,arrowpos=0.5,arrowlength=6.667,arrowwidth=4,arrowinset=0.2](48,42)(64,58)
    \Line[arrow,arrowpos=0.5,arrowlength=6.667,arrowwidth=4,arrowinset=0.2,flip](272,74)(256,58)
    \Line[arrow,arrowpos=0.5,arrowlength=6.667,arrowwidth=4,arrowinset=0.2,flip](272,42)(256,58)
    \Text(32,74)[lb]{\Huge{\Black{$2$}}}
    \Text(32,42)[lb]{\Huge{\Black{$1$}}}
    \Arc[clock](112.467,58.467)(47.535,176.946,-0.563)
    \Arc[double,sep=8,clock](112,58)(48,-0,-180)
    \Arc[clock](208,60.667)(48.074,-176.82,-363.18)
    \Arc[clock](208,58)(48,-0,-180)
  \end{picture}
 }

     \def\Ysix#1{
  \begin{picture}(224,157) (20,55)
    \SetWidth{4.0}
    \SetColor{Black}
    \Line(144,13)(176,-3)
    \Line(176,-3)(144,109)
    \Text(224,125)[lb]{\Huge{\Black{$1$}}}
    \Text(224,-3)[lb]{\Huge{\Black{$2$}}}
    \SetWidth{6.0}
    \Line[dash,dashsize=10](112,141)(112,-19)
    \SetWidth{4.0}
    \Line[arrow,arrowpos=0.5,arrowlength=6.667,arrowwidth=4,arrowinset=0.2](48,77)(64,61)
    \Line[arrow,arrowpos=0.5,arrowlength=6.667,arrowwidth=4,arrowinset=0.2](48,45)(64,61)
    \Line[arrow,arrowpos=0.5,arrowlength=6.667,arrowwidth=4,arrowinset=0.2,flip](208,125)(176,125)
    \Line[arrow,arrowpos=0.5,arrowlength=6.667,arrowwidth=4,arrowinset=0.2,flip](208,-3)(176,-3)
    \Text(32,77)[lb]{\Huge{\Black{$2$}}}
    \Text(32,45)[lb]{\Huge{\Black{$1$}}}
    \Line(64,61)(144,109)
    \Line[double,sep=8](64,61)(144,13)
    \Line(144,109)(176,125)
    \SetColor{White}
    \Vertex(160,61){16}
    \SetWidth{4.0}
    \SetColor{Black}
    \Line(176,125)(144,13)
  \end{picture}
 }

%% file: APP_IteratedIntegrals.tex
\section{ Hypergeometric functions through iterated integrals }
\label{sec:itint}
In this appendix we define a class of iterated integrals as also introduced in Section~\ref{sec:revunit}. 
First let us define the integral
\bea
\label{eq:itintdef1}
\mathcal{F}_{\vec{a}}(c;\delta)&=&\int\limits_0^{\delta} dt\, t^{a-1} \,(1-c t)^{-b}\nonumber\\
&=&\frac{\delta^a}{a} {}_{2}F_1(a,b;a+1;c \bar{ z})\nonumber\\
&=&\frac{\delta^a}{a}\sum_{n=0}^\infty \frac{(a)_n (b)_n}{(a+1)_n} \frac{(c\bar{ z})^n}{n!}\,,
\label{eq:2f1itint}
\eea
where we have abbreviated for later convenience $\vec{a}=\tiny{\left(\begin{array}{c}a\\ b\end{array}\right)}$.
We have made use of Gauss' hypergeometric function with the third argument being the first argument increased by one. Next, we define recursively the $n^{th}$ iterated integral by
\beq
\label{eq:itintdef2}
\mathcal{F}_{\vec{a}_n,\dots,\vec{a}_1} (x_n,\dots,x_1;\delta)=\int_0^\delta dt\, t^{a_n-1}\,(1-x_n t)^{-b_n}\,\mathcal{F}_{\vec{a}_{n-1},\dots,\vec{a}_1} (x_{n-1},\dots,x_1;t)\,.
\eeq
The integration kernel $t^{a-1}(1-c t)^{-b}$ has the same form for every iteration step with indices $a,b$ and argument $c$ changing. Next, we derive a hypergeometric series representation for these iterated integrals.
To simplify the expressions we rewrite eq.~\eqref{eq:2f1itint} and introduce a function f that is implicitly given by
\bea
\mathcal{F}_{\vec{a}}(c;\delta)&=&\sum_{n=0}^\infty f(a,b,c,n) \delta^{a+n}.
\eea 
In the next step we integrate over the integration kernel and the $ \mathcal{F}_{\vec{a}}(c;t)$
\bea
\mathcal{F}_{\vec{a}_2,\vec{a}_1}(c_2,c_1;\delta)&=&\int\limits_0^{\delta} dt\, t^{a_2-1}\, (1-c_2 t)^{-b_2}\, \mathcal{F}_{\vec{a}}(c;t)\nonumber\\
&=&\sum_{n=0}^\infty \int\limits_0^{\delta} dt\, t^{a_2+a_1+n-1} \,(1-c_2 t)^{-b_2}\, f(a_1,b_1,c_1,n)\nonumber\\
&=&\sum_{n,m=0}^\infty \frac{\delta^{a_1+a_2+n+m}}{a_1+a_2+n} \frac{(a_1+a_2+n)_m (b_2)_m}{(a_1+a_2+n+1)_m} \frac{c_2^m}{m!} f(a_1,b_1,c_1,n)\,.\nonumber
\eea
Using the identity 
\beq
(a+n)_m=(a)_{n+m} \frac{\Gamma (a) }{\Gamma(a+n)}\,,
\eeq
we can write
\bea
\mathcal{F}_{\vec{a}_2,\vec{a}_1}(c_2,c_1;\delta)&=&\frac{\delta^{a_1+a_2}}{(a_1+a_2)a_1}\sum_{n,m=0}^\infty \frac{(a_1+a_2)_{m+n}}{(a_1+a_2+1)_{m+n}} \frac{(a_1)_n}{(a_1+1)}_n (b_2)_m (b_1)_n \frac{(c_2 \delta)^m}{m!} \frac{(c_1 \delta)^n}{n!} \nonumber\,.
\eea
We now proceed iteratively, and find the following series representation for the iterated integrals
\bea
\mathcal{F}_{\vec{a}_{n},\vec{a}_{n-1},\dots,\vec{a_1}}(c_n,\dots,c_1;\delta)&=&\frac{\delta^{A_n}}{\prod\limits_{i=1}^n A_i}\,\sum\limits_{k_1,\dots,k_n=0}^\infty \, \prod\limits_{i=1}^n \frac{(A_i)_{K_i}}{(A_i+1)_{K_i}} (b_i)_{k_i} \frac{(c_i \delta)^{k_i}}{k_i !}\,,
\eea 
with the abbreviations 
\beq
A_i=\sum_{n=1}^i a_n {\rm~~and~~} K_i=\sum_{n=1}^i k_n\,.
\eeq

Following the the same procedure as for the sum representation we can derive a general Mellin barnes representation for our iterated integrals by utilizing the Mellin-Barnes representation of the Gauss Hypergeometric function
\beq
{}_2 F_1(a,b;c;\delta)=\frac{1}{2\pi i}\frac{\Gamma(c)}{\Gamma(a)\Gamma(b)} \int_{-i\infty}^{i \infty} ds \frac{\Gamma(a+s)\Gamma(b+s)}{\Gamma(c+s)} \Gamma(-s)(-\delta)^s.
\eeq
This leads to
\bea
\mathcal{F}_{\vec{a}_{n},\vec{a}_{n-1},\dots,\vec{a_1}}(c_n,\dots,c_1;\delta)&=&\delta^{A_n} \int_{-i\infty}^{i \infty} \prod\limits_{i=1}^n  \frac{dk_i}{2\pi i}\frac{\Gamma(A_i+K_i)}{\Gamma(A_i+K_i+1)}\frac{\Gamma(b_i+k_i)}{\Gamma(b_i)} \Gamma(-k_i )(-c_i \delta)^{k_i}\,.\nonumber\\
\eea
These iterated integrals interpolate between multiple polylogarithms~\cite{Goncharov:1998kja} and hypergeometric functions. In the framework of the above definitions the multiple polylogarithm is given by 
\beq\bsp
\textrm{Li}&_{m_1,\dots,m_k}(x_1,\dots,x_k)\\
&=\mathcal{F}_{\scriptsize\underbrace{\vec{0},\dots,\vec{0}}_{m_1-1},\vec{1},\dots,\underbrace{\vec{0},\dots,\vec{0}}_{m_k-1},\vec{1}}(\underbrace{0,\dots,0}_{m_1-1},x_1,\dots,\underbrace{0,\dots,0}_{m_k-1};x_k\dots x_1)\,\prod_{i=1}^kx_i^{k-i+1}\,.
\esp\eeq
Here the indices of the iterated integrals only take the form $\tiny{\vec{0}=\left(\begin{array}{c}0\\0\end{array}\right)}$ and $\tiny{\vec{1}=\left(\begin{array}{c}1\\1\end{array}\right)}$.
Even for general indices we discover further similarities of these iterated integrals with multiple polylogarithms.
As in the case of polylogarithms this class of hypergeometric functions may be written as multiple nested sums
\bea
\label{eq:nesteditint}
\mathcal{F}_{\vec{a}_{n},\vec{a}_{n-1},\dots,\vec{a_1}}(c_n,\dots,c_1;\delta)&=&\sum\limits_{k_n\geq\dots\geq k_1=0}^\infty \delta^{A_n+k_n}\, \prod\limits_{i=1}^n \frac{1}{A_i+k_{i}} \frac{(b_i)_{k_i-k_{i-1}}  (c_i )^{k_i-k_{i-1}}}{(k_i-k_{i-1}) !}\,,\nonumber\\
\eea
where $k_0=0$. The representation~\eqref{eq:nesteditint} may be useful to expand the iterated integrals in terms of the dimensional regulator (see, e.g., ref.~\cite{Moch:2001zr}).

The definition of these function as iterated integrals implies that they form a shuffle algebra,
\beq
\mathcal{F}_{\vec{a}_i,\dots,\vec{a}_1}(c_i,\dots,c_1;\delta)\, \mathcal{F}_{\vec{a}_n,\dots,\vec{a}_{i+1}}(c_n,\dots,c_{i+1};\delta)=\sum_{\sigma\in \Sigma(i,n-i)} \mathcal{F}_{\vec{a}_{\sigma(n)},\dots,\vec{a}_{\sigma(1)}}(c_{\sigma(n)},\dots,c_{\sigma(1)};\delta),
\eeq
where $\Sigma(i,n-i)$ denotes the set of all shuffles of $n$ elements, i.e., the subset of the symmetric group $S_n$ defined by
\beq
\Sigma(i,n-i)=\{\sigma\in S_n \mid \sigma^{-1}(1)< \dots < \sigma^{-1}(i) \hspace{0.5cm} \text{and}  \hspace{0.5cm} \sigma^{-1}(i+1) <  \dots < \sigma^{-1}(n) \}.
\eeq
To illustrate an application of the shuffle-product for generalized iterated integrals, let us look at the following example.
We would like to integrate an iterated integral over a non-standard integration kernel.
\beq
\mathcal{I}=\int_0^{\delta} dt\, t^{a_2-1}\,(1-c_2 t)^{-b_2}\,(1-c_3 t)^{-b_3}\,\mathcal{F}_{\vec{a}_1}(c_1;t).
\eeq
To simplify the integral we make use of 
\beq
\mathcal{F}_{\tiny{\left( \begin{array}{c}a \\ a+1\end{array}\right)}}(1; \delta)=\frac{\delta^a}{a}(1-c \delta)^{-a}
\eeq
and find 
\beq
\mathcal{I}=b_3 \int_0^{\delta} dt\, t^{a_2-b_3-1}\, (1-c_2 t)^{-b_2}\,\mathcal{F}_{\tiny{\left( \begin{array}{c}b_3 \\ b_3+1\end{array}\right)}}(c_3; t)\mathcal{F}_{\tiny{\left( \begin{array}{c}a_1 \\ b_1\end{array}\right)}}(c_1; t)
\eeq
Next, we apply the shuffle product and find
\bea
\mathcal{I}&=&b_3 \int_0^{\delta} dt\, t^{a_2-b_3-1}\, (1-c_2 t)^{-b_2} 
\left[\mathcal{F}_{\tiny{\left( \begin{array}{c | c} b_3 & a_1 \\ b_3+ 1 & b_1\end{array}\right)}}(c_3,c_1; t)+\mathcal{F}_{\tiny{\left( \begin{array}{c | c} a_1 & b_3 \\ b_1 & b_3+1\end{array}\right)}}(c_1,c_3; t) \right]\nonumber\\
&=& b_3\, \mathcal{F}_{\tiny{\left( \begin{array}{c | c | c} a_2-b_3 & b_3 & a_1 \\ b_2 & b_3+ 1 & b_1\end{array}\right)}}(c_2,c_3,c_1; \delta)+b_3\, \mathcal{F}_{\tiny{\left( \begin{array}{c | c | c} a_2-b_3 & a_1 & b_3 \\ b_2 & b_1 & b_3+1\end{array}\right)}}(c_2,c_1,c_3; \delta) .
\eea
Further identities among iterated integrals can be derived using integration-by-parts or by partial fractioning products of integration kernels.
Further properties and parallels of generalized iterated integrals and generalized poly-logarithms are under investigation.

%% file: APP_NNMasters.tex
\section{ NNLO RV Master Integrals as hypergeometric functions}
\label{sec:NNMasters}
In this appendix we demonstrate how certain differential equations for master integrals appearing in physical cross-sections can be solved using iterated integrals as introduced in Appendix~\ref{sec:itint}. We consider the example of the master integrals contributing to the $RV$ Higgs boson cross-section at $NNLO$. The master integrals were introduced and evaluated as an expansion in  the dimensional regulator $\epsilon$ in ref.~\cite{Anastasiou:2002yz} and evaluated to even higher order in ref.~\cite{Anastasiou:2012kq}. Here we solve them to all orders in $\epsilon$ in terms of hypergeometric functions. 

The master integrals and the corresponding differential equations are given by
\beq
\mathcal{Y}_{1}=\scalebox{0.3}{\Yone{90}}=\int d\Phi_{2} \text{Bub}^*(s_{13}).
\eeq

\beq
\partial_\delta \mathcal{Y}_1=\frac{ (1-3 \epsilon )}{\delta}\mathcal{Y}_1.
\eeq
\beq
\mathcal{Y}_{5}=\scalebox{0.3}{\Yfive{90}}=\int d\Phi_{2} \text{Bub}^*(s_{12}).
\eeq
\vspace{1cm}
\beq
\partial_\delta \mathcal{Y}_5=\frac{ (1-2 \epsilon )}{\delta}\mathcal{Y}_5.
\eeq
\beq
\mathcal{Y}_{3}=\scalebox{0.3}{\Ythree{90}}=\int d\Phi_{2} \text{Tri}^*(s_{12}+s_{23}).
\eeq
\vspace{1cm}
\beq
\partial_\delta \mathcal{Y}_3=\frac{2 \epsilon  \delta}{1-\delta}\mathcal{Y}_3
-\frac{(1-3   \epsilon ) (1-2 \epsilon )}{(1-\delta) \delta \epsilon   }\mathcal{Y}_1
   +   \frac{(1-2 \epsilon )^2(1-\delta) ^{-1-\epsilon}}{\delta   \epsilon }\mathcal{Y}_5.
\eeq
\beq
\mathcal{Y}_{4}=\scalebox{0.3}{\Yfour{90}}=\int d\Phi_{2}\text{Box}^*(s_{12},s_{23},s_{13}) \frac{1}{s_{23}}.
\eeq
\vspace{1cm}
\beq
\partial_\delta \mathcal{Y}_4=-\frac{(1+2 \epsilon ) }{\delta}\mathcal{Y}_4
-\frac{(2   \delta-3) (1-3 \epsilon ) (1-2 \epsilon )}{(1-\delta)   \delta^3 \epsilon }\mathcal{Y}_1
+\frac{2 (1-2 \epsilon   )^2}{(1-\delta)^{1+\epsilon} \delta^2 \epsilon }\mathcal{Y}_5
-\frac{2 \epsilon }{(1-\delta) \delta}\mathcal{Y}_3.
\eeq
\beq
\mathcal{Y}_{6}=\scalebox{0.3}{\Ysix{90}}=\int d\Phi_{2} \text{Box}^*(s_{13},s_{23},s_{12}) .
\eeq
\vspace{1cm}
\beq
\partial_\delta \mathcal{Y}_6=-\frac{(1+4 \epsilon ) }{\delta}\mathcal{Y}_6
+\frac{2(1-3   \epsilon ) (1-2 \epsilon )}{(1-\delta) \delta^2 \epsilon }\mathcal{Y}_1
   -\frac{4  \epsilon }{(1-\delta) \delta}\mathcal{Y}_3.
\eeq
We have abbreviated $\partial_\delta=\frac{\partial}{\partial\delta}$. Solid lines represent scalar propagators. The phase-space integral is represented by the dashed line and the cut-propagators are the lines cut by the dashed line.  The cut propagator of the Higgs boson is depicted by the double-line. The complex conjugated one-loop integral is on the right-hand side of the phase-space cut. The differential equations were obtained using the methods described in Section~\ref{sec:revunit}. 

The system of differential equations is decoupled and can be solved as described in Section~\ref{sec:revunit}.
To solve the differential equations we require the following boundary conditions, which can be obtained from ref.~\cite{Anastasiou:2012kq},
\beq
\mathcal{Y}^S_1=\frac{ (4 \pi )^{ \epsilon -1} s^{-2 \epsilon } \Gamma (1-2
   \epsilon ) \Gamma (1-\epsilon )^2 \Gamma (\epsilon +1)}{\epsilon 
   \Gamma (2-3 \epsilon ) \Gamma (2-2 \epsilon )},
\eeq
\beq
\mathcal{Y}^S_5=\frac{ (4 \pi )^{ \epsilon -1} (-s)^{-\epsilon } s^{-\epsilon }
   \Gamma (1-\epsilon )^3 \Gamma (\epsilon +1)}{2\epsilon  \Gamma
   (2-2 \epsilon )^2},
\eeq
\beq
\mathcal{Y}^S_6=-\frac{ (4 \pi) ^{ \epsilon -1} (-s)^{-\epsilon }
   s^{-2-\epsilon } \Gamma (1-2 \epsilon ) \Gamma (1-\epsilon )^2
   \Gamma (\epsilon +1)^2}{\epsilon ^3 \Gamma (1-4 \epsilon )},
\eeq
and all other cases vanish.
 These boundary conditions are given by the leading (soft) term of three master integrals in the limit of $\delta\rightarrow 0$. They are obtained using the methods described in Section~\ref{sec:regions}. For convenience we will from now on set $s=1$.
 
 The first two differential equations are homogeneous and can be easily solved to give
\beq
\mathcal{Y}_1(\delta)=\delta^{1-3\epsilon} \mathcal{Y}^S_1.
\eeq
\beq
\mathcal{Y}_5(\delta)=\delta^{1-2\epsilon} \mathcal{Y}^S_5.
\eeq
We find that the homogeneous solution to the differential equation of master $\mathcal{Y}_3$ is vanishing and the inhomogeneous solution is according to eq.~\eqref{eq:inhom} given by
\beq\bsp
&\mathcal{Y}_3(\delta)=\\
&=(1-\delta)^{-2\epsilon} \frac{(1-2\epsilon)}{\epsilon}\int\limits_{0}^{\delta} d\delta^\prime \left((1-2 \epsilon ) (1-{\delta^\prime})^{\epsilon -1}
   {\delta^\prime}^{-2 \epsilon }\mathcal{Y}_5^S   -(1-3   \epsilon )  (1-{\delta^\prime})^{2 \epsilon -1}   {\delta^\prime}^{-3 \epsilon }\mathcal{Y}_1^S\right)\\
   &=\frac{(2 \epsilon -1)^2 (1-\delta)^{-2 \epsilon }}{\epsilon }\mathcal{F}_{\tiny{\left(\begin{array}{c} 1-2\epsilon \\ 1-\epsilon \end{array}\right)}}(1;\delta)\,\mathcal{Y}_5^S
-\frac{ (2 \epsilon   -1) (3 \epsilon -1) (1-\delta)^{-2 \epsilon }   }{\epsilon }\mathcal{F}_{\tiny{\left(\begin{array}{c} 1-3\epsilon \\ 1-2\epsilon \end{array}\right)}}(1;\delta)\,\mathcal{Y}_1^S\,,\\
\esp\eeq
with
\beq
\mathcal{F}_{\tiny{\left(\begin{array}{c} a \\b \end{array}\right)}}(1;\delta)=\frac{\delta ^a}{a} \, _2F_1(a,b;a+1;\delta ).
\eeq
To obtain this result we made use of eq.~\eqref{eq:itintdef1}. As we proceed to solve the remaining two master integrals we find that the inhomogeneous solution to their differential equation is in turn dependent on $\mathcal{Y}_3$. We are able to find solutions to the inhomogeneous equation making use of the definition of our iterated integrals in eq.~\eqref{eq:itintdef2}.

\bea
\mathcal{Y}_4(\delta)&=&\delta^{-1-2\epsilon} \int\limits_{0}^{\delta} d\delta^\prime  2 (1-3 \epsilon ) (1-2 \epsilon ) {\delta^\prime}^{2   \epsilon } (1-{\delta^\prime})^{-2 \epsilon -1} \mathcal{F}_{\tiny{\left(\begin{array}{c} 1-3\epsilon \\ 1-2\epsilon \end{array}\right)}}(1;\delta^\prime)\mathcal{Y}_1^S\nonumber\\
&-&\delta^{-1-2\epsilon} \int\limits_{0}^{\delta} d\delta^\prime2(1-2 \epsilon )^2   {\delta^\prime}^{2 \epsilon } (1-{\delta^\prime})^{-2 \epsilon -1}   \mathcal{F}_{\tiny{\left(\begin{array}{c} 1-2\epsilon \\ 1-\epsilon \end{array}\right)}}(1;\delta^\prime)\mathcal{Y}_5^S\nonumber\\
&-&\delta^{-1-2\epsilon} \int\limits_{0}^{\delta} d\delta^\prime\frac{(2 {\delta^\prime}-3) (1-3 \epsilon ) (1-2 \epsilon ) {\delta^\prime}^{-\epsilon   -1}}{(1-{\delta^\prime}) \epsilon }\mathcal{Y}_1^S\nonumber\\
&+&\delta^{-1-2\epsilon} \int\limits_{0}^{\delta} d\delta^\prime\frac{2 (1-2   \epsilon )^2 (1-{\delta^\prime})^{-\epsilon -1}}{\epsilon }\mathcal{Y}_5^S\nonumber\\
&=&2 \left(6 \epsilon   ^2-5 \epsilon +1\right) \delta ^{-2 \epsilon -1}   \mathcal{F}_{\tiny{\left(\begin{array}{c | c} 1+2\epsilon & 1-3\epsilon \\ 1+ 2\epsilon & 1-2\epsilon \end{array}\right)}}(1,1;\delta)\mathcal{Y}_1^S\nonumber\\
&-&2  (1-2 \epsilon )^2 \delta ^{-2   \epsilon -1} \mathcal{F}_{\tiny{\left(\begin{array}{c | c} 1+2\epsilon & 1-2\epsilon \\ 1+ 2\epsilon & 1-\epsilon \end{array}\right)}}(1,1;\delta)\mathcal{Y}_5^S\nonumber\\
&+&\frac{ \left(6 \epsilon ^2-5 \epsilon +1\right) \delta   ^{-2 \epsilon -1}}{\epsilon }\mathcal{F}_{\tiny{\left(\begin{array}{c} 1-\epsilon \\ 1\end{array}\right)}}(1;\delta)\mathcal{Y}_1^S\nonumber\\
&-&\frac{3  \left(6 \epsilon ^2-5   \epsilon +1\right) \delta ^{-3 \epsilon -1}}{\epsilon ^2}\mathcal{Y}_1^S\nonumber\\
&+&\frac{2 (1-2 \epsilon )^2 \left((1-\delta )^{-\epsilon   }-1\right)  \delta ^{-2 \epsilon   -1}}{\epsilon ^2}\mathcal{Y}_5^S
\eea
\bea
\mathcal{Y}_6(\delta)&=&\delta^{-1-4\epsilon}4 (1-3 \epsilon ) (1-2 \epsilon )  \int\limits_{0}^{\delta} d\delta^\prime{\delta^\prime}^{4   \epsilon } (1-{\delta^\prime})^{-2 \epsilon -1}  \mathcal{F}_{\tiny{\left(\begin{array}{c} 1-3\epsilon \\ 1-2\epsilon \end{array}\right)}}(1;\delta^\prime)\mathcal{Y}_1^S\nonumber\\
&-&\delta^{-1-4\epsilon} 4 (1-2 \epsilon )^2\int\limits_{0}^{\delta} d\delta^\prime   {\delta^\prime}^{4 \epsilon } (1-{\delta^\prime})^{-2 \epsilon -1}  \mathcal{F}_{\tiny{\left(\begin{array}{c} 1-2\epsilon \\ 1-\epsilon \end{array}\right)}}(1;\delta^\prime)\mathcal{Y}_5^S\nonumber\\
&+&\delta^{-1-4\epsilon}\frac{2  \left(6 \epsilon   ^2-5 \epsilon +1\right)}{ \epsilon}  \int\limits_{0}^{\delta} d\delta^\prime  \frac{ {\delta^\prime}^{\epsilon  }}{(1-{\delta^\prime}) }\mathcal{Y}_1^S\nonumber\\
&+&\delta^{-1-4\epsilon}\mathcal{Y}_6^S\nonumber\\
&=&4\left(6 \epsilon   ^2-5 \epsilon +1\right) \delta ^{-4 \epsilon -1}  \mathcal{F}_{\tiny{\left(\begin{array}{c|c} 1+4 \epsilon&1-3\epsilon \\ 1+2\epsilon & 1-2\epsilon \end{array}\right)}}(1,1;\delta)\mathcal{Y}_1^S\nonumber\\
&-&4  (1-2 \epsilon )^2 \delta ^{-4   \epsilon -1} \mathcal{F}_{\tiny{\left(\begin{array}{c|c} 1+4 \epsilon&1-2\epsilon \\ 1+2\epsilon & 1-\epsilon \end{array}\right)}}(1,1;\delta)\mathcal{Y}_5^S\nonumber\\
&+&\frac{2 \left(6 \epsilon ^2-5 \epsilon +1\right) \delta   ^{-4 \epsilon -1} }{\epsilon }\mathcal{F}_{\tiny{\left(\begin{array}{c} 1+\epsilon \\ 1\end{array}\right)}}(1;\delta)\mathcal{Y}_1^S\nonumber\\
&+&\delta ^{-1-4 \epsilon}\mathcal{Y}_6^S
\eea
The iterated integrals with two indices contributing to $\mathcal{Y}_4$ and $\mathcal{Y}_6$ can be written as
\bea
\mathcal{F}_{\tiny{\left(\begin{array}{c|c} a_2&a_1 \\ b_2 & b_1 \end{array}\right)}}(1,1,\delta)&=&\frac{\delta^{a_1+a_2}}{(a_1+a_2)a_1}F_{1,1}^{1,2}\left(\left.
\begin{array}{c}
a_2+a_1\\
a_2+a_1+1 \\
\end{array}
\right|\left.
\begin{array}{cccc}
a_1 &1 & b_1 &b_2\\
a_1+1& 1&-& -
\end{array}
\right| \delta, \delta \right)\\
&=&\frac{\delta^{a_1+a_2}}{(a_1+a_2)a_1}\sum\limits_{n,m=0}^\infty \frac{(a_2+a_1)_{n+m}}{(a_2+a_1+1)_{n+m}}\frac{(a_1)_{n}(b_1)_{n}}{(a_1+1)_{n} } (b_2)_{m}\frac{\delta^n}{n!}\frac{\delta^m}{m!} \nonumber\,,
\eea
where we introduced the Kamp\'{e} de F\'{e}riet function,
\beq\label{eq:KdF}
F_{p',q'}^{p,q}\left(\left.
\begin{array}{c}
\alpha_i\\
\alpha'_i \\
\end{array}
\right|\left.
\begin{array}{cccc}
\beta_i &\gamma_i \\
\beta'_i &\gamma'_i \\
\end{array}
\right| x, y \right) = \sum_{n,m=0}^\infty\frac{\prod_{i=1}^p(\alpha_i)_{n+m}\,\prod_{i=1}^q(\beta_i)_{n}\,(\gamma_i)_m}{\prod_{i=1}^{p'}(\alpha'_i)_{n+m}\,\prod_{i=1}^{q'}(\beta'_i)_{n}\,(\gamma'_i)_m}\,\frac{x^n}{n!}\,\frac{y^m}{m!}\,.
\eeq
Note that the generalized Kamp\'{e} de F\'{e}riet function $S_1$ of Section~\ref{sec:sigma_expansion} is a special case of eq.~\eqref{eq:KdF},
\beq
S_1(a_1;a_2,b_1;a_3,b_2;x_1,x_2) = F_{1,1}^{2,1}\left(\begin{array}{cc|cc|}
a_1 & a_2 & b_1&1\\
a_3&-&b_2 &1
\end{array}\,\,x,y\right)\,.
\eeq

%% file: APP_MatrixElements.tex
\section{Matrix-elements}\label{app:AB_funcs}
\begin{equation}\begin{split}
&A_{ggg}(s_{12},s_{23},s_{31}) = \\
&\text{Box}(s_{12},s_{23},s_{13}) N\frac{s_{12} s_{23}}{2 s_{13}}\Big(-s_{12} s_{23} \left(s_{12}+s_{23}+s_{13}\right)\epsilon \\
&\ +s_{13} \left(s_{12}{}^2+\left(s_{23}+s_{13}\right) s_{12}+s_{23}{}^2+s_{13}{}^2+s_{23} s_{13}\right)\Big)\\
&+\text{Box}(s_{12},s_{13},s_{23})N\frac{s_{12} s_{13}}{2 s_{23}}\Big(-s_{12} s_{13} \left(s_{12}+s_{23}+s_{13}\right)\epsilon\\
&\ +s_{23} \left(s_{12}{}^2+\left(s_{23}+s_{13}\right) s_{12}+s_{23}{}^2+s_{13}{}^2+s_{23} s_{13}\right)\Big)\\
&+\text{Box}(s_{13},s_{23},s_{12})N\frac{s_{23} s_{13}}{2 s_{12}}\Big(-s_{23} s_{13} \left(s_{12}+s_{23}+s_{13}\right)\epsilon\\
&\ +s_{12} \left(s_{12}{}^2+\left(s_{23}+s_{13}\right) s_{12}+s_{23}{}^2+s_{13}{}^2+s_{23} s_{13}\right)\Big)\\
&+A_{ggg}^{\text{Bub}}(s_{12},s_{23},s_{13}) + A_{ggg}^{\text{Bub}}(s_{12},s_{23},s_{13})+A_{ggg}^{\text{Bub}}(s_{12},s_{13},s_{23})
\end{split}\end{equation}
\begin{equation}\begin{split}
&A_{ggg}^{\text{Bub}}(s_{12},s_{23},s_{13}) = \\
&\text{Bub}(s_{12})\Big\{\frac{s_{23} s_{13} \epsilon }{(\epsilon -1) (2 \epsilon -3)}N_f+\frac{N}{s_{23} s_{13} \left(s_{23}+s_{13}\right){}^2 (\epsilon -1) \epsilon  (2 \epsilon -3)}\\
\times &\Big[-3 s_{23} s_{13} \left(s_{23}+s_{13}\right){}^2 \left(s_{12}{}^2+\left(s_{23}+s_{13}\right) s_{12}+s_{23}{}^2+s_{13}{}^2+s_{23} s_{13}\right)\\
&+\left(s_{23}+s_{13}\right){}^2 \left(\left(3 s_{23}{}^2+11 s_{13} s_{23}+3 s_{13}{}^2\right) s_{12}{}^2+11 s_{23} s_{13} \left(s_{23}{}^2+s_{13}
   s_{23}+s_{13}{}^2\right)\right) \epsilon \\
&+s_{12} \left(s_{23}+s_{13}\right){}^3 \left(3 s_{23}{}^2+11 s_{13} s_{23}+3 s_{13}{}^2\right) \epsilon\\
&-s_{23} s_{13} \left(s_{23}+s_{13}\right){}^2 \left(12 s_{23}{}^2+13 s_{13} s_{23}+12 s_{13}{}^2\right) \epsilon ^2 \\
&-s_{12}{}^2 \left(11 s_{23}{}^4+31 s_{13} s_{23}{}^3+34 s_{13}{}^2 s_{23}{}^2+31 s_{13}{}^3 s_{23}+11 s_{13}{}^4\right) \epsilon ^2 \\
&-s_{12} \left(s_{23}+s_{13}\right) \left(11 s_{23}{}^4+31 s_{13} s_{23}{}^3+34 s_{13}{}^2 s_{23}{}^2+31 s_{13}{}^3 s_{23}+11 s_{13}{}^4\right)\epsilon ^2\\
&+s_{23} s_{13} \left(s_{23}+s_{13}\right){}^2 \left(4 s_{23}{}^2+5 s_{13} s_{23}+4 s_{13}{}^2\right) \epsilon ^3\\
&+4 s_{12}{}^2 \left(3 s_{23}{}^4+5 s_{13} s_{23}{}^3+3 s_{13}{}^2 s_{23}{}^2+5 s_{13}{}^3 s_{23}+3 s_{13}{}^4\right) \epsilon ^3 \\
&+4 s_{12} \left(s_{23}+s_{13}\right) \left(3 s_{23}{}^4+5 s_{13} s_{23}{}^3+3 s_{13}{}^2 s_{23}{}^2+5 s_{13}{}^3 s_{23}+3 s_{13}{}^4\right) \epsilon^3\\
&-4 s_{12} \left(s_{23}+s_{13}\right){}^2 \left(s_{12}+s_{23}+s_{13}\right) \left(s_{23}{}^2-s_{13} s_{23}+s_{13}{}^2\right) \epsilon ^4
\Big]\Big\}
\end{split}\end{equation}
\begin{equation}\begin{split}
&B_{ggg}(s_{12},s_{23},s_{13}) = \\
&\text{Box}(s_{12},s_{23},s_{13})\frac{s_{12} s_{23} N}{2 s_{13} (2 \epsilon -1)}\Big(- s_{13}\left(s_{12}+s_{23}\right) \left(s_{12}+s_{13}\right) \\
&\ +\left(2 \left(s_{12}+s_{23}\right) s_{13}{}^2+s_{12} \left(2 s_{12}+s_{23}\right) s_{13}-s_{12} s_{23} \left(s_{12}+s_{23}\right)\right) \epsilon\Big) \\
&+\text{Box}(s_{12},s_{13},s_{23})\frac{s_{12} s_{13}N}{2 s_{23} (2 \epsilon -1)}\Big(-s_{23} \left(s_{12}+s_{23}\right) \left(s_{12}+s_{13}\right) \\
&\ +\left(-s_{12} s_{13}{}^2-\left(s_{12}-2 s_{23}\right) \left(s_{12}+s_{23}\right) s_{13}+2 s_{12} s_{23} \left(s_{12}+s_{23}\right)\right) \epsilon\Big) \\
&+\text{Box}(s_{13},s_{23},s_{12})\frac{s_{23} s_{13}N}{2 s_{12} (2 \epsilon -1)}\Big(-s_{12} \left(s_{12}+s_{23}\right) \left(s_{12}+s_{13}\right) \\
&\ +\left(2 \left(s_{12}+s_{23}\right) s_{12}{}^2+s_{23} s_{13}{}^2+\left(s_{12}+s_{23}\right) \left(2 s_{12}+s_{23}\right) s_{13}\right) \epsilon\Big) \\
&+B_{ggg}^{\text{Bub},1}(s_{12},s_{23},s_{13})+B_{ggg}^{\text{Bub},2}(s_{12},s_{23},s_{13})+B_{ggg}^{\text{Bub},2}(s_{12},s_{13},s_{23})
\end{split}\end{equation}
\begin{equation}\begin{split}
&B^{\text{Bub},1}_{ggg}(s_{12},s_{23},s_{13}) =\\
&\text{Bub}(s_{12})\Big\{\frac{s_{23} s_{13} \epsilon }{(\epsilon -1) (2 \epsilon -3)}N_f+\frac{N}{s_{23} s_{13} \left(s_{23}+s_{13}\right){}^2 (\epsilon -1) \epsilon  (2 \epsilon -3)}\\
\times&\Big[-3 s_{23} \left(s_{12}+s_{23}\right) s_{13} \left(s_{12}+s_{13}\right) \left(s_{23}+s_{13}\right){}^2-3 s_{12} s_{23}{}^2 \left(s_{12}+s_{23}\right) \left(s_{23}+s_{13}\right){}^2 \epsilon \\
&-\left(s_{23}+s_{13}\right){}^2 \left(3 s_{12} s_{13}{}^3+\left(3 s_{12}-11 s_{23}\right) \left(s_{12}+s_{23}\right) s_{13}{}^2-s_{12} s_{23}\left(11 s_{12}+8 s_{23}\right) s_{13}\right) \epsilon\\
&+s_{12}{}^2 \left(5 s_{23}{}^4-5 s_{13} s_{23}{}^3-26 s_{13}{}^2 s_{23}{}^2-5 s_{13}{}^3 s_{23}+5 s_{13}{}^4\right) \epsilon ^2-13 s_{23}{}^2 s_{13}{}^2 \left(s_{23}+s_{13}\right){}^2 \epsilon ^2\\
&+s_{12} \left(5 s_{23}{}^5-31 s_{13}{}^2 s_{23}{}^3-31 s_{13}{}^3 s_{23}{}^2+5 s_{13}{}^5\right) \epsilon ^2 +5 s_{23}{}^2 s_{13}{}^2 \left(s_{23}+s_{13}\right){}^2 \epsilon ^3 \\
&-2 s_{12}{}^2 \left(s_{23}{}^4-s_{13} s_{23}{}^3-6 s_{13}{}^2 s_{23}{}^2-s_{13}{}^3 s_{23}+s_{13}{}^4\right) \epsilon ^3\\
&-2 s_{12} \left(s_{23}{}^5-7 s_{13}{}^2 s_{23}{}^3-7 s_{13}{}^3 s_{23}{}^2+s_{13}{}^5\right) \epsilon ^3
\Big]\Big\}
\end{split}\end{equation}
\begin{equation}\begin{split}
&B^{\text{Bub},2}_{ggg}(s_{12},s_{23},s_{13}) =\\
&\text{Bub}(s_{23})\frac{N}{s_{12} s_{13} \left(s_{12}+s_{13}\right) (\epsilon -1) \epsilon}\Big[s_{12} \left(s_{12}+s_{23}\right) s_{13} \left(s_{12}+s_{13}\right){}^2 \\
&-\left(s_{12}+s_{13}\right){}^2 \left(s_{23} s_{13}{}^2+\left(3 s_{12}{}^2+3 s_{23} s_{12}+s_{23}{}^2\right) s_{13}-s_{12} s_{23}\left(s_{12}+s_{23}\right)\right) \epsilon\\
&+s_{12}{}^3 \left(2 s_{12}+s_{23}\right) s_{13} \epsilon ^2-s_{12}{}^3 s_{23} \left(s_{12}+s_{23}\right) \epsilon ^2 \\
&+s_{23} s_{13}{}^4 \epsilon ^2+\left(s_{12}+s_{23}\right) \left(2 s_{12}+s_{23}\right) s_{13}{}^3 \epsilon ^2+4 s_{12}{}^2 \left(s_{12}+s_{23}\right)
   s_{13}{}^2 \epsilon ^2\Big]
\end{split}\end{equation}
\begin{equation}\begin{split}
&A_{q\bar{q}g}(s_{12},s_{23},s_{13})= \\
&-\text{Box}(s_{12},s_{23},s_{13})\frac{s_{23}N}{2 s_{13} (2 \epsilon -1)}\Big(s_{12} \left(s_{12}+s_{23}\right) (\epsilon -1) \epsilon+s_{13}{}^2 (2 \epsilon -1)+s_{12} s_{13} (\epsilon -1) \epsilon\Big)\\
&+\text{Box}(s_{12},s_{13},s_{23})\frac{s_{13}N}{2 s_{23} (2 \epsilon -1)}\Big(s_{12} \left(s_{12}+s_{23}+s_{13}\right) \epsilon ^2-2 s_{23} s_{13} \epsilon +s_{23} s_{13}\Big)\\
&+\text{Box}(s_{13},s_{23},s_{12})\frac{s_{23} s_{13}}{2 N s_{12} (2 \epsilon -1)}\Big(\left(s_{12}+s_{23}+s_{13}\right) \epsilon ^2+2 s_{13} \epsilon -s_{13}\Big) \\
&+\text{Bub}(s_{12})\Big\{-\frac{2 s_{13} (\epsilon -1)}{s_{12} (2 \epsilon -3)}N_f+\frac{N}{2 s_{12} s_{23} s_{13} \left(s_{23}+s_{13}\right){}^2 (\epsilon -1) \epsilon  (2 \epsilon -3)}\Big[\\
&\ +12 s_{23} s_{13}{}^2 \left(s_{23}+s_{13}\right){}^2+s_{23} \left(s_{23}+s_{13}\right){}^2 \left(-31 s_{13}{}^2+6 s_{12} s_{13}+6 s_{12} \left(s_{12}+s_{23}\right)\right) \epsilon \\
&\ +27 s_{23} s_{13}{}^2 \left(s_{23}+s_{13}\right){}^2 \epsilon ^2-2 s_{12} \left(s_{23}+s_{13}\right) \left(8 s_{23}{}^3+13 s_{13} s_{23}{}^2-4s_{13}{}^2 s_{23}-3 s_{13}{}^3\right) \epsilon ^2 \\
&\ +s_{12}{}^2 \left(-16 s_{23}{}^3-26 s_{13} s_{23}{}^2+8 s_{13}{}^2 s_{23}+6 s_{13}{}^3\right) \epsilon ^2-8 s_{23} s_{13}{}^2 \left(s_{23}+s_{13}\right){}^2 \epsilon ^3 \\
&\ +2 s_{12}{}^2 \left(7 s_{23}{}^3+9 s_{13} s_{23}{}^2-7 s_{13}{}^2 s_{23}-5 s_{13}{}^3\right) \epsilon ^3\\
&\ +2 s_{12} \left(s_{23}+s_{13}\right) \left(7s_{23}{}^3+9 s_{13} s_{23}{}^2-7 s_{13}{}^2 s_{23}-5 s_{13}{}^3\right) \epsilon ^3 \\
&\ -4 s_{12} \left(s_{23}-s_{13}\right) \left(s_{23}+s_{13}\right){}^2 \left(s_{12}+s_{23}+s_{13}\right)\Big]+\frac{s_{13} \left(2 \epsilon ^2-\epsilon +2\right)}{2 N s_{12} \epsilon }\Big\}\\
&+\text{Bub}(s_{23})\Big\{\frac{s_{13}}{N s_{12} \left(s_{12}+s_{13}\right) \epsilon }\Big(s_{23} \epsilon ^2+s_{12} \left(\epsilon ^2+2 \epsilon -1\right)+s_{13} \left(\epsilon ^2+2 \epsilon -1\right)\Big)\\
&+\frac{\left(s_{12}+s_{23}+s_{13}\right) \left(s_{12}+s_{13}-s_{12}\epsilon\right)}{s_{13} \left(s_{12}+s_{13}\right)}N\Big\}\\
&+\text{Bub}(s_{13})\Big\{\frac{\left(s_{12}+2 s_{23}\right) \epsilon ^2+2 s_{13} \left(\epsilon ^2+2 \epsilon -1\right)}{2 N s_{12} \epsilon }+\frac{\left(2 s_{12}+s_{23}+2 s_{13}\right) \epsilon }{2 s_{23}}N\Big\}
\end{split}\end{equation}

%% file: XSthreshold.tex
\subsection{$gg$ initial state}

\bea
\hat{\eta}_{gg}^{(2;1)}(\delta;\eps)&=&N^2 \Bigg[\frac{8 \left(\epsilon
   ^3+2 \epsilon ^2-3 \epsilon +1\right)^2}{\delta (1-2 \epsilon )^2
   (\epsilon -1) \epsilon ^5}+\frac{16 \left(\epsilon ^3+2 \epsilon ^2-3 \epsilon
   +1\right)^2}{(1-2 \epsilon )^2 \epsilon ^5}\nonumber\\
   &&\left.+\frac{4 \delta \left(\epsilon ^3+2 \epsilon
   ^2-3 \epsilon +1\right) }{(\epsilon -1)^2 \epsilon ^5 (\epsilon +1) (2 \epsilon   -3) (2 \epsilon -1)^3}\right.\nonumber\\
   &&\times\left(16 \epsilon ^9-26 \epsilon ^8-93 \epsilon
   ^7+327 \epsilon ^6-353 \epsilon ^5+338 \epsilon ^3-315 \epsilon ^2+120
   \epsilon -18\right)\nonumber\\
   &&\left.+\frac{4 \delta^2 \left(\epsilon ^3+2 \epsilon ^2-3 \epsilon
   +1\right) }{3 (\epsilon -1)^2 \epsilon ^5
   (\epsilon +1) (2 \epsilon -3) (2 \epsilon -1)^3}\right.\nonumber\\
   &&\times\left. \left(32 \epsilon ^{10}-58 \epsilon ^9-117 \epsilon ^8+639
   \epsilon ^7-1250 \epsilon ^6+857 \epsilon ^5\right.\right.\nonumber\\
   &&\left.\left.+646 \epsilon ^4-1414 \epsilon
   ^3+929 \epsilon ^2-288 \epsilon +36\right)\right.\nonumber\\
   &&\left.+\frac{\delta^3}{6 \epsilon ^5 (\epsilon +2)
   \left(\epsilon ^2-1\right)^2 \left(4 \epsilon ^2-8 \epsilon
   +3\right)^3}\right.\nonumber\\
   &&\times\left(512 \epsilon ^{18}+240 \epsilon ^{17}-6156
   \epsilon ^{16}+8504 \epsilon ^{15}+8669 \epsilon ^{14}-93660 \epsilon
   ^{13}+207893 \epsilon ^{12}\right.\nonumber\\
   &&\left.+118784 \epsilon ^{11}-907053 \epsilon
   ^{10}+751815 \epsilon ^9+694791 \epsilon ^8-1418680 \epsilon ^7+614080
   \epsilon ^6\right.\nonumber\\
   &&\left.+312651 \epsilon ^5-506068 \epsilon ^4+301338 \epsilon ^3-109476
   \epsilon ^2+24408 \epsilon -2592\right)\Bigg]\nonumber\\
   &&+N N_f\Bigg[-\frac{4 \delta \left(\epsilon ^2-1\right) \left(\epsilon ^3+2
   \epsilon ^2-3 \epsilon +1\right)}{(\epsilon -1)^2 \epsilon ^2 (\epsilon +1)
   (2 \epsilon -3) (2 \epsilon -1)^3}\nonumber\\
   &&-\frac{4 \delta^2 \left(\epsilon ^3+2
   \epsilon ^2-3 \epsilon +1\right)}{\epsilon ^2 (2 \epsilon -3) (2 \epsilon
   -1)^3}\nonumber\\
   &&-\frac{\delta^3 }{\epsilon ^2 (\epsilon +2) \left(\epsilon ^2-1\right)^2 \left(4
   \epsilon ^2-8 \epsilon +3\right)^3}\nonumber\\
   &&\times\left(8 \epsilon ^{11}+6 \epsilon ^{10}-91
   \epsilon ^9+\epsilon ^8+310 \epsilon ^7-161 \epsilon ^6-273 \epsilon ^5\right.\nonumber\\
   &&\left.+287
   \epsilon ^4-98 \epsilon ^3-109 \epsilon ^2+132 \epsilon
   -36\right)\Bigg]\nonumber\\
   &&+\frac{N_f^2 \delta^3}{2
   \left(4 \epsilon ^2-8 \epsilon +3\right)^3}\nonumber\\
      &&+\mathcal{O}(\delta^4)
\eea
\bea
\hat{\eta}_{gg}^{(3;1)}(\delta;\eps)=&&N^2\Bigg[\frac{6 \left(\epsilon ^3+2 \epsilon ^2-3 \epsilon +1\right) \left(5
   \epsilon ^3-16 \epsilon ^2+15 \epsilon -4\right)}{(1-2 \epsilon )^2
   (\epsilon -1)^2 \epsilon ^4 (\epsilon +1) (2 \epsilon -3)}\nonumber\\
      &&+\frac{2 \left(\epsilon ^3+2
   \epsilon ^2-3 \epsilon +1\right) \delta}{(1-2 \epsilon )^2 (\epsilon -2) (\epsilon -1)^2   \epsilon ^4 (\epsilon +1) (\epsilon +2) (2 \epsilon -3) (3 \epsilon   -1)}\nonumber\\
   	&&\left(39 \epsilon ^7-207 \epsilon ^6+211   \epsilon ^5+651 \epsilon ^4-1778 \epsilon ^3+1644 \epsilon ^2-632 \epsilon   +72\right) \nonumber\\
      &&+\frac{ \delta^2}{(1-2 \epsilon )^2 (3-2 \epsilon )^2   (\epsilon -3) (\epsilon -1)^2 \epsilon ^4 (\epsilon +1)^2 (\epsilon +2)   (\epsilon +3) (3 \epsilon -2) (3 \epsilon -1)}\nonumber\\
      	&&\times\left(198 \epsilon ^{15}-503 \epsilon ^{14}-2103 \epsilon
   ^{13}+13014 \epsilon ^{12}-17395 \epsilon ^{11}-89713 \epsilon ^{10}\right.\nonumber\\
   &&\left.+284685
   \epsilon ^9+5833 \epsilon ^8-910913 \epsilon ^7+1149456 \epsilon ^6-144594
   \epsilon ^5-797563 \epsilon ^4\right.\nonumber\\
   &&\left.+776466 \epsilon ^3-332844 \epsilon ^2+70488
   \epsilon -6048\right)\nonumber\\
   &&+\frac{  \delta^3}{9 (1-2 \epsilon )^2 (\epsilon -4) (\epsilon -3) (\epsilon -2)   (\epsilon -1)^2 \epsilon ^4 (\epsilon +1)^2 (\epsilon +2) \epsilon +3)   (\epsilon +4)}\nonumber\\
	&&\times\frac{1}{ (2 \epsilon -3) (3 \epsilon -2) (3 \epsilon   -1)}\left(243 \epsilon ^{18}-135 \epsilon ^{17}-2870 \epsilon
   ^{16}+1307 \epsilon ^{15}\right.\nonumber\\
&&\left.   -111723 \epsilon ^{14}+139225 \epsilon
   ^{13}
  +2358695 \epsilon ^{12}-3617247 \epsilon ^{11}-15121876 \epsilon
   ^{10}\right.\nonumber\\
&&\left.   +32597314 \epsilon ^9+23389539 \epsilon ^8-107267656 \epsilon
   ^7+79086832 \epsilon ^6+29647704 \epsilon ^5\right.\nonumber\\
&&\left.   -80204568 \epsilon ^4+54419616
   \epsilon ^3-18635616 \epsilon ^2+3272832 \epsilon -228096\right)\Bigg]\nonumber\\
      &&   +N_fN \Bigg[-\frac{2 \left(\epsilon ^3+2
   \epsilon ^2-3 \epsilon +1\right)}{(1-2 \epsilon )^2 (\epsilon -1)^2
   \epsilon ^3 (2 \epsilon -3)}\nonumber\\
   &&+\frac{2   \left(\epsilon ^3+2 \epsilon ^2-3 \epsilon +1\right) \left(-9 \epsilon
   ^5+43 \epsilon ^3-2 \epsilon ^2-28 \epsilon +8\right) \delta}{(1-2
   \epsilon )^2 (\epsilon -2) (\epsilon -1)^2 \epsilon ^3 (\epsilon +1)
   (\epsilon +2) (2 \epsilon -3) (3 \epsilon -1)}\nonumber\\
      &&+\frac{ \delta^2}{(1-2 \epsilon   )^2 (3-2 \epsilon )^2 (\epsilon -3) (\epsilon -1)^2 \epsilon ^3 (\epsilon   +1)^2 (\epsilon +2) (\epsilon +3) (3 \epsilon -2) (3 \epsilon -1)}\nonumber\\
      	&&\times\left(-90 \epsilon ^{13}-193 \epsilon ^{12}+1488 \epsilon
   ^{11}+2484 \epsilon ^{10}-8117 \epsilon ^9-6944 \epsilon ^8+20203 \epsilon
   ^7\right.\nonumber\\
   &&\left.+1086 \epsilon ^6-19033 \epsilon ^5+8697 \epsilon ^4+3593 \epsilon
   ^3-5106 \epsilon ^2+1764 \epsilon -216\right)\nonumber\\
   &&-\frac{ \delta^3}{9 (1-2 \epsilon )^2 (\epsilon -4) (\epsilon -3)   (\epsilon -2) (\epsilon -1)^2 \epsilon ^2 (\epsilon +2) (\epsilon +3)   (\epsilon +4) (2 \epsilon -3)}\nonumber\\
      	&&\times\frac{1}{ (3 \epsilon -2) (3 \epsilon   -1)}\left(189 \epsilon ^{13}-30 \epsilon ^{12}-6607
   \epsilon ^{11}+2723 \epsilon ^{10}+77751 \epsilon ^9\right.\nonumber\\
   &&\left.-61152 \epsilon
   ^8-353061 \epsilon ^7+470019 \epsilon ^6+375376 \epsilon ^5-1036856
   \epsilon ^4+721968 \epsilon ^3\right.\nonumber\\
   &&\left.-253392 \epsilon ^2+48384 \epsilon
   -6912\right)\Bigg] \nonumber\\
      &&+\frac{2 N_f^2 \delta^2
   \left(\epsilon ^4+\epsilon ^3-11 \epsilon ^2-9 \epsilon +18\right)}{(1-2
   \epsilon )^2 (3-2 \epsilon )^2 (\epsilon -3) (\epsilon -1)^2 (\epsilon +2)
   (\epsilon +3) (3 \epsilon -2) (3 \epsilon -1)}\nonumber\\
      &&+\mathcal{O}(\delta^4)
\eea
\bea
\hat{\eta}_{gg}^{(4;1)}(\delta;\eps)=&&N^2 \Bigg[\frac{\delta \left(-\epsilon ^5+111
   \epsilon ^4-495 \epsilon ^3+725 \epsilon ^2-396 \epsilon +72\right)}{(1-2
   \epsilon )^2 (3-2 \epsilon )^2 (\epsilon -1) \epsilon ^2 (\epsilon +1)^2 (4
   \epsilon -1)}\nonumber\\
      &&+\frac{\delta^2 \left(-14 \epsilon ^6+26 \epsilon ^5+78 \epsilon
   ^4+286 \epsilon ^3-940 \epsilon ^2+780 \epsilon -192\right)}{(1-2 \epsilon
   )^2 (\epsilon -2) (\epsilon -1) \epsilon ^2 (\epsilon +1)^2 (\epsilon +2)
   (2 \epsilon -3) (4 \epsilon -1)}\nonumber\\
      &&+\frac{2 \delta^3 }{(1-2 \epsilon )^2 (\epsilon -3)   (\epsilon -2)^2 (\epsilon -1)^2 \epsilon ^2 (\epsilon +1)^2 (\epsilon +2)^2   (\epsilon +3) (2 \epsilon -3)}\nonumber\\
      &&\times\frac{1}{ (4 \epsilon -3) (4 \epsilon   -1)}\left(44 \epsilon ^{12}+181 \epsilon ^{11}-2053
   \epsilon ^{10}-157 \epsilon ^9+20965 \epsilon ^8-13979 \epsilon ^7\right.\nonumber\\
   &&\left.-90403
   \epsilon ^6+118041 \epsilon ^5+110903 \epsilon ^4-318934 \epsilon ^3+258264
   \epsilon ^2-95616 \epsilon +13896\right)\Bigg]\nonumber\\
      &&+N N_f \Bigg[-\frac{2 \delta \left(\epsilon ^3+8 \epsilon ^2-17 \epsilon
   +6\right)}{(1-2 \epsilon )^2 (3-2 \epsilon )^2 (\epsilon -1)^2 \epsilon 
   (\epsilon +1) (4 \epsilon -1)}\nonumber\\
      &&+\frac{\delta^2 \left(-14 \epsilon ^4+24 \epsilon ^3+70 \epsilon
   ^2-84 \epsilon +28\right)}{(1-2 \epsilon )^2 (\epsilon -2) (\epsilon -1)^2
   \epsilon  (\epsilon +1) (\epsilon +2) (2 \epsilon -3) (4 \epsilon
   -1)}\nonumber\\
      &&-\frac{2 \delta^3 \left(16
   \epsilon ^7-29 \epsilon ^6-154 \epsilon ^5+104 \epsilon ^4-240 \epsilon
   ^3+897 \epsilon ^2-750 \epsilon +204\right)}{(1-2 \epsilon )^2 (\epsilon
   -3) (\epsilon -2) (\epsilon -1)^2 \epsilon  (\epsilon +1) (\epsilon +2)
   (\epsilon +3) (2 \epsilon -3) (4 \epsilon -3) (4 \epsilon
   -1)}\Bigg]\nonumber\\
      &&-\frac{N_f^2 \delta}{(1-2
   \epsilon )^2 (3-2 \epsilon )^2 (\epsilon -1)^2 (4 \epsilon -1)}\nonumber\\
      &&+\mathcal{O}(\delta^4)
\eea
\bea
\hat{\eta}_{gg}^{(4;2)}(\delta;\eps)=&&-N^2 \Bigg[-\frac{4 \left(\epsilon ^3+2
   \epsilon ^2-3 \epsilon +1\right)}{\delta \epsilon ^5 (2 \epsilon
   -1)}\nonumber\\
	&&-\frac{4
   \left(\epsilon ^4-7 \epsilon ^2+7 \epsilon -2\right)}{\epsilon ^5 (2
   \epsilon -1)}\nonumber\\
	&&-\frac{2 \delta \left(8 \epsilon ^9-24 \epsilon ^8-8 \epsilon ^7+266
   \epsilon ^6-529 \epsilon ^5+157 \epsilon ^4+406 \epsilon ^3-417 \epsilon
   ^2+147 \epsilon -18\right)}{\epsilon ^5 \left(16 \epsilon ^5-36 \epsilon
   ^4+4 \epsilon ^3+33 \epsilon ^2-20 \epsilon +3\right)}\nonumber\\
	&&-\frac{2 \delta^2 }{3 \epsilon ^5 \left(16 \epsilon ^5-36 \epsilon ^4+4 \epsilon   ^3+33 \epsilon ^2-20 \epsilon +3\right)}\left(8 \epsilon ^{10}-26
   \epsilon ^9+79 \epsilon ^8+185 \epsilon ^7\right.\nonumber\\
   &&\left.-1199 \epsilon ^6+1426 \epsilon
   ^5+220 \epsilon ^4-1435 \epsilon ^3+1048 \epsilon ^2-318 \epsilon
   +36\right)\nonumber\\
      &&-\frac{\delta^3 }{6 (\epsilon -1)   \epsilon ^5 (\epsilon +1) (\epsilon +2) (2 \epsilon -3) (2 \epsilon -1) (4   \epsilon -3) (4 \epsilon -1)}\nonumber\\
	&&\times\left(32 \epsilon ^{13}-40 \epsilon ^{12}+580
   \epsilon ^{11}+1074 \epsilon ^{10}-8773 \epsilon ^9+2443 \epsilon ^8+25643
   \epsilon ^7\right.\nonumber\\
      &&\left.-29791 \epsilon ^6-540 \epsilon ^5+23686 \epsilon ^4-21946
   \epsilon ^3+11196 \epsilon ^2-3276 \epsilon +432\right)\Bigg]\nonumber
   \eea
   \bea
  \phantom{\hat{\eta}_{gg}^{(4;2)}(\delta;\eps)=}
      &&-N N_f \Bigg[\frac{2 \delta}{\epsilon ^2 \left(16 \epsilon ^3-36 \epsilon
   ^2+20 \epsilon -3\right)}-\frac{2
   \delta^2}{\epsilon ^2 \left(16 \epsilon ^3-36 \epsilon ^2+20 \epsilon
   -3\right)}\nonumber\\
	&&+\frac{\delta^3 \left(4 \epsilon
   ^2-6 \epsilon +3\right)}{\epsilon ^2 \left(64 \epsilon ^5-256 \epsilon
   ^4+380 \epsilon ^3-260 \epsilon ^2+81 \epsilon -9\right)}\Bigg]\nonumber\\
      &&+\mathcal{O}(\delta^4)
\eea
\bea
\hat{\eta}_{gg}^{(4;3)}(\delta;\eps)=&&N^2 \Bigg[\frac{2 \delta
   \left(\epsilon ^6+54 \epsilon ^4-248 \epsilon ^3+363 \epsilon ^2-198
   \epsilon +36\right)}{3 (1-2 \epsilon )^2 (3-2 \epsilon )^2 \epsilon ^3
   (\epsilon +1)^2 (4 \epsilon -1)}\nonumber\\
	&&-\frac{4 \delta^2 \left(3 \epsilon ^7+20 \epsilon
   ^6+45 \epsilon ^5-11 \epsilon ^4-184 \epsilon ^3+270 \epsilon ^2-113
   \epsilon +12\right)}{3 \epsilon ^3 (\epsilon +1)^2 (\epsilon +2) (2
   \epsilon -3) (2 \epsilon -1)^3 (4 \epsilon -1)}\nonumber\\
	&&+\frac{4 \delta^3 }{3 (1-2 \epsilon )^2 (\epsilon -2) (\epsilon -1) \epsilon ^3   (\epsilon +1)^2 (\epsilon +2)^2 (\epsilon +3) (2 \epsilon -3) (4 \epsilon   -3) (4 \epsilon -1)}\nonumber\\
		&&\times\left(12 \epsilon ^{11}+133 \epsilon ^{10}+464
   \epsilon ^9-387 \epsilon ^8-3954 \epsilon ^7-756 \epsilon ^6+11661 \epsilon
   ^5\right.\nonumber\\
      &&\left.-2062 \epsilon ^4-14519 \epsilon ^3+15852 \epsilon ^2-5976 \epsilon
   +828\right)\Bigg]\nonumber\\
      &&+N N_f \Bigg[\frac{4 \delta
   \left(\epsilon ^4-\epsilon ^3-5 \epsilon ^2+9 \epsilon -3\right)}{3 (1-2
   \epsilon )^2 (3-2 \epsilon )^2 (\epsilon -1) \epsilon ^2 (\epsilon +1) (4
   \epsilon -1)}\nonumber\\
	&&-\frac{4 \delta^2 \left(3
   \epsilon ^5+17 \epsilon ^4+8 \epsilon ^3-23 \epsilon ^2+16 \epsilon
   -3\right)}{3 (\epsilon -1) \epsilon ^2 (\epsilon +1) (\epsilon +2) (2
   \epsilon -3) (2 \epsilon -1)^3 (4 \epsilon -1)}\nonumber\\
	&&+\frac{4
   \delta^3 \left(6 \epsilon ^7+47 \epsilon ^6+134 \epsilon ^5+82 \epsilon
   ^4-152 \epsilon ^3+111 \epsilon ^2-30 \epsilon +6\right)}{3 (\epsilon -1)
   \epsilon ^2 (\epsilon +1) (\epsilon +2) (\epsilon +3) (2 \epsilon -3) (2
   \epsilon -1)^3 (4 \epsilon -3) (4 \epsilon -1)}\Bigg]\nonumber\\
      &&+\frac{2 N_f^2 \delta}{3 \epsilon  (4 \epsilon
   -1) \left(4 \epsilon ^2-8 \epsilon +3\right)^2}\nonumber\\
      &&+\mathcal{O}(\delta^4)
\eea
\bea
\hat{\eta}_{gg}^{(5;1)}(\delta;\eps)=&&N^2 \Bigg[\frac{2 \left(5 \epsilon
   ^2-11 \epsilon +4\right)}{\epsilon ^4 \left(4 \epsilon ^3-4 \epsilon ^2-5
   \epsilon +3\right)}\nonumber\\
	&&-\frac{4 \delta \left(5 \epsilon ^6+41
   \epsilon ^5-118 \epsilon ^4-214 \epsilon ^3+632 \epsilon ^2-352 \epsilon
   +36\right)}{3 (\epsilon -2) \epsilon ^4 (\epsilon +1) (\epsilon +2) (2
   \epsilon -3) (2 \epsilon -1) (5 \epsilon -1)}\nonumber\\
	&&+\frac{4 \delta^2 }{3 (\epsilon -3) (\epsilon -2) (\epsilon -1) \epsilon ^4   (\epsilon +1) (\epsilon +2) (\epsilon +3) (2 \epsilon -3) (2 \epsilon -1)   (5 \epsilon -2) (5 \epsilon -1)}\nonumber\\
   	&&\times\left(75 \epsilon ^{10}+437 \epsilon
   ^9-2802 \epsilon ^8-3723 \epsilon ^7+25611 \epsilon ^6-5853 \epsilon
   ^5-72228 \epsilon ^4\right.\nonumber\\
   	&&\left.+99655 \epsilon ^3-53904 \epsilon ^2+12588 \epsilon
   -1008\right)\nonumber\\
	&&-\frac{8 \delta^3 }{3 (\epsilon -4) (\epsilon -3)   (\epsilon -2) (\epsilon -1) \epsilon ^4 (\epsilon +1) (\epsilon +2)   (\epsilon +3) (\epsilon +4) (2 \epsilon -1) (5 \epsilon -3) }\nonumber\\
	&&\times\frac{1}{(5 \epsilon -2)   (5 \epsilon -1)}\left(60 \epsilon ^{12}+127 \epsilon ^{11}-1881
   \epsilon ^{10}-5056 \epsilon ^9+22950 \epsilon ^8+58677 \epsilon ^7\right.\nonumber\\
   	&&\left.-154497
   \epsilon ^6-244594 \epsilon ^5+748092 \epsilon ^4-646274 \epsilon ^3+259068
   \epsilon ^2-48480 \epsilon +3168\right)\Bigg]\nonumber
   \eea
   \bea
   \phantom{\hat{\eta}_{gg}^{(5;1)}(\delta;\eps)=}
      &&+N N_f \Bigg[-\frac{2}{3 \epsilon ^3 \left(4 \epsilon ^3-12 \epsilon ^2+11
   \epsilon -3\right)} 
  -\frac{4 \delta}{3 \epsilon ^3 \left(4 \epsilon ^2-8 \epsilon
   +3\right)}\nonumber\\
   &&
    +\frac{4 \delta^2 (3
   \epsilon -1)}{3 \epsilon ^3 \left(20 \epsilon ^3-48 \epsilon ^2+31 \epsilon
   -6\right)}\Bigg]\nonumber\\
   &&+\mathcal{O}(\delta^4)
\eea
\bea
\hat{\eta}_{gg}^{(6;1)}(\delta;\eps)=&&N^2 \Bigg[+\frac{2 (\epsilon -1)}{3
   \delta \epsilon ^5}-\frac{4 (\epsilon -1)}{3 \epsilon
   ^5}+\frac{2 \delta \left(8 \epsilon ^2-7 \epsilon
   +1\right)}{\epsilon ^5 (6 \epsilon -1)}\nonumber\\
      &&-\frac{4 \delta^2 \left(9 \epsilon ^2-7 \epsilon
   +1\right)}{3 \epsilon ^5 (6 \epsilon -1)}-\frac{\delta^3 (3 \epsilon -1) \left(\epsilon ^4-20 \epsilon ^3+35
   \epsilon ^2-21 \epsilon +4\right)}{6 (1-2 \epsilon )^2 \epsilon ^5 (6
   \epsilon -1)}\nonumber\\
      &&+\frac{\delta^4 (\epsilon -2) (3
   \epsilon -1)}{3 (1-2 \epsilon )^2 \epsilon ^2 (6 \epsilon
   -1)}-\frac{\delta^5 (\epsilon -2) (3 \epsilon -1)}{6 (1-2 \epsilon
   )^2 \epsilon ^2 (6 \epsilon -1)}\Bigg]\nonumber\\
\eea

\subsection{$gq$ initial state}

\bea
\hat{\eta}_{qg}^{(2;1)}(\delta;\eps)=&&N^2   \Bigg[\frac{\left(\epsilon
   ^3+2 \epsilon ^2-3 \epsilon +1\right)^2}{(1-2 \epsilon )^2
   (\epsilon -1) \epsilon ^5}+\frac{\delta
   \left(\epsilon ^3+2 \epsilon ^2-3 \epsilon +1\right) }{2
   (\epsilon -1) \epsilon ^4 (2 \epsilon -1)^3}\nonumber\\
   &&\times\left(6
   \epsilon ^4+3 \epsilon ^3-37 \epsilon ^2+28 \epsilon -8\right)\nonumber\\
   &&+\frac{\delta^2 }{32 (\epsilon -1)^2   \epsilon ^5 (\epsilon +1) (2 \epsilon -1)^3}\left(84 \epsilon ^{11}+196
   \epsilon ^{10}-847 \epsilon ^9-906 \epsilon ^8\right.\nonumber\\
   &&\left.+3420 \epsilon
   ^7-850 \epsilon ^6-2033 \epsilon ^5+1712 \epsilon ^4-800 \epsilon
   ^3+424 \epsilon ^2-176 \epsilon +32\right)\nonumber\\
   &&+\frac{\delta^3 }{48   (\epsilon -1)^2 \epsilon ^4 (\epsilon +1) (2 \epsilon -3) (2   \epsilon -1)^3}\nonumber\\
   &&\times\left(156 \epsilon ^{12}+184 \epsilon
   ^{11}-2179 \epsilon ^{10}-284 \epsilon ^9+10512 \epsilon ^8-6458
   \epsilon ^7-11661 \epsilon ^6\right.\nonumber\\
   &&\left.+7782 \epsilon ^5+2440 \epsilon
   ^4+408 \epsilon ^3-3780 \epsilon ^2+2112 \epsilon -384\right)\Bigg]\nonumber\\
   &&-\frac{\delta \left(2 \epsilon ^2+5 \epsilon -2\right)
   \left(\epsilon ^3+2 \epsilon ^2-3 \epsilon +1\right)}{2 \epsilon
   ^3 (2 \epsilon -1)^3}\nonumber\\
   &&-\frac{\delta^2 \left(20 \epsilon ^7+56 \epsilon ^6-139
   \epsilon ^5-220 \epsilon ^4+417 \epsilon ^3-186 \epsilon ^2+16
   \epsilon +4\right)}{16 (\epsilon -1) \epsilon ^3 (2 \epsilon
   -1)^3}\nonumber\\
   &&-\frac{\delta^3 }{24 (\epsilon -1)
   \epsilon ^3 (\epsilon +1) (2 \epsilon -3) (2 \epsilon
   -1)^3}\left(44
   \epsilon ^{10}+106 \epsilon ^9-456 \epsilon ^8-651 \epsilon
   ^7\right.\nonumber\\
   &&\left.+1734 \epsilon ^6+347 \epsilon ^5-1419 \epsilon ^4+348 \epsilon
   ^3+5 \epsilon ^2+86 \epsilon -32\right)\nonumber\\
   &&+\frac{1}{N^2}\Bigg[\frac{\delta^2 (\epsilon -1) \left(2 \epsilon ^2+5
   \epsilon -2\right)^2}{32 \epsilon ^2 (2 \epsilon -1)^3}\nonumber\\
   &&+\frac{\delta^3 (\epsilon -1) \left(4 \epsilon ^6+12
   \epsilon ^5-27 \epsilon ^4-68 \epsilon ^3+42 \epsilon ^2+6
   \epsilon -4\right)}{16 \epsilon ^2 (2 \epsilon -3) (2 \epsilon
   -1)^3}\Bigg]\nonumber\\
&&+\mathcal{O}(\delta^4)
\eea
\bea
\hat{\eta}_{qg}^{(3;1)}(\delta;\eps)=&&N^2 \Bigg[\frac{3 \left(\epsilon ^4-7 \epsilon ^2+7 \epsilon
   -2\right)}{4 (1-2 \epsilon )^2 \epsilon ^5 (2 \epsilon
   -3)}+\frac{ \delta}{8 (1-2 \epsilon
   )^2 (\epsilon -1) \epsilon ^4 (\epsilon +1) (2 \epsilon -3) (3
   \epsilon -1)}\nonumber\\
   &&\times\left(-8 \epsilon ^8-48 \epsilon ^7+95
   \epsilon ^6+209 \epsilon ^5-629 \epsilon ^4+761 \epsilon ^3-602
   \epsilon ^2+274 \epsilon -60\right)\nonumber\\
   &&+\frac{   \delta^2}{8 (1-2 \epsilon )^2 (\epsilon -1) \epsilon ^5   (\epsilon +1) (\epsilon +2) (2 \epsilon -3) (3 \epsilon -2) (3   \epsilon -1)}\nonumber\\
   &&\times\left(-20 \epsilon ^{12}-110 \epsilon
   ^{11}+390 \epsilon ^{10}+1195 \epsilon ^9-3099 \epsilon ^8-1743
   \epsilon ^7+8329 \epsilon ^6\right.\nonumber\\
   &&\left.-6890 \epsilon ^5+1824 \epsilon
   ^4+312 \epsilon ^3-20 \epsilon ^2-144 \epsilon +48\right)\nonumber\\
   &&+\frac{\delta^3}{24 (1-2   \epsilon )^2 (\epsilon -1)^2 \epsilon ^4 (\epsilon +1)^2   (\epsilon +2) (\epsilon +3) (2 \epsilon -3) (3 \epsilon -2) (3   \epsilon -1)}\nonumber\\
   &&\times\left(-30
   \epsilon ^{15}-241 \epsilon ^{14}+218 \epsilon ^{13}+4307
   \epsilon ^{12}+1359 \epsilon ^{11}-24229 \epsilon ^{10}+162
   \epsilon ^9\right.\nonumber\\
   &&\left.+66820 \epsilon ^8-35825 \epsilon ^7-67308 \epsilon
   ^6+53626 \epsilon ^5+16711 \epsilon ^4-19534 \epsilon ^3\right.\nonumber\\
   &&\left.-4892
   \epsilon ^2+6744 \epsilon -1728\right)\Bigg] \nonumber\\
   &&+N N_f   \Bigg[-\frac{(\epsilon -1) \left(\epsilon ^3+2
   \epsilon ^2-3 \epsilon +1\right)}{(1-2 \epsilon )^2 \epsilon ^4
   (2 \epsilon -3)}\nonumber\\
   &&-\frac{(\epsilon -1) \left(4
   \epsilon ^4+\epsilon ^3-27 \epsilon ^2+20 \epsilon -6\right)
   \delta}{2 (1-2 \epsilon )^2 \epsilon ^3 \left(6 \epsilon ^2-11
   \epsilon +3\right)}\nonumber\\
   &&-\frac{(\epsilon -1) \left(6
   \epsilon ^8+7 \epsilon ^7-39 \epsilon ^6+5 \epsilon ^5+45
   \epsilon ^4+8 \epsilon ^2-12 \epsilon +4\right) \delta^2}{2
   (1-2 \epsilon )^2 \epsilon ^4 (\epsilon +1) (2 \epsilon -3) (3
   \epsilon -2) (3 \epsilon -1)}\nonumber\\
   &&+\frac{ \delta^3}{18 (1-2
   \epsilon )^2 \epsilon ^3 (\epsilon +1) (2 \epsilon -3) (3
   \epsilon -2) (3 \epsilon -1)}\left(-18 \epsilon ^9-24 \epsilon ^8+147 \epsilon
   ^7\right.\nonumber\\
   &&\left.+49 \epsilon ^6-407 \epsilon ^5+10 \epsilon ^4+442 \epsilon
   ^3+43 \epsilon ^2-146 \epsilon +48\right)\Bigg]\nonumber\\
   &&+\frac{\left(2 \epsilon ^2-\epsilon +2\right)
   \left(\epsilon ^3+2 \epsilon ^2-3 \epsilon +1\right)}{4 (1-2
   \epsilon )^2 \epsilon ^5}\nonumber\\
   &&+\frac{\left(8 \epsilon ^9-34
   \epsilon ^8-41 \epsilon ^7+263 \epsilon ^6-345 \epsilon ^5+288
   \epsilon ^4-281 \epsilon ^3+227 \epsilon ^2-99 \epsilon
   +18\right) \delta}{4 (1-2 \epsilon )^2 (\epsilon -1)^2
   \epsilon ^4 \left(6 \epsilon ^2-11 \epsilon
   +3\right)}\nonumber\\
   &&+\frac{ \delta^2}{8 (1-2 \epsilon )^2 (\epsilon   -2) (\epsilon -1)^2 \epsilon ^5 (\epsilon +1) (2 \epsilon -3) (3   \epsilon -2) (3 \epsilon -1)}\nonumber\\
   &&\times\left(24 \epsilon ^{14}-132 \epsilon ^{13}-35 \epsilon
   ^{12}+1306 \epsilon ^{11}-2058 \epsilon ^{10}-882 \epsilon
   ^9+2752 \epsilon ^8+2134 \epsilon ^7\right.\nonumber\\
   &&\left.-5866 \epsilon ^6+3030
   \epsilon ^5+599 \epsilon ^4-1512 \epsilon ^3+960 \epsilon ^2-320
   \epsilon +48\right)\nonumber\\
   &&+\frac{\delta^3}{72 (1-2 \epsilon )^2 (\epsilon -3)   (\epsilon -2) (\epsilon -1)^2 \epsilon ^4 (\epsilon +1) (\epsilon   +2) (2 \epsilon -3) (3 \epsilon -2) (3 \epsilon   -1)}\nonumber\\
   &&\times\left(72 \epsilon ^{16}-408 \epsilon ^{15}-735 \epsilon
   ^{14}+7064 \epsilon ^{13}-1282 \epsilon ^{12}-41535 \epsilon
   ^{11}+26560 \epsilon ^{10}\right.\nonumber\\
   &&\left.+120391 \epsilon ^9-101538 \epsilon
   ^8-196013 \epsilon ^7+291319 \epsilon ^6-107643 \epsilon ^5+2036
   \epsilon ^4\right.\nonumber\\
   &&\left.+14096 \epsilon ^3-25488 \epsilon ^2+15696 \epsilon
   -3456\right) \nonumber
   \eea
   \bea
   \phantom{\hat{\eta}_{qg}^{(3;1)}(\delta;\eps)}
   &&+\frac{N_f}{N} \Bigg[+\frac{(\epsilon -1)^2 \left(2 \epsilon ^2+5 \epsilon
   -2\right) \delta}{2 (1-2 \epsilon )^2 \epsilon ^2 \left(6
   \epsilon ^2-11 \epsilon +3\right)}+\frac{(\epsilon -1)^2 \left(3 \epsilon ^4+3 \epsilon ^3-15
   \epsilon ^2+4 \epsilon +1\right) \delta^2}{2 (1-2 \epsilon )^2
   \epsilon ^2 (2 \epsilon -3) (3 \epsilon -2) (3 \epsilon
   -1)}\nonumber\\
   &&+\frac{\left(9
   \epsilon ^7-59 \epsilon ^5+53 \epsilon ^4-13 \epsilon ^3+48
   \epsilon ^2-54 \epsilon +16\right) \delta^3}{18 (1-2 \epsilon
   )^2 \epsilon ^2 (2 \epsilon -3) (3 \epsilon -2) (3 \epsilon
   -1)}\Bigg]\nonumber\\
   &&+\frac{1}{N^2}\Bigg[-\frac{(\epsilon -1) \left(4 \epsilon ^4+8 \epsilon
   ^3-5 \epsilon ^2+12 \epsilon -4\right) \delta}{8 (1-2 \epsilon
   )^2 \epsilon ^3 (3 \epsilon -1)}\nonumber\\
   &&+\frac{\left(-6
   \epsilon ^7+5 \epsilon ^6+49 \epsilon ^5-39 \epsilon ^4+11
   \epsilon ^3-13 \epsilon ^2+\epsilon +2\right) \delta^2}{8 (1-2
   \epsilon )^2 \epsilon ^3 \left(9 \epsilon ^2-9 \epsilon
   +2\right)}\nonumber\\
   &&+\frac{\left(-18 \epsilon ^9+18
   \epsilon ^8+145 \epsilon ^7-321 \epsilon ^6-232 \epsilon ^5+394
   \epsilon ^4+74 \epsilon ^3-266 \epsilon ^2+148 \epsilon
   -32\right) \delta^3}{72 (1-2 \epsilon )^2 (\epsilon -1)
   \epsilon ^3 (3 \epsilon -2) (3 \epsilon -1)}\Bigg]\nonumber\\
      &&+\mathcal{O}(\delta^4)
\eea
\bea
\hat{\eta}_{qg}^{(4;1)}(\delta;\eps)=&&N^2 \Bigg[-\frac{3 (\epsilon -2) \left(\epsilon
   ^2+7 \epsilon -6\right) \delta}{8 (1-2 \epsilon )^2 \epsilon
   ^2 (\epsilon +1) (2 \epsilon -3) (4 \epsilon -1)}\nonumber\\
   &&+\frac{\left(-11 \epsilon ^7-116 \epsilon ^6+128
   \epsilon ^5+248 \epsilon ^4-495 \epsilon ^3+312 \epsilon ^2+42
   \epsilon -60\right) \delta^2}{16 (\epsilon -1) \epsilon ^3
   (\epsilon +2) (2 \epsilon -3) (4 \epsilon -1) \left(2 \epsilon
   ^2+\epsilon -1\right)^2}\nonumber\\
   &&+\frac{ \delta^3}{8 (\epsilon -2) (\epsilon -1) \epsilon   ^3 (\epsilon +2) (\epsilon +3) (2 \epsilon -3) (4 \epsilon -3) (4   \epsilon -1) \left(2 \epsilon ^2+\epsilon   -1\right)^2}\nonumber\\
   &&\times\left(28 \epsilon ^{10}+23 \epsilon
   ^9+303 \epsilon ^8-740 \epsilon ^7-3850 \epsilon ^6+7977 \epsilon
   ^5\right.\nonumber\\
   &&\left.+183 \epsilon ^4-6276 \epsilon ^3+2616 \epsilon ^2+264 \epsilon
   -192\right)\Bigg]\nonumber\\
   &&+N N_f \Bigg[\frac{\left(\epsilon ^3+6 \epsilon ^2-13 \epsilon +6\right)
   \delta}{2 (1-2 \epsilon )^2 \epsilon  (\epsilon +1) (2
   \epsilon -3) (4 \epsilon -1)}\nonumber\\
   &&-\frac{(\epsilon -1) \left(3
   \epsilon ^4+16 \epsilon ^3+31 \epsilon ^2+2 \epsilon -10\right)
   \delta^2}{4 (1-2 \epsilon )^2 \epsilon ^2 (\epsilon +1)
   (\epsilon +2) (2 \epsilon -3) (4 \epsilon
   -1)}\nonumber\\
   &&-\frac{(\epsilon -1)
   \left(3 \epsilon ^5+27 \epsilon ^4+32 \epsilon ^3-26 \epsilon
   ^2-24 \epsilon +16\right) \delta^3}{2 (1-2 \epsilon )^2
   \epsilon ^2 (\epsilon +1) (\epsilon +2) (\epsilon +3) (2 \epsilon
   -3) (4 \epsilon -3) (4 \epsilon -1)}\Bigg]\nonumber\\
   &&-\frac{\left(2 \epsilon ^7+8
   \epsilon ^6-45 \epsilon ^5+91 \epsilon ^4-113 \epsilon ^3+54
   \epsilon ^2+3 \epsilon -6\right) \delta}{4 (1-2 \epsilon )^2
   (\epsilon -1) \epsilon ^3 (\epsilon +1) (2 \epsilon -3) (4
   \epsilon -1)}\nonumber\\
   &&+\frac{\left(6 \epsilon ^9+8 \epsilon ^8-31 \epsilon ^7-137
   \epsilon ^6+238 \epsilon ^5-23 \epsilon ^4-157 \epsilon ^3+230
   \epsilon ^2-134 \epsilon +24\right) \delta^2}{8 (1-2 \epsilon
   )^2 (\epsilon -1)^2 \epsilon ^3 (\epsilon +1) (\epsilon +2) (2
   \epsilon -3) (4 \epsilon -1)}\nonumber\\
   &&+\frac{ \delta^3}{8 (1-2 \epsilon )^2 (\epsilon -3)   (\epsilon -2) (\epsilon -1)^2 \epsilon ^2 (\epsilon +2) (\epsilon   +3) (2 \epsilon -3) (4 \epsilon -3) (4 \epsilon   -1)}\nonumber\\
   &&\times\left(12 \epsilon ^{10}+16 \epsilon ^9-636 \epsilon ^8+2247
   \epsilon ^7-2751 \epsilon ^6-2183 \epsilon ^5\right.\nonumber\\
   &&\left.+18007 \epsilon
   ^4-35676 \epsilon ^3+31832 \epsilon ^2-13052 \epsilon
   +1992\right)\nonumber
      \eea
   \bea\phantom{\hat{\eta}_{qg}^{(4;3)}(\delta;\eps)=}
   &&+\frac{N_f}{N} \Bigg[\frac{\left(\epsilon
   ^3+7 \epsilon ^2-8 \epsilon +2\right) \delta}{2 (1-2 \epsilon
   )^2 \epsilon ^2 \left(8 \epsilon ^2-14 \epsilon
   +3\right)}\nonumber\\
   &&+\frac{\left(-3 \epsilon ^4+2 \epsilon ^3+19 \epsilon ^2-6
   \epsilon -2\right) \delta^2}{4 (1-2 \epsilon )^2 (\epsilon -2)
   \epsilon ^2 (2 \epsilon -3) (4 \epsilon -1)}\nonumber\\
   &&+\frac{\left(-7 \epsilon ^5+28
   \epsilon ^4+18 \epsilon ^3-89 \epsilon ^2+68 \epsilon -16\right)
   \delta^3}{2 (1-2 \epsilon )^2 (\epsilon -3) (\epsilon -2)
   \epsilon ^2 (2 \epsilon -3) (4 \epsilon -3) (4 \epsilon
   -1)}\Bigg]\nonumber\\
   &&+\frac{1}{N^2}\Bigg[-\frac{\left(2 \epsilon ^5+13 \epsilon ^4-21
   \epsilon ^3+26 \epsilon ^2-18 \epsilon +4\right) \delta}{8
   (1-2 \epsilon )^2 (\epsilon -1) \epsilon ^3 (4 \epsilon
   -1)}\nonumber\\
   &&+\frac{\left(6 \epsilon ^6-7 \epsilon ^5-30 \epsilon ^4+27
   \epsilon ^3-40 \epsilon ^2+10 \epsilon +4\right) \delta^2}{16
   (1-2 \epsilon )^2 (\epsilon -2) (\epsilon -1) \epsilon ^3 (4
   \epsilon -1)}\nonumber\\
   &&+\frac{\left(14
   \epsilon ^7-63 \epsilon ^6+6 \epsilon ^5+140 \epsilon ^4-261
   \epsilon ^3+278 \epsilon ^2-152 \epsilon +32\right)
   \delta^3}{8 (1-2 \epsilon )^2 (\epsilon -3) (\epsilon -2)
   (\epsilon -1) \epsilon ^3 (4 \epsilon -3) (4 \epsilon
   -1)}\Bigg]\nonumber\\
      &&+\mathcal{O}(\delta^4)
\eea
\bea
\hat{\eta}_{qg}^{(4;2)}(\delta;\eps)=&&-N^2 \Bigg[\frac{\epsilon ^3+2
   \epsilon ^2-3 \epsilon +1}{2 \epsilon ^5-4 \epsilon
   ^6}\nonumber\\
   &&+\frac{\delta \left(-6 \epsilon ^4-5
   \epsilon ^3+33 \epsilon ^2-22 \epsilon +6\right)}{4 \epsilon ^4
   \left(8 \epsilon ^2-6 \epsilon +1\right)}\nonumber\\
   &&+\frac{\delta^2 \left(-6 \epsilon ^7-14 \epsilon
   ^6+28 \epsilon ^5+47 \epsilon ^4+8 \epsilon ^3-5 \epsilon ^2-10
   \epsilon +4\right)}{8 \epsilon ^5 \left(8 \epsilon ^3+2 \epsilon
   ^2-5 \epsilon +1\right)}\nonumber\\
   &&+\frac{\delta^3 \left(-24 \epsilon ^9-46 \epsilon ^8+206
   \epsilon ^7+287 \epsilon ^6-550 \epsilon ^5-373 \epsilon ^4+542
   \epsilon ^3+144 \epsilon ^2-186 \epsilon +48\right)}{24 (\epsilon
   -1) \epsilon ^4 (\epsilon +1) (2 \epsilon -1) (4 \epsilon -3) (4
   \epsilon -1)}\Bigg]\nonumber\\
   &&-\frac{\delta \left(2
   \epsilon ^3+3 \epsilon ^2-7 \epsilon +2\right)}{4 \epsilon ^3
   \left(8 \epsilon ^2-6 \epsilon +1\right)}\nonumber\\
   &&-\frac{\delta^2 \left(2 \epsilon ^4+2
   \epsilon ^3-11 \epsilon ^2+3 \epsilon +1\right)}{8 \epsilon ^3
   \left(8 \epsilon ^2-6 \epsilon +1\right)}\nonumber\\
   &&-\frac{\delta^3 \left(8 \epsilon ^6+10 \epsilon
   ^5-70 \epsilon ^4-41 \epsilon ^3+45 \epsilon ^2+28 \epsilon
   -16\right)}{24 \epsilon ^3 \left(32 \epsilon ^3-48 \epsilon ^2+22
   \epsilon -3\right)}\nonumber\\
      &&+\mathcal{O}(\delta^4)
\eea
\bea
\hat{\eta}_{qg}^{(4;3)}(\delta;\eps)=&&N^2   \Bigg[\frac{3 (\epsilon -2)^2 (\epsilon -1)}{16
   \epsilon ^5 \left(4 \epsilon ^2-8 \epsilon +3\right)^2}+\frac{\left(25 \epsilon ^4-98
   \epsilon ^3+125 \epsilon ^2-76 \epsilon +36\right) \delta}{8
   \epsilon ^4 (\epsilon +1) (4 \epsilon -1) \left(4 \epsilon ^2-8
   \epsilon +3\right)^2}\nonumber\\
   &&+\frac{ \delta^2}{96 (3-2 \epsilon )^2   (\epsilon -1) \epsilon ^5 (\epsilon +1)^2 (\epsilon +2) (2   \epsilon -1)^3 (4 \epsilon -1)}\nonumber\\
   &&\times\left(12 \epsilon
   ^{13}+120 \epsilon ^{12}+87 \epsilon ^{11}-2121 \epsilon
   ^{10}+2977 \epsilon ^9+2461 \epsilon ^8-8519 \epsilon ^7+11725
   \epsilon ^6\right.\nonumber\\
   &&\left.-15113 \epsilon ^5+9895 \epsilon ^4-900 \epsilon
   ^3-480 \epsilon ^2+144\right)\nonumber\\
   &&-\frac{ \delta^3}{24 (\epsilon -1) \epsilon ^4 (\epsilon   +1)^2 (\epsilon +2) (\epsilon +3) (2 \epsilon -3) (2 \epsilon   -1)^3 (4 \epsilon -3) (4 \epsilon -1)}\nonumber\\
   &&\times\left(18 \epsilon ^{13}+189 \epsilon ^{12}+438
   \epsilon ^{11}-681 \epsilon ^{10}-3133 \epsilon ^9+2448 \epsilon
   ^8+9364 \epsilon ^7-14461 \epsilon ^6\right.\nonumber\\
   &&\left.+3369 \epsilon ^5+8725
   \epsilon ^4-9576 \epsilon ^3+1668 \epsilon ^2+1152 \epsilon
   -576\right)\Bigg]\nonumber\\
    &&+N N_f \Bigg[-\frac{(\epsilon -2) (\epsilon -1)^2}{2 \epsilon ^4
   \left(4 \epsilon ^2-8 \epsilon
   +3\right)^2}+\frac{\left(-11 \epsilon ^4+32 \epsilon ^3-37
   \epsilon ^2+28 \epsilon -12\right) \delta}{3 \epsilon ^3
   (\epsilon +1) (4 \epsilon -1) \left(4 \epsilon ^2-8 \epsilon
   +3\right)^2}\nonumber\\
   &&+\frac{(\epsilon -1) \left(45
   \epsilon ^6+161 \epsilon ^5-271 \epsilon ^4-161 \epsilon ^3+82
   \epsilon ^2+12 \epsilon -24\right) \delta^2}{12 (3-2 \epsilon
   )^2 \epsilon ^4 (\epsilon +1) (\epsilon +2) (2 \epsilon -1)^3 (4
   \epsilon -1)}\nonumber\\
   &&-\frac{(\epsilon -1)
   \left(10 \epsilon ^5+163 \epsilon ^4+287 \epsilon ^3-4 \epsilon
   ^2-108 \epsilon +48\right) \delta^3}{3 \epsilon ^3 (\epsilon
   +1) (\epsilon +2) (\epsilon +3) (2 \epsilon -3) (2 \epsilon -1)^3
   (4 \epsilon -3) (4 \epsilon -1)}\Bigg]\nonumber\\
   &&+N_f^2 \Bigg[\frac{(\epsilon -1)^3}{3 \epsilon ^3 \left(4 \epsilon ^2-8
   \epsilon +3\right)^2}-\frac{2 \delta (\epsilon -1)^3}{3 (1-2 \epsilon
   )^2 (3-2 \epsilon )^2 \epsilon ^2 (4 \epsilon
   -1)}\nonumber\\
   &&+\frac{\delta^2 (\epsilon -2) (\epsilon
   -1)^4}{6 (3-2 \epsilon )^2 \epsilon ^3 (2 \epsilon -1)^3 (4
   \epsilon -1)}\Bigg]\nonumber\\
   &&+\frac{(\epsilon -1) \left(2 \epsilon ^3-5 \epsilon ^2+4
   \epsilon -4\right)}{8 (1-2 \epsilon )^2 \epsilon ^5 (2 \epsilon
   -3)}+\frac{\left(22 \epsilon ^5-53 \epsilon ^4+75 \epsilon ^3-82
   \epsilon ^2+44 \epsilon -24\right) \delta}{12 (1-2 \epsilon
   )^2 \epsilon ^4 (\epsilon +1) (2 \epsilon -3) (4 \epsilon
   -1)}\nonumber\\
   &&+\frac{ \delta^2}{48 \epsilon ^5 (\epsilon +1)
   (\epsilon +2) (2 \epsilon -3) (2 \epsilon -1)^3 (4 \epsilon
   -1)}\left(6 \epsilon ^{10}+81
   \epsilon ^9+201 \epsilon ^8\right.\nonumber\\
   &&\left.-682 \epsilon ^7-278 \epsilon ^6+683
   \epsilon ^5+\epsilon ^4+380 \epsilon ^3-104 \epsilon ^2-48
   \epsilon +48\right)\nonumber\\
   &&-\frac{ \delta^3}{24
   \epsilon ^4 (\epsilon +2) (\epsilon +3) (2 \epsilon -1)^3 (4
   \epsilon -3) (4 \epsilon -1)}\left(18 \epsilon ^9+219 \epsilon ^8+825 \epsilon ^7\right.\nonumber\\
   &&\left.+950
   \epsilon ^6-1192 \epsilon ^5-1668 \epsilon ^4+1276 \epsilon
   ^3-1256 \epsilon ^2+720 \epsilon -192\right)\nonumber\\
   &&+\frac{N_f}{N}\Bigg[   -\frac{\left(2
   \epsilon ^2-\epsilon +2\right) (\epsilon -1)^2}{6 (1-2 \epsilon
   )^2 \epsilon ^4 (2 \epsilon -3)}+\frac{\delta \left(2 \epsilon
   ^2-\epsilon +2\right) (\epsilon -1)^2}{3 (1-2 \epsilon )^2
   \epsilon ^3 \left(8 \epsilon ^2-14 \epsilon
   +3\right)}\nonumber\\
   &&-\frac{\delta^2 \left(2 \epsilon ^3-5 \epsilon ^2+4
   \epsilon -4\right) (\epsilon -1)^3}{12 \epsilon ^4 (2 \epsilon
   -1)^3 \left(8 \epsilon ^2-14 \epsilon +3\right)}\Bigg]\nonumber
      \eea
   \bea
   \phantom{\hat{\eta}_{gg}^{(5;1)}(\delta;\eps)=}
   &&+\frac{1}{N^2}\Bigg[\frac{(\epsilon -1) \left(-2 \epsilon ^2+\epsilon
   -2\right)^2}{48 (1-2 \epsilon )^2 \epsilon ^5}-\frac{(\epsilon -1) \left(-2 \epsilon ^2+\epsilon
   -2\right)^2 \delta}{24 (1-2 \epsilon )^2 \epsilon ^4 (4
   \epsilon -1)}\nonumber\\
   &&+\frac{\left(7 \epsilon
   ^8+15 \epsilon ^7+194 \epsilon ^6-308 \epsilon ^5+225 \epsilon
   ^4-139 \epsilon ^3+82 \epsilon ^2-36 \epsilon +8\right)
   \delta^2}{96 (\epsilon -1) \epsilon ^5 (2 \epsilon -1)^3 (4
   \epsilon -1)}\nonumber\\
   &&+\frac{\left(-3 \epsilon ^6-22 \epsilon
   ^5-12 \epsilon ^4+28 \epsilon ^3+2 \epsilon ^2-9 \epsilon
   +2\right) \delta^3}{8 (\epsilon -1) \epsilon ^2 (2 \epsilon
   -1)^3 (4 \epsilon -3) (4 \epsilon -1)}\Bigg]\nonumber\\
     &&+\mathcal{O}(\delta^4)
\eea
\bea
\hat{\eta}_{qg}^{(5;1)}(\delta;\eps)=&&N^2 \Bigg[\frac{(\epsilon -2) (\epsilon
   -1)}{4 \epsilon ^5 \left(4 \epsilon ^2-8 \epsilon
   +3\right)}\nonumber\\
      &&-\frac{\delta \left(6 \epsilon
   ^5+33 \epsilon ^4-140 \epsilon ^3+131 \epsilon ^2-38 \epsilon
   +24\right)}{12 \epsilon ^4 \left(20 \epsilon ^4-24 \epsilon ^3-21
   \epsilon ^2+20 \epsilon -3\right)}\nonumber\\
      &&+\frac{\delta^2
   \left(54 \epsilon ^9+33 \epsilon ^8-258 \epsilon ^7-920 \epsilon
   ^6+2135 \epsilon ^5-78 \epsilon ^4-1544 \epsilon ^3+440 \epsilon
   ^2+120 \epsilon -48\right)}{12 (\epsilon -2) \epsilon ^5
   (\epsilon +1) (\epsilon +2) (2 \epsilon -3) (2 \epsilon -1) (5
   \epsilon -2) (5 \epsilon -1)}\nonumber\\
      &&-\frac{\delta^3}{6 (\epsilon -3) (\epsilon -2) (\epsilon -1)   \epsilon ^4 (\epsilon +2) (\epsilon +3) (2 \epsilon -1) (5   \epsilon -3) (5 \epsilon -2) (5 \epsilon -1)}\nonumber\\
	&&\times \left(9 \epsilon ^{10}-105 \epsilon
   ^9+78 \epsilon ^8+1886 \epsilon ^7-5553 \epsilon ^6+929 \epsilon
   ^5+14430 \epsilon ^4\right.\nonumber\\
      &&\left.-22054 \epsilon ^3+13788 \epsilon ^2-4272
   \epsilon +576\right)\Bigg]\nonumber\\
      &&+N N_f \Bigg[-\frac{(\epsilon -1)^2}{3 \epsilon ^4 \left(4
   \epsilon ^2-8 \epsilon +3\right)}+\frac{\delta \left(\epsilon ^2-3 \epsilon
   +2\right)}{3 \epsilon ^3 \left(20 \epsilon ^3-44 \epsilon ^2+23
   \epsilon -3\right)}\nonumber\\
      &&+\frac{\delta^2 (\epsilon
   -1) \left(11 \epsilon ^3+4 \epsilon ^2-8 \epsilon +2\right)}{3
   \epsilon ^4 (2 \epsilon -3) (2 \epsilon -1) (5 \epsilon -2) (5
   \epsilon -1)}\Bigg]\nonumber\\
      &&+\frac{(\epsilon -1) \left(2 \epsilon
   ^2-\epsilon +2\right)}{12 \epsilon ^5 (2 \epsilon -1)}\nonumber\\
      &&+\frac{\delta
   \left(-5 \epsilon ^4-14 \epsilon ^3+15 \epsilon ^2+2 \epsilon
   -4\right)}{12 \epsilon ^4 \left(10 \epsilon ^3-17 \epsilon ^2+8
   \epsilon -1\right)}\nonumber\\
      &&+\frac{\delta^2 (\epsilon +1) \left(5 \epsilon ^6+25
   \epsilon ^5-179 \epsilon ^4+240 \epsilon ^3-160 \epsilon ^2+56
   \epsilon -8\right)}{12 (\epsilon -2) (\epsilon -1) \epsilon ^5 (2
   \epsilon -1) (5 \epsilon -2) (5 \epsilon -1)}\nonumber\\
      &&+\frac{\delta^3
   \left(-3 \epsilon ^7+63 \epsilon ^6-267 \epsilon ^5+296 \epsilon
   ^4+34 \epsilon ^3-209 \epsilon ^2+106 \epsilon -16\right)}{2
   (\epsilon -3) (\epsilon -2) (\epsilon -1) \epsilon ^3 (2 \epsilon
   -1) (5 \epsilon -3) (5 \epsilon -2) (5 \epsilon
   -1)}\nonumber\\
      &&+\mathcal{O}(\delta^4)
\eea
\bea
\hat{\eta}_{qg}^{(6;1)}(\delta;\eps)=&&N^2 \Bigg[\frac{\epsilon -1}{12 \epsilon ^5}+\frac{\delta}{6 \epsilon ^4 (6 \epsilon
   -1)}-\frac{\delta^2 (\epsilon +1)
   \left(3 \epsilon ^3+12 \epsilon ^2-11 \epsilon +2\right)}{24
   \epsilon ^5 \left(12 \epsilon ^2-8 \epsilon
   +1\right)}\nonumber\\
      &&+\frac{\delta^3 (\epsilon -1)}{4 \epsilon ^2 \left(12
   \epsilon ^2-8 \epsilon +1\right)}-\frac{\delta^4 (\epsilon -1)}{8 \epsilon ^2 \left(12 \epsilon
   ^2-8 \epsilon +1\right)}\Bigg]\nonumber\\
      &&+\mathcal{O}(\delta^4)
\eea

\subsection{$q\bar{q}$ initial state}
\bea
\hat{\eta}_{q\bar{q}}^{(2;1)}(\delta;\eps)=&&+\frac{N^2 \delta^3
   \left(2 \epsilon ^2+5 \epsilon -13\right)^2 (\epsilon -1)^2}{8
   \epsilon ^2 \left(4 \epsilon ^2-8 \epsilon
   +3\right)^3}\nonumber\\
   &&-\frac{N N_f \delta^3 \left(2 \epsilon ^2+5
   \epsilon -13\right) (\epsilon -1)^3}{\epsilon ^2 \left(4 \epsilon
   ^2-8 \epsilon +3\right)^3}\nonumber\\
   &&+\frac{2 N_f^2 \delta^3
   (\epsilon -1)^4}{\epsilon ^2 \left(4 \epsilon ^2-8 \epsilon
   +3\right)^3}\nonumber\\
   &&-\frac{\delta^3 \left(10 \epsilon ^3+17 \epsilon
   ^2-85 \epsilon +52\right) (\epsilon -1)^2}{4 (3-2 \epsilon )^2
   \epsilon ^3 (2 \epsilon -1)^3}\nonumber\\
  &&+\frac{N_f \delta^3 (5 \epsilon -4)
   (\epsilon -1)^3}{N (3-2 \epsilon )^2 \epsilon ^3 (2 \epsilon
   -1)^3}\nonumber\\
   &&+\frac{\delta^3 (4-5 \epsilon )^2 (\epsilon -1)^2}{8 N^2 \epsilon
   ^4 (2 \epsilon -3) (2 \epsilon -1)^3}\nonumber\\
      &&+\mathcal{O}(\delta^4)
\eea
\bea
\hat{\eta}_{q\bar{q}}^{(3;1)}(\delta;\eps)=&&N^2 \Bigg[
-\frac{\delta^2 (\epsilon +1) \left(2 \epsilon
   ^2+5 \epsilon -13\right)}{4 (1-2 \epsilon )^2 \epsilon  (2
   \epsilon -3) (3 \epsilon -2)}\nonumber\\
   &&+\frac{\delta^3 \left(-72 \epsilon ^8+2 \epsilon ^7+837
   \epsilon ^6-963 \epsilon ^5-2119 \epsilon ^4+5379 \epsilon
   ^3-4674 \epsilon ^2+1914 \epsilon -312\right)}{36 (1-2 \epsilon
   )^2 (\epsilon -1) \epsilon ^3 (2 \epsilon -3) (3 \epsilon -2) (3
   \epsilon -1)} \Bigg]\nonumber\\
      &&+\frac{1}{N^2}\Bigg[\frac{\delta^2 \left(5 \epsilon ^2-9
   \epsilon +4\right)}{4 (1-2 \epsilon )^2 \epsilon ^2 (3 \epsilon
   -2)}\nonumber\\
   &&+\frac{\delta^3
   \left(9 \epsilon ^8+15 \epsilon ^7-58 \epsilon ^6-99 \epsilon
   ^5+241 \epsilon ^4-76 \epsilon ^3-103 \epsilon ^2+80 \epsilon
   -16\right)}{6 (1-2 \epsilon )^2 \epsilon ^4 \left(9 \epsilon ^3-7
   \epsilon +2\right)}\Bigg]\nonumber\\
      &&+N N_f \Bigg[\frac{\delta^2 \left(\epsilon
   ^2-1\right)}{(1-2 \epsilon )^2 \epsilon  (2 \epsilon -3) (3
   \epsilon -2)}-\frac{2 \delta^3 \left(3 \epsilon ^4+7 \epsilon
   ^3-21 \epsilon ^2+15 \epsilon -4\right)}{3 \epsilon ^3 \left(36
   \epsilon ^4-108 \epsilon ^3+107 \epsilon ^2-43 \epsilon
   +6\right)}\Bigg]\nonumber
      \eea
   \bea\phantom{\hat{\eta}_{qg}^{(4;3)}(\delta;\eps)=}
      &&+\frac{N_f}{N} \Bigg[\frac{\delta^2
   (\epsilon -1)^2}{(1-2 \epsilon )^2 \epsilon  (2 \epsilon -3) (3
   \epsilon -2)}-\frac{2 \delta^3 \left(3 \epsilon ^5+10 \epsilon
   ^4-20 \epsilon ^3+11 \epsilon -4\right)}{3 \epsilon ^3 \left(36
   \epsilon ^5-72 \epsilon ^4-\epsilon ^3+64 \epsilon ^2-37 \epsilon
   +6\right)}\Bigg]\nonumber\\
&&+\frac{\delta^2 \left(-2 \epsilon
   ^4+7 \epsilon ^3+5 \epsilon ^2-24 \epsilon +12\right)}{4 (1-2
   \epsilon )^2 \epsilon ^2 \left(6 \epsilon ^2-13 \epsilon
   +6\right)}\nonumber\\
      &&+\frac{\delta^3 }{36 (1-2 \epsilon   )^2 (\epsilon -1) \epsilon ^4 (\epsilon +1) (2 \epsilon -3) (3   \epsilon -2) (3 \epsilon -1)}\nonumber\\
      &&\times\left(36 \epsilon
   ^{10}+200 \epsilon ^9+31 \epsilon ^8-1710 \epsilon ^7+2162
   \epsilon ^6\right.\nonumber\\
   &&\left.-1094 \epsilon ^5-387 \epsilon ^4+2730 \epsilon
   ^3-3276 \epsilon ^2+1608 \epsilon -288\right)\nonumber\\
&&+\mathcal{O}(\delta^4)
\eea
\bea
\hat{\eta}_{q\bar{q}}^{(4;1)}(\delta;\eps)=&&+N^2 \Bigg[\frac{\delta \epsilon  (\epsilon +1)^2}{8 (1-2
   \epsilon )^2 (\epsilon -1)^2 (4 \epsilon
   -1)}\nonumber\\
      &&-\frac{\delta^2 \left(\epsilon ^6+3 \epsilon ^5-16
   \epsilon ^4+4 \epsilon ^3+11 \epsilon ^2-9 \epsilon +2\right)}{4
   (1-2 \epsilon )^2 (\epsilon -2) (\epsilon -1)^2 \epsilon ^2 (4
   \epsilon -1)}\nonumber\\
      &&+\frac{\delta^3
   \left(3 \epsilon ^8-23 \epsilon ^7+10 \epsilon ^6+348 \epsilon
   ^5-1126 \epsilon ^4+1442 \epsilon ^3-956 \epsilon ^2+328 \epsilon
   -48\right)}{4 (\epsilon -3) (\epsilon -2)^2 (\epsilon -1)
   \epsilon ^4 (2 \epsilon -1) (4 \epsilon -3) (4 \epsilon
   -1)}\Bigg]\nonumber\\
      &&+\frac{\delta
   \epsilon  (\epsilon +1)}{4 (1-2 \epsilon )^2 (\epsilon -1) (4
   \epsilon -1)}\nonumber\\
  &&+\frac{\delta^2 \left(-2 \epsilon ^4-7 \epsilon ^3+33
   \epsilon ^2-21 \epsilon +4\right)}{4 (1-2 \epsilon )^2 (\epsilon
   -2) (\epsilon -1) \epsilon  (4 \epsilon -1)}\nonumber\\
    	&&+\frac{\delta^3 \left(11 \epsilon ^9+\epsilon
   ^8-266 \epsilon ^7+395 \epsilon ^6+799 \epsilon ^5-1548 \epsilon
   ^4+400 \epsilon ^3+600 \epsilon ^2-464 \epsilon +96\right)}{4
   (\epsilon -3) (\epsilon -2) (\epsilon -1) \epsilon ^4 (\epsilon
   +1) (\epsilon +2) (2 \epsilon -1) (4 \epsilon -3) (4 \epsilon
   -1)}\nonumber\\
      &&+\frac{1}{N^2}\Bigg[\frac{\delta \epsilon }{8 (1-2
   \epsilon )^2 (4 \epsilon -1)}\nonumber\\
      &&+\frac{\delta^2 \left(-\epsilon ^4-7 \epsilon ^3+\epsilon
   ^2+3 \epsilon -1\right)}{4 (1-2 \epsilon )^2 \epsilon ^2
   (\epsilon +1) (4 \epsilon -1)}\nonumber\\
      &&+\frac{\delta^3 \left(4 \epsilon ^7+32 \epsilon ^6+39
   \epsilon ^5-32 \epsilon ^4-33 \epsilon ^3+19 \epsilon ^2+8
   \epsilon -4\right)}{2 \epsilon ^4 (\epsilon +1)^2 (\epsilon +2)
   (2 \epsilon -1) (4 \epsilon -3) (4 \epsilon
   -1)}\Bigg]\nonumber\\
      &&+\mathcal{O}(\delta^4)
\eea
\bea
\hat{\eta}_{q\bar{q}}^{(4;2)}(\delta;\eps)=&&-\frac{1}{N^2}\Bigg[\frac{\delta^2 (5 \epsilon -4)}{4 \epsilon ^2 \left(8
   \epsilon ^2-6 \epsilon +1\right)}-\frac{\delta^3 \left(-8
   \epsilon ^7-18 \epsilon ^6+34 \epsilon ^5+21 \epsilon ^4-46
   \epsilon ^3+\epsilon ^2+22 \epsilon -8\right)}{4 \epsilon ^4
   (\epsilon +1) (2 \epsilon -1) (4 \epsilon -3) (4 \epsilon
   -1)}\Bigg]\nonumber\\
      &&-\frac{N_f}{N} \Bigg[\frac{\delta^2 (\epsilon
   -1)}{\epsilon  \left(16 \epsilon ^3-36 \epsilon ^2+20 \epsilon
   -3\right)}+\frac{\delta^3 \left(-4
   \epsilon ^4+3 \epsilon ^3+6 \epsilon ^2-7 \epsilon
   +2\right)}{\epsilon ^3 \left(64 \epsilon ^4-192 \epsilon ^3+188
   \epsilon ^2-72 \epsilon +9\right)}\Bigg]\nonumber\\
      &&-\frac{\delta^2 \left(-2 \epsilon ^2-5 \epsilon
   +13\right)}{4 \epsilon  \left(16 \epsilon ^3-36 \epsilon ^2+20
   \epsilon -3\right)}\nonumber\\
      &&-\frac{\delta^3 \left(-32 \epsilon ^6+8
   \epsilon ^5+280 \epsilon ^4-384 \epsilon ^3-48 \epsilon ^2+225
   \epsilon -78\right)}{12 \epsilon ^3 \left(64 \epsilon ^4-192
   \epsilon ^3+188 \epsilon ^2-72 \epsilon
   +9\right)}\nonumber\\
      &&+\mathcal{O}(\delta^4)
\eea
\bea
\hat{\eta}_{q\bar{q}}^{(4;3)}(\delta;\eps)=&&N^2 \Bigg[\frac{\delta (\epsilon +1)^2}{8 (1-2 \epsilon
   )^2 (\epsilon -1) (4 \epsilon -1)}\nonumber\\
   &&-\frac{\delta^2
   (\epsilon +1) \left(3 \epsilon ^4+7 \epsilon ^3-14 \epsilon ^2+7
   \epsilon -1\right)}{4 (\epsilon -1) \epsilon ^2 (2 \epsilon -1)^3
   (4 \epsilon -1)}\nonumber\\
  &&+\frac{\delta^3 \left(6 \epsilon
   ^9+20 \epsilon ^8-99 \epsilon ^7-89 \epsilon ^6+686 \epsilon
   ^5-1070 \epsilon ^4+816 \epsilon ^3-346 \epsilon ^2+80 \epsilon
   -8\right)}{4 (\epsilon -2) (\epsilon -1) \epsilon ^4 (2 \epsilon
   -1)^3 (4 \epsilon -3) (4 \epsilon -1)}\Bigg]\nonumber\\
      &&+\frac{\delta (\epsilon +1)}{4 (1-2 \epsilon )^2 (4
   \epsilon -1)}\nonumber\\
   &&-\frac{\delta^2 \left(6 \epsilon ^3+19 \epsilon ^2-14
   \epsilon +2\right)}{4 \epsilon  (2 \epsilon -1)^3 (4 \epsilon
   -1)}\nonumber\\
	&&+\frac{\delta^3
   \left(12 \epsilon ^9+119 \epsilon ^8+277 \epsilon ^7-214 \epsilon
   ^6-548 \epsilon ^5+534 \epsilon ^4+76 \epsilon ^3-260 \epsilon
   ^2+112 \epsilon -16\right)}{4 \epsilon ^4 (\epsilon +1) (\epsilon
   +2) (2 \epsilon -1)^3 (4 \epsilon -3) (4 \epsilon
   -1)}\nonumber\\
  &&+\frac{1}{N^2}\Bigg[\frac{\delta (\epsilon -1)}{8 (1-2 \epsilon )^2 (4
   \epsilon -1)}\nonumber\\
   &&+\frac{\delta^2 \left(-3 \epsilon ^5-9
   \epsilon ^4+12 \epsilon ^3+5 \epsilon ^2-6 \epsilon +1\right)}{4
   \epsilon ^2 (\epsilon +1) (2 \epsilon -1)^3 (4 \epsilon
   -1)}\nonumber\\
   &&+\frac{\delta^3 (\epsilon -1) \left(3 \epsilon ^8+39
   \epsilon ^7+156 \epsilon ^6+177 \epsilon ^5-59 \epsilon ^4-116
   \epsilon ^3+26 \epsilon ^2+24 \epsilon -8\right)}{4 \epsilon ^4
   (\epsilon +2) (4 \epsilon -3) (4 \epsilon -1) \left(2 \epsilon
   ^2+\epsilon -1\right)^2}\Bigg]\nonumber\\
   &&+\mathcal{O}(\delta^4)
\eea
\bea
\hat{\eta}_{q\bar{q}}^{(5;1)}(\delta;\eps)=
      &&\frac{\delta (\epsilon +1)}{-20
   \epsilon ^3+34 \epsilon ^2-16 \epsilon +2}\nonumber\\
      &&+\frac{\delta^2 \left(11 \epsilon ^5+4 \epsilon ^4-96
   \epsilon ^3+100 \epsilon ^2-39 \epsilon +6\right)}{2 \epsilon ^2
   \left(50 \epsilon ^5-205 \epsilon ^4+284 \epsilon ^3-169 \epsilon
   ^2+44 \epsilon -4\right)}\nonumber\\
      &&+\frac{\delta^3 \left(-9 \epsilon ^9+9 \epsilon
   ^8+202 \epsilon ^7-538 \epsilon ^6+29 \epsilon ^5+1110 \epsilon
   ^4-1414 \epsilon ^3+793 \epsilon ^2-218 \epsilon
   +24\right)}{(\epsilon -3) (\epsilon -2) (\epsilon -1) \epsilon ^4
   (2 \epsilon -1) (5 \epsilon -3) (5 \epsilon -2) (5 \epsilon
   -1)}\nonumber\\
   &&+\frac{1}{N^2}\Bigg[-\frac{\delta}{20 \epsilon ^2-14 \epsilon
   +2}\nonumber\\
&&+\frac{\delta^2 \left(11 \epsilon ^4+40 \epsilon
   ^3-6 \epsilon ^2-18 \epsilon +5\right)}{2 \epsilon ^2 \left(50
   \epsilon ^4-5 \epsilon ^3-36 \epsilon ^2+17 \epsilon
   -2\right)}\nonumber\\
	&&-\frac{\delta^3 \left(9 \epsilon ^8+87 \epsilon ^7+196
   \epsilon ^6-83 \epsilon ^5-238 \epsilon ^4+120 \epsilon ^3+55
   \epsilon ^2-46 \epsilon +8\right)}{\epsilon ^4 (\epsilon +1)
   (\epsilon +2) (2 \epsilon -1) (5 \epsilon -3) (5 \epsilon -2) (5
   \epsilon -1)}  \Bigg]\nonumber\\
      &&+\mathcal{O}(\delta^4)
\eea
\bea
\hat{\eta}_{q\bar{q}}^{(6;1)}(\delta;\eps)=&&\frac{1}{N^2}\Bigg[\frac{\delta}{24 \epsilon ^2-16 \epsilon +2}+\frac{\delta^2 \left(-2 \epsilon ^2-2 \epsilon
   +1\right)}{24 \epsilon ^4-16 \epsilon ^3+2 \epsilon
   ^2}+\frac{\delta^3 \left(3 \epsilon ^4+6 \epsilon ^3-3 \epsilon
   +1\right)}{72 \epsilon ^6-48 \epsilon ^5+6 \epsilon
   ^4}\Bigg]\nonumber\\
\eea